\documentclass[aps,prd,preprint,eqsecnum,nofootinbib,groupedaddress,floatfix]{revtex4}
\usepackage{graphicx}
\usepackage{dcolumn}
\usepackage{bm}
\usepackage{amsmath}
\usepackage{amssymb}
\usepackage{amsfonts}
\usepackage[dvips]{color}
\newcommand{\half}{\frac{1}{2}}

\newcommand{\der}{\partial}

\newcommand{\Tr}{\mbox{\rm Tr}}

\newcommand{\pd}[2]{\frac{\partial #1}{\partial #2}}

\newcommand{\bfp}{\bm{p}}

\newcommand{\bfx}{\bm{x}}

\newcommand{\bfJ}{\bm{J}}

\newcommand{\bfP}{\bm{P}}

\newcommand{\bfX}{\bm{X}}

\newcommand{\bftau}{\bm{\tau}}
\begin{document}

\preprint{KYUSHU-HET-75}
\preprint{SAGA-HE-216-04}

\title{Effective Theory Approach to the Skyrme model \\
and Application to Pentaquarks}
\author{Koji Harada}
\email{koji1scp@mbox.nc.kyushu-u.ac.jp}
\author{Yohei Mitsunari}
\email{mitsunari@higgs.phys.kyushu-u.ac.jp}
\affiliation{Department of Physics, Kyushu University\\
Fukuoka 810-8581 Japan}
\author{Nao-aki Yamashita}
\email{naoaki@higgs.phys.kyushu-u.ac.jp}
\affiliation{Department of Physics, Saga University\\
Saga 840-8502 Japan}

\date{\today}

\begin{abstract}
 The Skyrme model is reconsidered from an effective theory point of
 view. From the most general chiral Lagrangian up to including terms of
 order $p^4$, $N_c$ and $\delta m^2$ ($\delta m\equiv m_s-m$), new
 interactions, which have never been considered before, appear upon
 collective coordinate quantization. We obtain the parameter set best
 fitted to the observed low-lying baryon masses, by performing the
 second order perturbative calculations with respect to $\delta m$.  We
 calculate the masses and the decay widths of the other members of
 (mainly) anti-decuplet pentaquark states. The formula for the decay
 widths is reconsidered and its baryon mass dependence is clarified.
\end{abstract}


\maketitle

\section{Introduction}

Evidence of a new baryonic resonance state, called $\Theta^{+}(1540)$,
has been claimed recently by Nakano
\textit{et.~al.}\cite{Nakano:2003qx}, with $S=+1$ and a very narrow
width $\Gamma \le 15$ MeV. Several other experimental groups have
confirmed the existence\cite{Barmin:2003vv, Stepanyan:2003qr,
Barth:2003es}. It appears in the recent version of \textit{Reviews of
Particle Physics}\cite{Eidelman:2004wy} with the *** rating, though its
parity has not been established.  Evidences of less certain exotic
pentaquark states, $\phi(1860)$\cite{Alt:2003vb} and
$\Theta_c^{+}$\cite{Aktas:2004qf, Chekanov:2004kn} have also been
claimed.  The discovery of the pentaquarks is expected to lead us to a
deeper understanding of strong interactions at low energies. In reality,
it stimulates new ideas and reconsideration of old theories and
experimental data.

The discovery was motivated by a paper by Diakonov, Petrov and
Polyakov\cite{Diakonov:1997mm}.  They predicted the masses and the
widths of the anti-decuplet, of which $\Theta^{+}$ is presumed to be a
member, within the framework of the ``chiral quark-soliton
model''($\chi$QSM)\footnote{The $\chi$QSM has its own scenario based on
instantons. For our purpose, however, it is useful to regard it as a
version of the Skyrme model\cite{Skyrme:1961vq} with specific symmetry
breaking interactions. See \cite{Diakonov:2004ie} for a recent review on
the background of the $\chi$QSM.}. See Ref.~\cite{Diakonov:2000pa} for a
review of the $\chi$QSM. The chiral quark-soliton model prediction is
reexamined in Refs.~\cite{Praszalowicz:2003ik, Ellis:2004uz}. See also
Refs.~\cite{Manohar:1984ys, Chemtob:1985ar, Praszalowicz:1987em} for
earlier papers on the anti-decuplet in the Skyrme model.

Jaffe and Wilczek\cite{Jaffe:2003sg}, on the other hand, proposed a
quark model picture of pentaquarks based on diquark correlation. See
Ref.~\cite{Jaffe:2004ph} for a review of this approach and other
interesting aspects.

Since Witten pointed out that baryons may be considered as
solitons\cite{Witten:1979kh} in the large-$N_c$
limit\cite{'tHooft:1973jz}, and showed that the soliton (``Skyrmion'')
has the right spin-statistics\cite{Witten:1983tx} thanks to the
Wess-Zumino-Witten (WZW) term\cite{Wess:1971yu, Witten:1983tw}, much
effort has been done to explore the consequences. But the results do not
look very successful. In order to fit the results to the observed values
of masses, coupling constants, such as the pion decay constant, which
appear in the chiral Lagrangian must be very different from the
experimental values. For example, in the $SU(3)$ Skyrme model, the pion
decay constant becomes typically one third of the experimental value to
reproduce the correct mass splitting\cite{Praszalowicz:1985bt}. It also
predicted an anti-decuplet\cite{Manohar:1984ys, Chemtob:1985ar,
Praszalowicz:1987em}, which many did not believe to exist at that time.

But $\Theta^{+}$ has been discovered! It is time to take a serious look
at the Skyrme model again.

One of the most important aspects of $\Theta^{+}$ is its
narrowness. Several analyses of older data indicate that the width may
be less than 1 MeV\cite{Nussinov:2003ex, Arndt:2003xz, Workman:2004yd,
Cahn:2003wq}. Interestingly, the Skyrme model is believed to be capable
to explain the narrowness. It is claimed that the width becomes very
narrow due to a strong cancellation\cite{Praszalowicz:2003tc}.

A natural question is: Is this a general result of the Skyrme model, or
a ``model-dependent'' one? What is the most general Skyrme model? The
Skyrme model is nothing but a model. But, as Witten emphasized, the
soliton picture of baryons is a general consequence of the large-$N_c$
limit of QCD.  Assuming that the large-$N_c$ QCD bears a close
resemblance to the real QCD, we may consider an effective theory (not
just a model) of baryons based on the soliton picture, which may be
called as the ``Skyrme-Witten large-$N_c$ effective theory.''

Our key observation is that the Skyrme model conventionally starts with
the particular chiral Lagrangian, which consists of the kinetic term,
the Skyrme term (which stabilizes the soliton), the WZW term and the
leading $SU(3)$ breaking term,
\begin{eqnarray}
 S_{Skyrme}&=&\frac{F_\pi^2}{16}\int d^4x 
  \Tr\left(\der_\mu U \der^\mu U^\dagger\right)
  +\frac{1}{32e^2}\int d^4x
  \Tr\left(
      \left[
       U^\dagger \der_\mu U,U^\dagger\der_\nu U
      \right]^2
     \right)
  + N_c\Gamma[U] \nonumber \\
 &&{}+\frac{F_\pi^2 B}{8}\int d^4x
  \Tr\left({\cal M}^\dagger U+{\cal M}U^\dagger\right),
\end{eqnarray}
where ${\cal M}$ is the quark mass matrix\footnote{We do not consider
the isospin breaking in this paper.}
\begin{equation}
 {\cal M}=\left(
 \begin{array}{ccc}
  m&0&0 \\
  0&m&0 \\
  0&0&m_s
 \end{array}\right).
\end{equation}
and
\begin{equation}
 \Gamma[U]=\frac{i}{240\pi^2}\int_Q d\Sigma^{ijklm}\Tr
\bigg[
 \left(\der_i U\right)U^\dagger
 \left(\der_j U\right)U^\dagger
 \left(\der_k U\right)U^\dagger
 \left(\der_l U\right)U^\dagger
 \left(\der_m U\right)U^\dagger
\bigg]
\end{equation}
is the WZW term. The chiral perturbation theory\cite{Gasser:1983yg,
Gasser:1984gg} ($\chi$PT) is however an effective field theory with
infinitely many operators\footnote{Long time ago, Kindo and
Yukawa\cite{Kindo:1987dc} considered the Skyrme model in the framework
of the $\chi$PT context, but their work did not seem to attract much
attention at that time.}. We should keep it in mind that there are
(infinitely) many other terms and the expansion must be systematic.

In this paper, we explore such an ``effective theory'' approach, i.e.,
an approach based on a systematic expansion of the operators based on
the symmetry and power-counting. The parameters are determined by
fitting the results to the experimental values. Once the parameters are
fixed, the rest are the predictions. As an application of our approach,
we give the predictions to the masses and the decay widths of the other
(mainly) anti-decuplet baryons.

Generalizations of the Skyrme model have been considered by several
authors, by including vector mesons\cite{Park:1991bv, Park:1991fb,
Weigel:1995cz} or the radial modes\cite{Weigel:1998vt,
Weigel:2004px}. Inclusion of heavier vector mesons may correspond to the
inclusion of higher order terms in the usual pseudoscalar Lagrangian,
while the inclusion of the radial modes is beyond the scope of the
present paper.

The paper is organized as follows: In Sec.~\ref{Sec:Ham}, we derive the
collective coordinate quantized Hamiltonian based on the effective
theory approach. The Hamiltonian contains several $SU(3)$ flavor
symmetry breaking interactions which have never been considered in the
literature. In Sec.~\ref{Sec:Mass}, we consider the eigenstates of the
Hamiltonian, with the eigenvalues being the baryon masses. If the
symmetry breaking terms were absent, the eigenstates form the flavor
$SU(3)$ representations. The symmetry breaking terms mix the
representations. We calculate the masses up to the second order in the
perturbation theory with respect to the symmetry breaking parameter
$\delta m\equiv m_s-m$. The masses are represented as functions of the
parameters of the theory. In Sec.~\ref{Sec:Num}, we numerically fit the
calculated masses to the experimental values to determine the
parameters. After determining the parameters, we obtain the masses of
mainly anti-decuplet baryons, ${\rm N}^{\prime}$, an excited nucleon
with $I(J^P)=\half(\half^+)$, and $\Sigma^{\prime}$ with $1
(\half^+)$. The decay widths are calculated in Sec.~\ref{Sec:Widths},
after reconsidering the derivation of the formula for the widths in the
collective coordinate quantization. Unfortunately, our calculation of
the widths suffers from large ambiguity. We summarize our results and
discuss some issues in Sec.~\ref{Sec:Sum}. The notations, conventions,
and the derivations of several useful mathematical formulae are
delegated to Appendix~\ref{Sec:Math}.  Various matrix elements used in
the calculations are summarized in Appendix~\ref{Sec:Tables}. A lot of
tables are given there. The reason why we present them is that most of
them have never appeared in the literature and it requires much labor to
calculate them. The ``traditional'' approach to the Skyrme model is
reconsidered from the new perspective in Appendix~\ref{Sec:Trad}. The
results obtained in Appendix~\ref{Sec:Trad} may be viewed as an evidence
that our basic strategy is right. Finally in Appendix~\ref{Sec:DPP}, we
perform a parallel analysis with the $\chi$QSM symmetry breaking terms,
\begin{equation}
 H^{DPP}_1
  =\alpha D_{88}^{(8)}+\beta Y +\frac{\gamma}{\sqrt{3}}D_{8i}^{(8)}F^i,
  \label{DPP}
\end{equation}
without considering how these interactions are derived. The parameters
$\alpha$, $\beta$, and $\gamma$ are determined in a similar way, and the
decay widths are calculated too.  

\section{Collective Hamiltonian}
\label{Sec:Ham}

\subsection{Chiral Lagrangian up to including ${\cal O}(p^4)$ and ${\cal
O}(N_c)$}

Effective field theories are not just models. They represent very
general principles such as analyticity, unitarity, cluster decomposition
of quantum field theory and the symmetries of the
systems\cite{Weinberg:1978kz}. The chiral perturbation theory
($\chi$PT), for example, represents the low-energy behavior of QCD (at
least) in the meson sector.

Although baryons in the large-$N_c$ limit behave like solitons, it is
not very clear in what theory they appear. A natural candidate is the
$\chi$PT, because, as emphasized above, it is a very general framework
in which the low-energy QCD is represented. It seems that if baryons may
appear as solitons, they should appear in the $\chi$PT, with infinitely
many operators.

At low-energies, only a few operators are important in the $\chi$PT
Lagrangian. We may systematically expand the results with respect to the
typical energy/momentum scale, $p$. This is the usual power counting in
the $\chi$PT, and we assume it is the case even in the soliton (i.e.,
baryon) sector.

To summarize, a general Skyrme-Witten soliton theory may be a systematic
expansion of the soliton sector of the $\chi$PT, with respect to $N_c$
and $p$. Therefore, our starting point is the $SU(3)$ $\chi$PT
action (without external gauge fields) up to including ${\cal
O}(p^4)$,
\begin{equation}
 S^{\chi{\rm PT}}=S_0+S_1+{\cal O}(p^6),
\end{equation}
\begin{eqnarray}
 S_0&=&\frac{F_0^2}{16}\int d^4x 
  \Tr\left(\der_\mu U \der^\mu U^\dagger\right) 
  +\frac{F_0^2 B_0}{8}\int d^4x
  \Tr\left({\cal M}^\dagger U+{\cal M}U^\dagger\right) 
  +N_c\Gamma[U], \\
 S_1&=&L_1\int d^4x
  \left[\Tr\left(\der_\mu U\der^\mu U^\dagger\right)\right]^2 
  +L_2 \int d^4x
  \Tr\left(\der_\mu U^\dagger \der_\nu U\right)
  \Tr\left(\der^\mu U^\dagger \der^\nu U\right) \nonumber \\
 &&{}+L_3\int d^4x
  \Tr\left(\der_\mu U^\dagger \der^\mu U 
      \der_\nu U^\dagger \der^\nu U\right) 
  +L_4B_0\int d^4x
  \Tr\left(\der_\mu U^\dagger \der^\mu U\right)
  \Tr\left({\cal M}^\dagger U+{\cal M}U^\dagger\right) \nonumber \\
 &&{}+L_5B_0\int d^4x
  \Tr\left[\der_\mu U^\dagger \der^\mu U
      \left({\cal M}^\dagger U+U^\dagger {\cal M}\right)\right]
      +L_6B_0^2\int d^4x
      \left[\Tr\left({\cal M}^\dagger U+{\cal M}U^\dagger\right)\right]^2
      \nonumber \\
 &&{}+L_7B_0^2\int d^4x
  \left[\Tr\left({\cal M}^\dagger U-{\cal M}U^\dagger\right)\right]^2
  +L_8B_0^2\int d^4x
  \Tr\left({\cal M}^\dagger U{\cal M}^\dagger U
      +{\cal M}U^\dagger{\cal M}U^\dagger\right), \nonumber \\
\end{eqnarray}
where $L_i \ (i=1,\cdots, 8)$ are dimensionless constants. The
definition of these parameters is the same as that of
Ref.~\cite{Gasser:1984gg} except for $F_0$. Our normalization of $F_0$
is more popular in the Skyrme model literature. The $N_c$ dependence of
them is known\cite{Gasser:1984gg, Peris:1994dh},
\begin{eqnarray}
 B_0,2L_1-L_2,L_4,L_6,L_7 &\cdots& {\cal O}(N_c^0), \\
 F_0^2,L_2,L_3,L_5,L_8 &\cdots& {\cal O}(N_c^1).
\end{eqnarray}
In the following, we keep only the operators whose coefficients are of
order $N_c$. Experimentally, these constants are not very accurately
known. We further assume that the constants $L_1$, $L_2$ and $L_3$ have
the ratio,
\begin{equation}
 L_1:L_2:L_3=1:2:-6,
\end{equation}
which is consistent with the experimental values, $L_1=0.4\pm 0.3$,
$2L_1-L_2=-0.6\pm 0.5$, and $L_3=-3.5\pm 1.1$ (times
$10^{-3}$)\cite{Pich:1995bw}.  Note that vector meson dominance also
implies this ratio\cite{Ecker:1988te}. This assumption simplifies the
analysis greatly, due to the identity,
\begin{equation}
 \Tr\left(ABAB\right)
  =-2\Tr\left(A^2B^2\right)+\half\Tr\left(A^2\right)\Tr\left(B^2\right)
  +\left(\Tr\left(AB\right)\right)^2,
\end{equation}
which holds for any $3\times3$ traceless matrices $A$ and $B$. By using
it, the $L_1$, $L_2$, and $L_3$ terms are made up to a single
expression,
\begin{equation}
 \frac{1}{32e^2}\Tr
  \left(
   \left[
    U^\dagger \der_\mu U,U^\dagger\der_\nu U
   \right]^2
  \right),
\end{equation}
where we introduced $L_2=1/(16e^2)$. This term is nothing but the Skyrme
term\footnote{We do not know who first noticed this fact. Probably this
is widely known. We learned it from Ref.~\cite{Diakonov:2000pa}.}.  (If
we would not assume these exact ratios among $L_1$, $L_2$ and $L_3$, we
would have extra terms which lead to the terms quartic in time
derivatives of the collective coordinates, and would make the
quantization a bit harder. Because we consider the case in which the
``rotation'' is slow enough, such terms could be ignored.)

We thus end up with the action,
\begin{eqnarray}
 S[U]&=&\frac{F_0^2}{16}\int d^4x 
  \Tr\left(\der_\mu U \der^\mu U^\dagger\right) 
  +\frac{1}{32e^2}\int d^4x
  \Tr\left(
      \left[
       U^\dagger \der_\mu U,U^\dagger\der_\nu U
      \right]^2
     \right)  
  +N_c\Gamma[U] \nonumber \\
 &&+\frac{F_0^2 B_0}{8}\int d^4x
  \Tr\left({\cal M}^\dagger U+{\cal M}U^\dagger\right) 
  +L_5B_0\int d^4x
  \Tr\left(\der_\mu U^\dagger \der^\mu U
      \left({\cal M}^\dagger U
       + U^\dagger {\cal M}\right)\right) \nonumber \\
 &&{}+L_8B_0^2\int d^4x
  \Tr\left({\cal M}^\dagger U{\cal M}^\dagger U
      +{\cal M}U^\dagger{\cal M}U^\dagger\right),
  \label{action}
\end{eqnarray}
which is up to including ${\cal O}(N_c)$ and ${\cal O}(p^4)$ terms.

It is important to note that, though (\ref{action}) is our starting
point, we always need to keep in mind that there are infinitely many
higher order contributions. It is the symmetry and the power counting
that actually matter. The discussion given in this section (and in
Appendix~\ref{Sec:Trad}) may be considered as a heuristic derivation,
which is however very convenient because it explicitly shows the
(leading) orders of $N_c$, $m$ and $\delta m$ for various parameters.

In the usual $\chi$PT, one loop quantum effects of mesons can be
incorporated consistently to this order and the physical parameters such
as the pion decay constant $F_\pi$ are so defined as to include one-loop
corrections. In our analysis, we ignore all the quantum effects of
mesons, but there are still tree-level contributions to physical
parameters from the higher order terms. For the decay constants, we have
\begin{eqnarray}
 F_\pi&=&F_0\left(1+(2m)K_6\right), \\
 F_{\rm K}&=&F_0\left(1+(m+m_s)K_6\right), \\
 F_\eta&=&F_0\left(1+\frac{2}{3}(m+2m_s)K_6\right),
\end{eqnarray}
where
\begin{equation}
 K_6=\frac{16B_0}{F_0^2}L_5.
\end{equation}
Meson masses are obtained by looking at the quadratic terms in the
Lagrangian when expanded around $U=1$,
\begin{eqnarray}
 M_\pi^2&=&B_0(2m)\left(1+(2m)K_3\right), \\
 M_{\rm K}^2&=&B_0(m+m_s)\left(1+(m+m_s)K_3\right), \\
 M_\eta^2&=&\frac{2}{3}B_0(m+2m_s)\left(1+\frac{2}{3}(m+2m_s)K_3\right)
  +K_5,
\end{eqnarray}
where
\begin{eqnarray}
 K_3&=&\frac{32B_0}{F_0^2}(2L_8-L_5), \\
 K_5&=&(m_s-m)^2\frac{512}{9}\frac{B_0^2}{F_0^2}L_8.
\end{eqnarray}

\subsection{Collective coordinate quantization}

In this subsection, we derive the Hamiltonian which describes the
baryons by using the collective coordinate quantization. In this
treatment, we consider the soliton as a ``rigid rotator'' and do not
consider the breathing degrees of freedom, though such ``radial''
excitations should be important if we consider other states, such as
those with negative parity\footnote{We assume that the pentaquark states
have positive parity. Otherwise they would  not be ``rotational'' modes of the
soliton, and our analysis in this paper would not make sense
at all.}.

There are two important criticisms against the above mentioned
treatment. The first is an old one that the flavor $SU(3)$ symmetry
breaking is so large that the perturbation theory does not work. The
so-called ``bound-state'' approach has been advocated by Callan and
Klebanov\cite{Callan:1985hy} and it was recently reconsidered after the
discovery of the $\Theta^{+}$ resonance\cite{Itzhaki:2003nr}. It is Yabu
and Ando\cite{Yabu:1987hm} who showed that the ``exact'' treatment of
the symmetry breaking term gives good results even though the collective
coordinate quantization is employed. Later, it was shown that the
perturbation theory is capable to reproduce the qualitatively equivalent
results if one includes mixings with an enough number of
representations\cite{Praszalowicz:1987em,Park:1989wz}. The second is
claimed by Cohen\cite{Cohen:2003yi,Cohen:2004xp} that, in the
large-$N_c$ limit, the ``rotation'' is not slow enough for the
collective treatment to be justified. Diakonov and
Petrov\cite{Diakonov:2003ei} emphasized that due to the WZW term, the
``rotation'' is slow enough even in the large-$N_c$ limit.

The collective coordinate quantization with the flavor $SU(3)$ symmetry
is different from the one with the isospin $SU(2)$\cite{Adkins:1983ya,
Adkins:1983hy}, due to the existence of the WZW term. See
Refs.~\cite{Witten:1983tx, Guadagnini:1983uv, Mazur:1984yf, Jain:1984gp,
Praszalowicz:1985bt}.

The rotational collective coordinates are introduced as a time-dependent
$SU(3)$-valued variable $A(t)$ through
\begin{equation}
 U(t,\bfx)=A(t)U_c(\bfx)A^\dagger(t),
  \label{UUc}
\end{equation}
where $U_c(\bfx)$ is the classical hedgehog soliton ansatz,
\begin{equation}
 U_c(\bfx)=
  \left(
   \begin{array}{cc}
    \exp\left(i\bftau\cdot\hat{\bfx}F(r)\right)& 
     \begin{array}{c}
      0\\
      0
     \end{array} \\
    \begin{array}{cc}
     0\ &\  0\\
    \end{array}& 1
   \end{array}
  \right),
\end{equation}
with the baryon number (topological charge) $B=1$. The profile function
$F(r)$ satisfies the boundary conditions $F(0)=\pi$ and
$F(\infty)=0$. By substituting Eq.~(\ref{UUc}) into Eq.~(\ref{action}),
one obtains the Lagrangian
\begin{equation}
 L=-M_{cl}+\half \omega^\alpha I_{\alpha\beta}(A)\omega^\beta
   +\frac{N_c}{2\sqrt{3}}\omega^8
   -V(A),
   \label{lag}
\end{equation}
where we have introduced the ``angular velocity'' $\omega^\alpha(t) \
(\alpha=1,\cdots,8)$ by
\begin{equation}
 A^\dagger (t)\dot{A}(t)
  =\frac{i}{2}\sum_{\alpha=1}^8 \lambda_\alpha \omega^\alpha(t),
\end{equation}
with $\lambda_\alpha \ (\alpha=1,\cdots, 8)$ being the usual Gell-Mann
matrices. 

The first term represents the rest mass energy of the classical
soliton. In the ``traditional'' approach, the $A$-independent part is a
functional of the profile function $F(r)$. The classical mass $M_{cl}$
is given by minimizing (the minus of) it by varying $F(r)$ subject to
the boundary condition. One might think that the parameters in the
$\chi$PT action (\ref{action}) have already been given and $M_{cl}$
would be determined completely in terms of these parameters. It is
however wrong because the $\chi$PT action contains infinitely many terms
and therefore there are infinitely many contributions from higher
orders. We do not know all of those higher order couplings and thus
practically we cannot calculate $M_{cl}$ at all. A more physical
procedure is to fit it to the experimental value. This is our basic
strategy in the effective theory approach. The operators are determined
by the $\chi$PT action, reflecting the fundamental principles and
symmetries of QCD, while the coefficients are fitted to the experimental
values. Because we do not know the higher order contributions, the
number of parameters is different from that of (\ref{action}). For the
comparison with the ``traditional'' approach, see
Appendix~\ref{Sec:Trad}, which also serves as a ``derivation'' of the
terms discussed below.

The most important feature of the Lagrangian (\ref{lag}) is that the
``inertia tensor'' $I_{\alpha\beta}(A)$ depends on
$A$\footnote{A mechanical analogy is a top with a thick axis
and a round pivot, rotating on a smooth floor, for which the moment of
inertia depends on how the axis tilts.}. From the symmetry of the ansatz
and the structure of the symmetry breaking, it has the following form,
\begin{equation}
 I_{\alpha\beta}(A)=I_{\alpha\beta}^0+I'_{\alpha\beta}(A), 
\end{equation}
with
\begin{equation}
  I_{\alpha\beta}^0=
  \left\{
   \begin{array}{lcl}
    I_1\delta_{\alpha\beta} & \quad & \alpha,\beta\in{\cal I} \\
    I_2\delta_{\alpha\beta} & \quad & \alpha,\beta\in{\cal J} \\
    0 & \quad & \mbox{\rm otherwise}
   \end{array}
  \right.
  \label{Izero}
\end{equation}
and 
\begin{eqnarray}
 I'_{\alpha\beta}(A)&=&
\left\{
 \begin{array}{lcl}
  \overline{x}\delta_{\alpha\beta} 
   D_{88}^{(8)}(A)& \qquad &(\alpha,\beta\in{\cal I})\\
  \overline{y} d_{\alpha\beta\gamma}
   D_{8\gamma}^{(8)}(A) &\qquad &
    \begin{array}{l}
     (\alpha\in {\cal I},\ \beta\in{\cal J}\\
     \mbox{\rm or }\  
      \alpha\in {\cal J},\ \beta\in {\cal I})\\ 
    \end{array} \\
  \overline{z}\delta_{\alpha\beta}
   D_{88}^{(8)}(A)
   +\overline{w} d_{\alpha\beta\gamma} 
   D_{8\gamma}^{(8)}(A) &  \qquad&(\alpha,\beta\in {\cal J}) \\ 
  0 & \qquad &(\alpha=8\  \mbox{\rm or}\  \beta=8)
 \end{array}
\right.\label{Iprime}
\end{eqnarray}
where $ {\cal I}=\{1,2,3\}$ and ${\cal J}=\{4,5,6,7\}$.
We denote the $SU(3)$ representation matrix in the
adjoint (octet) representation as
\begin{equation}
 D_{\alpha\beta}^{(8)}(A)=\half\Tr
\left(
 A^\dagger \lambda_\alpha A\lambda_\beta
\right) \quad
(\alpha,\beta=1,\cdots,8)
\label{octetrep}
\end{equation}
and $d_{\alpha\beta\gamma}$ is the usual symmetric tensor.

The parameters $I_1$, $I_2$, $\overline{x}$, $\overline{y}$,
$\overline{z}$, and $\overline{w}$ are to be determined. Note that $I_1$
and $I_2$ are of ${\cal O}(1)$ while $\overline{x}$, $\overline{y}$,
$\overline{z}$, and $\overline{w}$ are of ${\cal O}(\delta m)$.

The third term in the Lagrangian (\ref{lag}) comes from the WZW term and
gives rise to a first-class constraint which selects possible
representations. The last term is a potential term,
\begin{eqnarray}
 V(A)&=&V^{(1)}(A)+V^{(2)}(A), \label{potential}\\
 V^{(1)}(A)&=&\frac{\gamma}{2} \left(1-D_{88}^{(8)}(A)\right), 
  \label{pot1}\\
 V^{(2)}(A)&=&v\left(
      1-\sum_{\alpha\in{\cal I}} \left(D_{8\alpha}^{(8)}(A)\right)^2
      -\left(D_{88}^{(8)}(A)\right)^2
     \right),
 \label{pot2}
\end{eqnarray}
where $\gamma$ is of ${\cal O}(\delta m)$ and $v$ of ${\cal O}(\delta
m^2)$. Note that $V^{(2)}(A)$ is of higher order in $\delta m$ than any
other operators. The reason why we include it is that it is the leading
order in $N_c$. Equivalently, we assume
\begin{equation}
 \frac{\delta m}{\Lambda}< \frac{1}{N_c^2}
\end{equation}
with some relevant mass scale $\Lambda$.

Collective quantization of the theory is a standard procedure. The only
difference comes from the fact that the ``inertia tensor''
$I_{\alpha\beta}$ depends on the ``coordinates'' $A$. The operator
ordering must be cared about and we adopt the standard one. The kinetic
term now involves the {\em inverse} of the ``inertial tensor,'' and we
expand it up to including the terms of order $\delta m$. We obtain the
following Hamiltonian,
\begin{eqnarray}
 H&=&M_{cl}+H_0+H_1+H_2, \\
 H_0&=&\frac{1}{2I_1}\sum_{\alpha\in{\cal I}} \left(F_\alpha\right)^2
  +\frac{1}{2I_2}\sum_{\alpha\in{\cal J}}\left(F_\alpha\right)^2, \\
 H_1&=&x D_{88}^{(8)}(A)\sum_{\alpha\in{\cal I}}\left(F_\alpha\right)^2 
  +y
  \left[
   \sum_{\alpha\in {\cal I},\beta\in{\cal J}}+
   \sum_{\alpha\in {\cal J},\beta\in{\cal I}}
  \right]
  \sum_{\gamma=1}^8 
  d_{\alpha\beta\gamma}F_\alpha D_{8\gamma}^{(8)}(A)F_\beta
  \nonumber \\
 &&{}+z\sum_{\alpha \in {\cal J}}F_\alpha D_{88}^{(8)}(A) F_\alpha
  +w\sum_{\alpha,\beta \in{\cal J}}\sum_{\gamma=1}^8
  d_{\alpha\beta\gamma}F_\alpha D_{8\gamma}^{(8)}(A)F_\beta
  \nonumber \\
 &&{}+\frac{\gamma}{2} \left(1-D^{(8)}_{88}(A)\right), \\
 H_2&=&v\left(
      1-\sum_{\alpha\in{\cal I}} \left(D_{8\alpha}^{(8)}(A)\right)^2
      -\left(D_{88}^{(8)}(A)\right)^2
     \right),
\end{eqnarray}
where
\begin{equation}
  x=-\frac{\overline{x}}{2I_1^2},\ \ y=-\frac{\overline{y}}{2I_1I_2},\ \
  z=-\frac{\overline{z}}{2I_2^2},\ \ w=-\frac{\overline{w}}{2I_2^2},
\end{equation}
and $F_\alpha \ (\alpha=1,\cdots, 8)$ are the $SU(3)$ generators,
\begin{equation}
  [F_\alpha,F_\beta]=i\sum_{\gamma=1}^8 f_{\alpha\beta\gamma} F_\gamma,
   \label{F}
\end{equation}
where $f_{\alpha\beta\gamma}$ is the totally anti-symmetric
structure constant of $SU(3)$. Note that they act on $A$ {\em from the
right}.

The WZW term leads to the first-class constraint, giving the
auxiliary condition to physical states $\Psi(A)$,
\begin{equation}
 Y_R\Psi(A)\equiv-\frac{2}{\sqrt{3}}F_8\Psi(A)=\frac{N_c}{3}\Psi(A).
  \label{WZWconstraint}
\end{equation}
See \cite{Witten:1983tx,Guadagnini:1983uv,Mazur:1984yf,Jain:1984gp} for
more details. In the following, we set $N_c=3$.

The novel feature is the existence of the interactions quadratic in
$F_\alpha$.  Generalizations of the Skyrme model have been considered by
several authors, but they have never considered such complicated
interactions as given above.

It is also important to note that we do not get any interactions linear
in $F_\alpha$. The reason can be traced back to the fact that the action
does not include the terms linear in time derivative, except for the WZW
term. The absence of such terms comes from the time reversal invariance
of QCD\footnote{The vacuum angle $\theta$ is assumed to be zero.}.

\section{Mixing among Representations and the Masses of Baryons}
\label{Sec:Mass}

\subsection{Symmetric case}

In the absence of the $SU(3)$ flavor symmetry breaking interactions, the
eigenstates of the collective Hamiltonian furnish the $SU(3)$
representations. The symmetric part $H_{sym}=M_{cl}+H_0$ may be written
as
\begin{equation}
 H_{sym}=M_{cl}+\half\left(\frac{1}{I_1}-\frac{1}{I_2}\right)C_2(SU(2))
  +\frac{1}{2I_2}\left(C_2(SU(3))-\frac{N_c^2}{12}\right)
  \label{Hsym}
\end{equation}
where $C_2(SU(2))$ is the spin $SU(2)$ quadratic Casimir operator,
\begin{equation}
 C_2(SU(2))=\sum_{\alpha\in{\cal I}}\left(F_\alpha\right)^2
\end{equation}
with the eigenvalue $J(J+1)$. The operator $C_2(SU(3))$ is the flavor
$SU(3)$ quadratic Casimir,
\begin{equation}
 C_2(SU(3))=\sum_{\alpha=1}^8 \left(F_\alpha\right)^2
\end{equation}
with the eigenvalue
\begin{equation}
 C_2(p,q)=\frac{1}{3}\left[p^2+q^2+pq+3(p+q)\right].
\end{equation}
where $(p,q)$ is the Dynkin index of the representation.  Note that we
have used the constraint $F_8=N_c/2\sqrt{3}$ in Eq.~(\ref{Hsym}).

The eigenstate is given by the $SU(3)$ representation matrix,
\begin{equation}
 \Psi^{(p,q)}
  \left(
   \genfrac{}{}{0pt}{1}{Y\ I}{I_3}
   ;\genfrac{}{}{0pt}{1}{Y_R\ J}{J_3}
  \right)(A) 
 \equiv\sqrt{\mbox{\rm dim}(p,q)} (-1)^{J_3-Y_R/2}
 \left\langle 
  Y,I,I_3\left|D^{(p,q)}(A)\right|Y_R,J,-J_3
 \right\rangle^*
 \label{eigenstate}
\end{equation}
with $Y_R=1$, where $\mbox{\rm dim}(p,q)$ is the dimension of
representation $(p,q)$ and is given by
\begin{equation}
 \mbox{\rm dim}(p,q)=(p+1)(q+1)\left(1+\frac{p+q}{2}\right).
\end{equation}
For the properties of the wave function (\ref{eigenstate}), see
Appendix~\ref{Sec:Math}.

By using them, we readily calculate the symmetric mass $M_{\cal R}$ of the
representation ${\cal R}$,
\begin{eqnarray}
 M_{\bm{8}}&=&M_{cl}+\frac{3}{8}\left[\frac{1}{I_1}+\frac{2}{I_2}\right], 
  \quad
 M_{\bm{10}}=M_{cl}+\frac{3}{8}\left[\frac{5}{I_1}+\frac{2}{I_2}\right], 
 \nonumber \\
 M_{\bm{\overline{10}}}
  &=&M_{cl}+\frac{3}{8}\left[\frac{1}{I_1}+\frac{6}{I_2}\right], 
  \quad
 M_{\bm{27}_d}=M_{cl}+\frac{1}{8}\left[\frac{3}{I_1}+\frac{26}{I_2}\right], 
 \nonumber \\
 M_{\bm{27}_q}&=&M_{cl}+\frac{1}{8}\left[\frac{15}{I_1}+\frac{14}{I_2}\right],
\end{eqnarray}
and so on. Because $\bm{27}$ contains the spin-$\half$ part and the
spin-$\frac{3}{2}$ part, the former is denoted as $\bm{27}_d$, the
latter as $\bm{27}_q$.

\subsection{Matrix elements of the symmetry breaking operators}
\label{symbreak}

The $SU(3)$ flavor symmetry breaking interactions mix the
representations and an eigenstate of the full Hamiltonian $H$ is a
linear combination of the (infinitely many) states of
representations. In this paper, we consider the perturbative expansion
up to including ${\cal O}(\delta m^2)$. 

To calculate the perturbative corrections, we need the matrix elements
of the symmetry breaking operators. Though the calculations are group
theoretical, they are considerably complicated because of the generators
$F_\alpha$. Several useful mathematical tools are summarized in
Appendix~\ref{Sec:Math}.

Note that the symmetry breaking operators conserve spin $SU(2)$, isospin
$SU(2)$ and hypercharge $U(1)$ symmetries. So that the matrix elements
are classified by $(I,Y)$, the magnitude of isospin and the hypercharge,
and $J$, the magnitude of spin. The $SU(3)$ representation and the
magnitude of spin are related by the constraint (\ref{WZWconstraint}).

We introduce the following notation,
\begin{eqnarray}
 {\cal O}_\gamma&\equiv& D_{88}^{(8)}(A),\quad
  {\cal O}_x\equiv D_{88}^{(8)}(A)
  \sum_{\alpha\in{\cal I}}\left(F_\alpha\right)^2, \\
 {\cal O}_y&\equiv&\left[
   \sum_{\alpha\in {\cal I},\beta\in{\cal J}}+
   \sum_{\alpha\in {\cal J},\beta\in{\cal I}}
  \right]\sum_{\gamma=1}^8
 d_{\alpha\beta\gamma}F_\alpha D_{8\gamma}^{(8)}(A)F_\beta, \\
 {\cal O}_z&\equiv&
  \sum_{\alpha \in {\cal J}}F_\alpha D_{88}^{(8)}(A) F_\alpha, \quad
  {\cal O}_w\equiv
  \sum_{\alpha,\beta \in{\cal J}}\sum_{\gamma=1}^8
  d_{\alpha\beta\gamma}F_\alpha D_{8\gamma}^{(8)}(A)F_\beta, \\
 {\cal O}_{v_1}&\equiv&\left(D_{88}^{(8)}(A)\right)^2, \quad
  {\cal O}_{v_2}\equiv
  \sum_{\alpha\in{\cal I}} \left(D_{8\alpha}^{(8)}(A)\right)^2,
\end{eqnarray}
and denote them by ${\cal O}_i$ collectively.

\subsubsection{$J=\half$}

The $SU(3)$ representations with spin $\half$ are $\bm{8}$,
$\bm{\overline{10}}$, $\bm{27}_d$, and so on. States with the same
$(I,Y)$ can mix.

Because all the symmetry breaking operators behave as
$\bm{8}$, the octet states can mix with $\bm{\overline{10}}$ and
$\bm{27}_d$ in the first order, and also with $\bm{\overline{35}}_d$ and
$\bm{64}_d$ in the second order. The anti-decuplet states can mix with
$\bm{8}$, $\bm{27}_d$ and $\bm{\overline{35}}_d$ in the first order and
also with $\bm{64}_d$ and $\bm{\overline{81}}_d$ in the second order. To
calculate the masses to second order, we do not need the second
order mixings, but they are used for the decay amplitudes.

The matrix elements for the $\bm{8}$ representation are given in
Table~\ref{88}. Note that the matrix elements of the operator ${\cal
O}_z$ are zero.  
\begin{table}[h]
 \caption{\label{88}$ \left\langle \bm{8}\left|{\cal
 O}_i\right|\bm{8}\right\rangle$} 
 \begin{ruledtabular}
  \begin{tabular}{cc|ccccccc}
   $\bm{8}$&$(I,Y)$ & $\gamma$ & $x$ & $y$ & $z$ & $w$ & $v_1$ & $v_2$ 
   \\ \hline
   $\mbox{\rm N}_8$ & 
   $(\half, +1)$ & $\frac{3}{10}$ & $\frac{3}{4}\frac{3}{10}$ &
   $\frac{\sqrt{3}}{20}$ & $0$ & $-\frac{\sqrt{3}}{5}$ & $\frac{1}{5}$ &
   $\frac{1}{5}$ \\
   $\Sigma_8$ & 
   $(1, 0)$ & $-\frac{1}{10}$ & $-\frac{3}{4}\frac{1}{10}$ &
   $\frac{3\sqrt{3}}{20}$ & $0$ & $\frac{3\sqrt{3}}{20}$ & $\frac{1}{10}$ &
   $\frac{13}{30}$ \\
   $\Xi_8$ & 
   $(\half, -1)$ & $-\frac{1}{5}$ & $-\frac{3}{4}\frac{1}{4}$ &
   $-\frac{\sqrt{3}}{5}$ & $0$ & $\frac{\sqrt{3}}{20}$ & $\frac{1}{10}$ &
   $\frac{1}{2}$ \\
   $\Lambda_8$ & 
   $(0, 0)$ & $\frac{1}{10}$ & $\frac{3}{4}\frac{1}{10}$ &
   $-\frac{3\sqrt{3}}{20}$ & $0$ & $-\frac{3\sqrt{3}}{20}$ & $\frac{1}{10}$ &
   $\frac{3}{10}$
  \end{tabular}
 \end{ruledtabular}
\end{table}

Similarly, the matrix elements for $\bm{\overline{10}}$
are given in Table~\ref{10b10b}.
\begin{table}[h]
 \caption{\label{10b10b}$ \left\langle \bm{\overline{10}}\left|{\cal
 O}_i\right|\bm{\overline{10}}\right\rangle$} 
 \begin{ruledtabular}
  \begin{tabular}{cc|ccccccc}
   $\bm{\overline{10}}$&$(I,Y)$ & $\gamma$ & $x$ & $y$ & $z$ & $w$ &
   $v_1$ & $v_2$  
   \\ \hline
   $\Theta_{\overline{10}}$ & 
   $(0, +2)$ & $\frac{1}{4}$ & $\frac{3}{4}\frac{1}{4}$ &
   $-\frac{5\sqrt{3}}{8}$ & $\frac{3}{4}$ & $-\frac{5\sqrt{3}}{8}$ &
   $\frac{1}{7}$ & $\frac{3}{14}$ \\
   $\mbox{\rm N}_{\overline{10}}$ & 
   $(\half, +1)$ & $\frac{1}{8}$ &
   $\frac{3}{4}\frac{1}{8}$ & $-\frac{5\sqrt{3}}{16}$ & $\frac{3}{8}$ & 
   $-\frac{5\sqrt{3}}{16}$ & $\frac{9}{56}$ & $\frac{17}{56}$ \\
   $\Sigma_{\overline{10}}$ & 
   $(1, 0)$ & $0$ & $0$ & $0$ & $0$ & $0$ & 
   $\frac{1}{7}$ & $\frac{8}{21}$ \\
   $\Xi_{\overline{10}}$ & 
   $(\frac{3}{2}, -1)$ & $-\frac{1}{8}$ & $-\frac{3}{4}\frac{1}{8}$ &
   $\frac{5\sqrt{3}}{16}$ & $-\frac{3}{8}$ & $\frac{5\sqrt{3}}{16}$ & 
   $\frac{5}{56}$ & $\frac{25}{56}$ 
  \end{tabular}
 \end{ruledtabular}
\end{table}

In order to calculate the masses to second order, we also need the
matrix elements off-diagonal in representation.  They are given in
Tables~\ref{10b8}, \ref{27d8}, \ref{27d10b}, and \ref{35bd10b}.
\begin{table}[h]
 \caption{\label{10b8}$ \left\langle \bm{\overline{10}}\left|{\cal
 O}_i\right|\bm{8}\right\rangle$} 
 \begin{ruledtabular}
  \begin{tabular}{c|ccccccc}
   $(I,Y)$ & $\gamma$ & $x$ & $y$ & $z$ & $w$ & $v_1$ & $v_2$ \\ \hline
   $(\half, +1)$ & $\frac{1}{2\sqrt{5}}$ & $\frac{3}{4} \frac{1}{2\sqrt{5}}$
   & $\frac{\sqrt{3}}{2\sqrt{5}}$ & $\frac{3}{4\sqrt{5}}$ &
   $-\frac{\sqrt{3}}{4\sqrt{5}}$ & $\frac{1}{4\sqrt{5}}$ & 
   $-\frac{1}{4\sqrt{5}}$ \\
   $(1,0)$ & $\frac{1}{2\sqrt{5}}$ & $\frac{3}{4} \frac{1}{2\sqrt{5}}$
   & $\frac{\sqrt{3}}{2\sqrt{5}}$ & $\frac{3}{4\sqrt{5}}$ &
   $-\frac{\sqrt{3}}{4\sqrt{5}}$ & $0$ & $-\frac{1}{3\sqrt{5}}$
  \end{tabular}
 \end{ruledtabular}
\end{table}

\begin{table}[h]
 \caption{\label{27d8}$ \left\langle \bm{\bm{27}}_d\left|{\cal
 O}_i\right|\bm{8}\right\rangle$} 
 \begin{ruledtabular}
  \begin{tabular}{c|ccccccc}
   $(I,Y)$ & $\gamma$ & $x$ & $y$ & $z$ & $w$ & $v_1$ & $v_2$ \\ \hline
   $(\half, +1)$ & 
   $\frac{\sqrt{6}}{10}$ & 
   $\frac{3}{4} \frac{\sqrt{6}}{10}$ & 
   $-\frac{\sqrt{2}}{5}$ & 
   $\frac{\sqrt{6}}{4}$ &
   $\frac{\sqrt{2}}{20}$ & 
   $\frac{17\sqrt{3}}{140\sqrt{2}}$ &
   $-\frac{13\sqrt{3}}{140\sqrt{2}}$ \\
   $(1,0)$ & 
   $\frac{1}{5}$ & 
   $\frac{3}{4} \frac{1}{5}$ & 
   $-\frac{2\sqrt{3}}{15}$ & 
   $\frac{1}{2}$ &
   $\frac{\sqrt{3}}{30}$ &
   $\frac{1}{70}$ &
   $-\frac{9}{70}$ \\
   $(\half, -1)$ &
   $\frac{\sqrt{6}}{10}$ & 
   $\frac{3}{4} \frac{\sqrt{6}}{10}$ & 
   $-\frac{\sqrt{2}}{5}$ & 
   $\frac{\sqrt{6}}{4}$ &
   $\frac{\sqrt{2}}{20}$ & 
   $-\frac{\sqrt{3}}{35\sqrt{2}}$ &
   $-\frac{\sqrt{3}}{7\sqrt{2}}$ \\
   $(0,0)$ &
   $\frac{3}{10}$ & 
   $\frac{3}{4} \frac{3}{10}$ & 
   $-\frac{\sqrt{3}}{5}$ & 
   $\frac{3}{4}$ &
   $\frac{\sqrt{3}}{20}$ & 
   $\frac{3}{35}$ &
   $-\frac{6}{35}$
  \end{tabular}
 \end{ruledtabular}
\end{table}

\begin{table}[h]
 \caption{\label{27d10b}$ \left\langle\bm{27}_d\left|{\cal
 O}_i\right| \bm{\overline{10}}\right\rangle$}
 \begin{ruledtabular}
  \begin{tabular}{c|ccccccc}
   $(I,Y)$ & $\gamma$ & $x$ & $y$ & $z$ & $w$ & $v_1$ & $v_2$ \\ \hline
   $(\half, +1)$ &
   $\frac{\sqrt{3}}{8\sqrt{10}}$ & 
   $\frac{3}{4} \frac{\sqrt{3}}{8\sqrt{10}}$ & 
   $\frac{11}{16\sqrt{10}}$ & 
   $\frac{\sqrt{3}}{2\sqrt{10}}$ &
   $-\frac{23}{16\sqrt{10}}$ & 
   $\frac{11\sqrt{3}}{56\sqrt{10}}$ &
   $-\frac{\sqrt{3}}{56\sqrt{10}}$ \\
   $(1,0)$ &
   $\frac{1}{4\sqrt{5}}$ & 
   $\frac{3}{4} \frac{1}{4\sqrt{5}}$ &
   $\frac{11}{8\sqrt{15}}$ & 
   $\frac{1}{\sqrt{5}}$ &
   $-\frac{23}{8\sqrt{15}}$ &
   $\frac{\sqrt{5}}{28}$ &
   $-\frac{3}{28\sqrt{5}}$ \\
   $(\frac{3}{2},-1)$ &
   $\frac{\sqrt{3}}{8\sqrt{2}}$ & 
   $\frac{3}{4} \frac{\sqrt{3}}{8\sqrt{2}}$ &
   $\frac{11}{16\sqrt{2}}$ & 
   $\frac{\sqrt{3}}{2\sqrt{2}}$ &
   $-\frac{23}{16\sqrt{2}}$ &
   $-\frac{\sqrt{3}}{56\sqrt{2}}$ &
   $-\frac{5\sqrt{3}}{56\sqrt{2}}$
  \end{tabular}
 \end{ruledtabular}
\end{table}

\begin{table}[h]
 \caption{\label{35bd10b}$ \left\langle\bm{\overline{35}}_d\left|{\cal
 O}_i\right| \bm{\overline{10}}\right\rangle$}
 \begin{ruledtabular}
  \begin{tabular}{c|ccccccc}
   $(I,Y)$ & $\gamma$ & $x$ & $y$ & $z$ & $w$ & $v_1$ & $v_2$ \\ \hline
   $(0, +2)$ &
   $\frac{3}{4\sqrt{7}}$ &
   $\frac{3}{4}\frac{3}{4\sqrt{7}}$ &
   $-\frac{\sqrt{3}}{8\sqrt{7}}$ &
   $\frac{9}{2\sqrt{7}}$ &
   $\frac{\sqrt{3}}{8\sqrt{7}}$ &
   $\frac{3}{8\sqrt{7}}$ &
   $-\frac{3}{8\sqrt{7}}$
   \\
   $(\half, +1)$ &
   $\frac{\sqrt{9}}{8\sqrt{14}}$ & 
   $\frac{3}{4} \frac{\sqrt{9}}{8\sqrt{14}}$ & 
   $-\frac{3\sqrt{3}}{16\sqrt{14}}$ & 
   $\frac{27}{4\sqrt{14}}$ &
   $\frac{3\sqrt{3}}{16\sqrt{14}}$ &
   $\frac{3}{8\sqrt{14}}$ &
   $-\frac{5}{8\sqrt{14}}$
   \\
   $(1,0)$ &
   $\frac{3}{4\sqrt{7}}$ & 
   $\frac{3}{4} \frac{3}{4\sqrt{7}}$ &
   $-\frac{\sqrt{3}}{8\sqrt{7}}$ & 
   $\frac{9}{2\sqrt{7}}$ &
   $\frac{\sqrt{3}}{8\sqrt{7}}$ &
   $\frac{1}{8\sqrt{7}}$ &
   $-\frac{11}{24\sqrt{7}}$
   \\
   $(\frac{3}{2},-1)$ &
   $\frac{3\sqrt{5}}{8\sqrt{14}}$ & 
   $\frac{3}{4} \frac{3\sqrt{5}}{8\sqrt{14}}$ &
   $-\frac{\sqrt{15}}{16\sqrt{14}}$ & 
   $\frac{9\sqrt{5}}{4\sqrt{14}}$ &
   $\frac{\sqrt{15}}{16\sqrt{14}}$ &
   $0$ &
   $-\frac{\sqrt{5}}{4\sqrt{14}}$
   \\
  \end{tabular}
 \end{ruledtabular}
\end{table}

Other matrix elements, which are necessary for the
calculations of the mixings, are collected in
Appendix~\ref{Sec:Tables}.

By using them, we can write down the perturbation theory results for the
masses and the representation mixings. For example, the nucleon (mainly
octet $(I,Y)=(\half, +1)$ state) mass is calculated as
\begin{eqnarray}
 M_{\rm N}
  &=&M_8+\frac{\gamma}{2}\left(1-\frac{3}{10}\right) 
  +\frac{9}{40}x+\frac{\sqrt{3}}{20}y-\frac{\sqrt{3}}{5}w 
  +\left(1-\frac{1}{5}-\frac{1}{5}\right)v \nonumber \\
 &&{}-\frac{\Delta H^2_{\overline{10}-8}}{M_{\overline{10}}-M_8} 
 -\frac{2}{3}\frac{\Delta H^2_{27_d-8}}{M_{27_d}-M_8},
\end{eqnarray}
where
\begin{eqnarray}
 \Delta H^2_{\overline{10}-8}&=&
  \left[
   \frac{\gamma}{2}\left(-\frac{1}{2\sqrt{5}}\right)
   +\frac{3}{4}\frac{x}{2\sqrt{5}}
   +\frac{\sqrt{3}}{2\sqrt{5}}y 
   +\frac{3}{4\sqrt{5}}z
   -\frac{\sqrt{3}}{4\sqrt{5}}w
  \right]^2, \\
 \Delta H^2_{27_d-8}&=&
  \left[
   \frac{\gamma}{2}\left(-\frac{3}{10}\right)
   +\frac{3}{4}\frac{3}{10}x
   -\frac{\sqrt{3}}{5}y 
   +\frac{3}{4}z
   +\frac{\sqrt{3}}{20}w
  \right]^2.
\end{eqnarray}
The masses for other states may be calculated similarly.

\subsubsection{$J=\frac{3}{2}$}

Spin $\frac{3}{2}$ states are composed of the representations $\bm{10}$,
$\bm{27}_q$, and so on. The matrix elements for the $\bm{10}$
representation are given in Tables~\ref{1010}. Note that the
matrix elements of ${\cal O}_z$ are zero.
\begin{table}[h]
 \caption{\label{1010}$ \left\langle \bm{10}\left|{\cal
 O}_i\right|\bm{10}\right\rangle$} 
 \begin{ruledtabular}
  \begin{tabular}{cc|ccccccc}
   $\bm{10}$&$(I,Y)$ & $\gamma$ & $x$ & $y$ & $z$ & $w$ & $v_1$ & $v_2$ 
   \\ \hline
   $\Delta_{10}$ & 
   $(\frac{3}{2}, +1)$ & 
   $\frac{1}{8}$ & 
   $\frac{15}{4}\frac{1}{8}$ &
   $\frac{5\sqrt{3}}{16}$ & $0$ & $\frac{\sqrt{3}}{16}$ & $\frac{9}{56}$ &
   $\frac{17}{56}$ \\
   $\Sigma_{10}$& 
   $(1, 0)$ & $0$ & $0$ & $0$ & $0$ & $0$ & 
   $\frac{3}{28}$ & $\frac{31}{84}$ \\
   $\Xi_{10}$ & 
   $(\half, -1)$ & $-\frac{1}{8}$ & $-\frac{15}{4}\frac{1}{8}$ &
   $-\frac{5\sqrt{3}}{16}$ & $0$ & $-\frac{\sqrt{3}}{16}$ & 
   $\frac{5}{56}$ & $\frac{25}{56}$ \\
   $\Omega$&$(0, -2)$ & 
   $-\frac{1}{4}$ & $-\frac{15}{4}\frac{1}{4}$ &
   $-\frac{5\sqrt{3}}{8}$ & $0$ & $-\frac{\sqrt{3}}{8}$ & 
   $\frac{3}{28}$ & $\frac{15}{28}$
  \end{tabular}
 \end{ruledtabular}
\end{table}

The decuplet states can mix with $\bm{27_q}$ and $\bm{35}$ in the first
order, and also with $\bm{\overline{35}}_q$, $\bm{64}_q$ and $\bm{81}$
in the second order.  In order to calculate the mass to second order, we
need the matrix elements of $\bm{10}$ with $\bm{27}_q$ and $\bm{35}_q$
representations. They are given in Tables~\ref{27q10} and \ref{3510}.
\begin{table}[h]
 \caption{\label{27q10}$ \left\langle \bm{27}_q\left|{\cal
 O}_i\right|\bm{10}\right\rangle$}
 \begin{ruledtabular}
  \begin{tabular}{c|ccccccc}
   $(I,Y)$ & $\gamma$ & $x$ & $y$ & $z$ & $w$ & $v_1$ & $v_2$ \\ \hline
   $(\frac{3}{2}, +1)$ &
   $\frac{\sqrt{15}}{8\sqrt{2}}$ & 
   $\frac{15}{4} \frac{\sqrt{15}}{8\sqrt{2}}$ & 
   $\frac{5\sqrt{5}}{16\sqrt{2}}$ & 
   $\frac{\sqrt{15}}{8\sqrt{2}}$ &
   $-\frac{5\sqrt{5}}{16\sqrt{2}}$ &
   $\frac{\sqrt{15}}{28\sqrt{2}}$ &
   $-\frac{\sqrt{15}}{14\sqrt{2}}$ \\
   $(1,0)$ &
   $\frac{1}{4}$ & 
   $\frac{15}{4} \frac{1}{4}$ & 
   $\frac{5}{8\sqrt{3}}$ & 
   $\frac{1}{4}$ &
   $-\frac{5}{8\sqrt{3}}$ &
   $\frac{1}{56}$ &
   $-\frac{9}{56}$ \\
   $(\half,-1)$ &
   $\frac{\sqrt{3}}{8\sqrt{2}}$ & 
   $\frac{15}{4} \frac{\sqrt{3}}{8\sqrt{2}}$ & 
   $\frac{5}{16\sqrt{2}}$ & 
   $\frac{\sqrt{3}}{8\sqrt{2}}$ &
   $-\frac{5}{16\sqrt{2}}$ &
   $-\frac{\sqrt{3}}{56\sqrt{2}}$ &
   $-\frac{5\sqrt{3}}{56\sqrt{2}}$
  \end{tabular}
 \end{ruledtabular}
\end{table}

\begin{table}[h]
 \caption{\label{3510} $ \left\langle \bm{35}\left|{\cal
 O}_i\right|\bm{10}\right\rangle$}
 \begin{ruledtabular}
  \begin{tabular}{c|ccccccc}
   $(I,Y)$ & $\gamma$ & $x$ & $y$ & $z$ & $w$ & $v_1$ & $v_2$ \\ \hline
   $(\frac{3}{2}, +1)$ &
   $\frac{5}{8\sqrt{14}}$ & 
   $\frac{15}{4}\frac{5}{8\sqrt{14}}$ & 
   $-\frac{25\sqrt{3}}{16\sqrt{14}}$ & 
   $\frac{15}{8\sqrt{14}}$ &
   $\frac{5\sqrt{3}}{16\sqrt{14}}$ &
   $\frac{5}{16\sqrt{14}}$ &
   $-\frac{5}{16\sqrt{14}}$ \\
   $(1,0)$ &
   $\frac{\sqrt{5}}{4\sqrt{7}}$ & 
   $\frac{15}{4} \frac{\sqrt{5}}{4\sqrt{7}}$ & 
   $-\frac{5\sqrt{15}}{8\sqrt{7}}$ & 
   $\frac{3\sqrt{5}}{4\sqrt{7}}$ &
   $\frac{\sqrt{15}}{8\sqrt{7}}$ &
   $\frac{\sqrt{5}}{16\sqrt{7}}$ &
   $-\frac{\sqrt{35}}{48}$ \\
   $(\half,-1)$ &
   $\frac{3\sqrt{5}}{8\sqrt{14}}$ & 
   $\frac{15}{4} \frac{3\sqrt{5}}{8\sqrt{14}}$ & 
   $-\frac{15\sqrt{15}}{16\sqrt{14}}$ & 
   $\frac{9\sqrt{5}}{8\sqrt{14}}$ &
   $\frac{3\sqrt{15}}{16\sqrt{14}}$ &
   $0$ &
   $-\frac{\sqrt{5}}{4\sqrt{14}}$ \\
   $(0,-2)$ &
   $\frac{\sqrt{5}}{4\sqrt{7}}$ &
   $\frac{15}{4}\frac{\sqrt{5}}{4\sqrt{7}}$ &
   $-\frac{5\sqrt{15}}{8\sqrt{7}}$ &
   $\frac{3\sqrt{5}}{4\sqrt{7}}$ &
   $\frac{\sqrt{15}}{8\sqrt{7}}$ &
   $-\frac{\sqrt{5}}{16\sqrt{7}}$ &
   $-\frac{3\sqrt{5}}{16\sqrt{7}}$
  \end{tabular}
 \end{ruledtabular}
\end{table}

\section{Numerical Determination of Parameters}
\label{Sec:Num}

We can calculate the baryon masses once the Skyrme model parameters are
given. In the effective theory approach, however, we
have to solve in the {\em opposite} direction. Namely, we need to
determine the Skyrme model parameters so as to best fit to the
experimental values of the baryon masses. In order to measure how good
the fitting is, we introduce the evaluation function
\begin{equation}
 \chi^2=\sum_i\frac{\left(M_i-M_i^{exp}\right)^2}{\sigma_i^2},
\end{equation}
where $M_i$ stands for the calculated mass of baryon $i$, and
$M_i^{exp}$, the corresponding experimental value. How accurately the
experimental values should be considered is measured by
$\sigma_i$. Because we neglect the isospin violation effect completely,
we use the average among the members of an isospin multiplet for the
mass and the range of variation within the isospin multiplet for the
$\sigma_i$. This is why our estimate of $\sigma_i$ for isospin singlets is
severe, while $\sigma_{\rm N}$ is considerably large though the masses of
proton and neutron are very accurately determined. At any rate, these
numbers should not be taken too seriously.

The sum is taken over the octet and decuplet baryons, as well as
$\Theta^{+}(1540)$ and $\phi(1860)$.  Note that we have nine parameters
to be determined. We need at least one more state than the low-lying
octet and the decuplet. In this sense, our effective theory cannot
predict the $\Theta^+$ mass. In our calculations, we use the values of
$M_i^{exp}$ and $\sigma_i$ given in Table~\ref{expmass}.

\begin{table}[h]
 \caption{\label{expmass}Experimental values of baryon masses and their
 deviations used in our calculation. In the last low, we also give the
 baryon masses calculated by using the best fit set of parameters
 (\ref{bestfit}).}
 \begin{ruledtabular}
  \begin{tabular}{c|cccccccccc}
   (MeV)&$\mbox{\rm N}$& $\Sigma$& $\Xi$& $\Lambda$& $\Delta$& ${\Sigma^*}$&
   ${\Xi^*}$& $\Omega$&$\Theta$& $\phi$\\
   \hline
   $M_i^{exp}$ &939 & 1193 & 1318 & 1116 & 1232 & 1385 & 1533 & 1672 & 1539 &
   1862 \\
   $\sigma_i$ &0.6& 4.0 & 3.2 & 0.01 & 2.0 & 2.2 & 1.6 & 0.3 & 1.6 & 2.0
   \\ 
   \hline
   $M_i$ &940 & 1180 & 1332 & 1116 & 1228 & 1389 & 1537 & 1672 & 1539 & 1862
  \end{tabular}
 \end{ruledtabular}
\end{table}

The problem is a multidimensional minimization of the function $\chi^2$
of nine variables. In general such a problem is very difficult, but in
our case, thanks to the fact that the function is a polynomial of the
variable in the perturbation theory, a stable numerical solution can be
obtained. Our method is basically Powell's one, but we tried several
minimization algorithms with the equivalent results. The best fit set of
parameters is the bottom point of a very shallow (and narrow) ``valley''
of the function, and $\chi^2$ does not change very much even if we
change the values of parameters in a certain way.

The best fit set of parameters is
\begin{eqnarray}
 M_{cl}&=& 389 \ \mbox{MeV},\ I_1^{-1}=174\ \mbox{MeV},\ 
  I_2^{-1}=585\ \mbox{MeV}, \  \gamma=832\ \mbox{MeV}, \nonumber \\
 x&=&27.8\ \mbox{MeV},\ y=-104\ \mbox{MeV},\ z=-306\ \mbox{MeV},\ 
  w=111\ \mbox{MeV},\nonumber \\
 v&=&-148\ \mbox{MeV},
  \label{bestfit}
\end{eqnarray}
and leads to the mass spectrum given in Table~\ref{expmass} with
$\chi^2=4.4\times 10^1$.

Note that best fit set of parameters is quite reasonable, though we do
not impose any constraint that the higher order (in $\delta m$)
parameters should be small. The parameter $\gamma$ is unexpectedly large
(even though it is of leading order in $N_c$), but considerably smaller
than the value ($\gamma=1573$ MeV) for the case (3) of Yabu and
Ando\cite{Yabu:1987hm}. The parameter $z$ seems also too large and we do
not know the reason.

Once we determine the best fit set of parameters, we can predicts the
masses of (mainly) anti-decuplet members,
\begin{equation}
M_{{\rm N}^{\prime}} =1711\ {\rm MeV},
 \quad M_{\Sigma^{\prime}}=1819\ {\rm MeV}.
 \label{massprediction}
\end{equation}

Compare with the $\chi$QSM prediction\cite{Ellis:2004uz},
\begin{equation}
 M_{{\rm N}^{\prime}} =1646\ \mbox{\rm MeV},
  \quad M_{\Sigma^{\prime}}=1754\ \mbox{\rm MeV}.
\label{EKPprediction}
\end{equation}

It is tempting to identify $\mbox{N}'$ with $\mbox{N}(1710)\
(***)$ state. On the other hand, for $\Sigma'$ there are two candidates,
$\Sigma(1779)\ (*)$ and $\Sigma(1880)\ (**)$. In any case, it is too
early to identify them.

The mixing coefficients of the eigenstates are also obtained. For the
(mainly) octet states, the coefficients are given in Table~\ref{8mix}.


\begin{table}[h]
\caption{\label{8mix} Mixing coefficients for the (mainly) octet
 states in the first order perturbation theory. The numbers in parentheses
 are those in the second order.}
 \begin{ruledtabular}
\begin{tabular}{c|cccc}
 $\quad{\cal R}_i\quad $ & N & $\Sigma$ & $\Xi$ & $\Lambda$ 
 \\ \hline
 $\bm{8}$ & 
 $1$ $(-0.056)$ & $1$ $(-0.051)$ & $1$ $(-0.014)$ & $1$ $(-0.021)$ \\
 $\bm{\overline{10}}$ & 
 $0.288$ $(0.036)$ & $0.288$ $(0.086)$ & $0$ $(0)$ & $0$ $(0)$ \\
 $\bm{27}_d$ & 
 $0.169$ $(0.087)$ & $0.138$ $(0.087)$ & $0.169$ $(0.036)$ & 
 $0.207$ $(0.075)$ \\
 $\bm{\overline{35}}_d$ & 
 $0$ $(0.085)$ & $0$ $(0.080)$ & $0$ $(0)$ & $0$ $(0)$ \\
 $\bm{64}_d$ & 
 $0$ $(0.038)$ & $0$ $(0.028)$ & $0$ $(0.038)$ & $0$ $(0.051)$ 
\end{tabular}
 \end{ruledtabular}
\end{table}


Similarly, for (mainly) decuplet and (mainly) anti-decuplet states, they
are given in Tables~\ref{10mix} and \ref{b10mix}.  The numbers in
parentheses are the second order contributions calculated by using the
parameters given in (\ref{bestfit}), which are used in
Sec~\ref{Sec:decay-num}.


\begin{table}[h]
\caption{\label{10mix} Mixing coefficients for the (mainly) decuplet
 states in the first order perturbation theory. The numbers in parentheses
 are those in the second order.}
 \begin{ruledtabular}
\begin{tabular}{c|cccc}
 $\quad{\cal R}_i\quad $ & $\Delta$ & $\Sigma^*$ & $\Xi^*$ & $\Omega$ 
 \\ \hline
 $\bm{10}$ & 
 $1$ $(-0.150)$ & $1$ $(-0.082)$ & $1$ $(-0.033)$ & $1$ $(-0.003)$ \\
 $\bm{27}_q$ & 
 $0.543$ $(0.122)$ & $0.397$ $(0.031)$ & $0.243$ $(-0.017)$ & $0$ $(0)$ \\
 $\bm{35}$ & 
 $0.065$ $(0.076)$ & $0.082$ $(0.072)$ & $0.087$ $(0.051)$ & 
 $0.082$ $(0.024)$ \\
 $\bm{\overline{35}}_q$ & 
 $0$ $(0.138)$ & $0$ $(0.049)$ & $0$ $(0)$ & $0$ $(0)$ \\
 $\bm{64}_q$ & 
 $0$ $(0.053)$ & $0$ $(0.050)$ & $0$ $(0.033)$ & $0$ $(0)$ \\
 $\bm{81}$ & 
 $0$ $(0.009)$ & $0$ $(0.013)$ & $0$ $(0.015)$ & $0$ $(0.013)$
\end{tabular}
 \end{ruledtabular}
\end{table}

\begin{table}[h]
\caption{\label{b10mix} Mixing coefficients for the (mainly)
 anti-decuplet states in the first order perturbation theory. The
 numbers in parentheses are those in the second order.}
 \begin{ruledtabular}
\begin{tabular}{c|cccc}
 $\quad{\cal R}_i\quad $ & $\Theta$ & $\mbox{N}'$ & $\Sigma'$ & $\phi$ 
 \\ \hline
 $\bm{8}$ & 
 $0$ $(0)$ & $-0.288$ $(-0.089)$ & $-0.288$ $(-0.157)$ & $0$ $(0)$ \\
 $\bm{\overline{10}}$ & 
 $1$ $(-0.061)$ & $1$ $(-0.160)$ & $1$ $(-0.235)$ & $1$ $(-0.285)$ \\
 $\bm{27}_d$ & 
 $0$ $(0)$ & $0.314$ $(0.304)$ & $0.513$ $(0.219)$ & $0.703$ $(-0.079)$ \\
 $\bm{\overline{35}}_d$ & 
 $0.351$ $(0.130)$ & $0.372$ $(0.193)$ & $0.352$ $(0.233)$ 
 & $0.277$ $(0.225)$ \\
 $\bm{64}_d$ & 
 $0$ $(0)$ & $0$ $(0.136)$ & $0$ $(0.203)$ & $0$ $(0.215)$ \\
 $\bm{\overline{81}}_d$ & 
 $0$ $(0.116)$ & $0$ $(0.128)$ & $0$ $(0.116)$ & $0$ $(0.082)$
\end{tabular}
 \end{ruledtabular}
\end{table}


The mixings are rather large. One may think that the perturbation theory
does not work. For the (mainly) octet states, the second order
contributions are much smaller than the first order ones, while for the
(mainly) decuplet and the (mainly) anti-decuplet states, the mixing with
$\bm{27}_d$ and $\bm{\overline{35}}_d$ are large and the magnitude of
the second order contributions are comparable with the first order
ones.

\section{Decay Widths}
\label{Sec:Widths}

In this section, we calculate the decay widths of various channels based
on the calculations done in the previous sections. Since our treatment
of the baryons is a quantum-mechanical one, the full-fledged field
theoretical calculation is impossible. What we actually do is a
perturbative evaluation of the decay operators in the collective
coordinate quantum mechanics.

\subsection{Formula for the decay width}

Since there seem to be confusions\cite{Jaffe:2004qj, Diakonov:2004ai,
Jaffe:2004dc} concerning the factors in the decay widths, we reconsider
the derivation of the formula. See \cite{Ellis:2004uz} for the
discussions of the calculations of the decay widths in the Skyrme model.

Decay of a baryon to another baryon with a pseudoscalar meson may be
described by the interactions of the type
\begin{equation}
 {\cal L}_{decay}=-ig\der_\mu \phi^\alpha J_5^{\alpha \mu},
  \label{Acoupling}
\end{equation}
where ${J_5}_\mu^\alpha$ is the baryon axial-vector current and
$\phi^\alpha$ is a pseudoscalar meson field. The coupling $g$ has the
dimension $(mass)^{-1}$ and usually related to the pion decay constant
$F_\pi$, $g\sim F_\pi^{-1}$.  In the nonrelativistic limit, the time
component may be dropped, and it is useful for us to work in the
Hamiltonian formulation,
\begin{equation}
 H_{decay}(t)=\int d^3x {\cal H}_{decay}
  =ig\int d^3x \der_k \phi^\alpha(x) J_5^{\alpha k}(x).
\end{equation}

In the leading order, the amplitude of the decay $B\rightarrow
B'\phi$ may be given  by
\begin{eqnarray}
 {\cal A}&=&\int dt 
  \langle \phi^\alpha(\bfp)B'(\bfP')|H_{decay}|B(\bfP)\rangle 
  \nonumber \\
 &=&ig\int d^4x 
  \langle \phi^\alpha(\bfp)
  \left|
   \der_k \phi^\beta(x)
  \right|
  0\rangle
  \left\langle 
   B'(\bfP')\left|{J_5}^{\beta k}(x)\right|B(\bfP)
  \right\rangle.
\end{eqnarray}
With the relativistic normalization of the state, the first matrix
element may be written as
\begin{equation}
 \langle \phi^\alpha(\bfp)
  \left|
   \der_k \phi^\beta(x)
  \right|
  0\rangle
  =ip^k e^{ipx}\delta^{\alpha\beta}.
\end{equation}
In our treatment of baryons, the state $|B(\bfP)\rangle$ has the wave
function
\begin{equation}
 \sqrt{2E_B(\bfP)}\Psi_B(A)e^{-iE_B(\bfP)t+i\bfP\cdot\bfX},
  \label{baryonwf}
\end{equation}
where $E_B(\bfP)=\sqrt{\bfP^2+M_B^2}$ and $\bfX$ is the position of the
baryon. The state satisfies the relativistic normalization,
\begin{equation}
 \left\langle B'(\bfP')|B(\bfP)\right\rangle
  =2E_B(\bfP)(2\pi)^3 \delta^3(\bfP'-\bfP),
\end{equation}
where the inner product is defined as the integration over $\bfX$ and
$A$.  The axial-vector current may be obtained from the $\chi$PT action
by using Noether's method. After replacing $U(x)$ with
$A(t)U_c(\bfx-\bfX)A^\dagger(t)$, we obtain the collective coordinate
quantum mechanical operator by representing the ``angular velocity''
$\omega^\alpha$ in terms of the generator $F_\alpha$ (``angular
momentum'') according to the usual rule. $J_5^{\alpha k}$ depends on
$F_\alpha$ but not on $t$ anymore.  Note that it depends on $\bfX$
through the combination $\bfx-\bfX$. Now the second matrix element may
be written as
\begin{eqnarray}
 &&\left\langle B'(\bfP')
  \left|
   {J_5}^{\beta k}(\bfx-\bfX)
  \right|
  B(\bfP)\right\rangle \nonumber \\
 &=& \sqrt{2E_{B'}2E_B} e^{i(E_{B'}-E_B)t}
  \int dA\Psi_{B'}^*(A)
  \left(\int d^3X e^{-i(\bfP'-\bfP)\cdot \bfX}
  {J_5}^{\beta k}(\bfx-\bfX)\right)\Psi_B(A).
\end{eqnarray}

By making a shift $\bfx\rightarrow \bfx+\bfX$ and integrating over
$\bfX$ and $t$, we get\footnote{${\cal M}$ corresponds to the invariant
amplitude in the relativistic field theory, though, in our formulation,
relativistic property has already been lost. The spin of the wave
function (\ref{baryonwf}) does not transform properly under boosts. The
decoupling of spin reflects the fact that baryons are now considered to
be (almost) static, i.e., $\left|\bfP\right| \ll M_B$ and
$\left|\bfP'\right| \ll M_{B'}$.}
\begin{eqnarray}
 {\cal A}
 &=&(2\pi)^4\delta^4(P'+p-P)\sqrt{2E_{B'}2E_B} 
  \int dA \Psi_{B'}^*(A)
  \left(g p^k
  \int d^3x e^{-i\bfp\cdot \bfx} {J_5}^{\alpha k}(\bfx)
  \right) \Psi_B(A)\nonumber \\
  &\equiv&(2\pi)^4\delta^4(P'+p-P){\cal M}.
\end{eqnarray}
In the nonrelativistic approximation, we put $e^{-i\bfp\cdot\bfx}
\approx 1$.  The integral of the axial-vector current,
\begin{equation}
 \frac{1}{\Lambda}{\cal O}^\alpha_{decay}
  =g p^k \int d^3x {J_5}^{\alpha k}(\bfx),
\end{equation}
depends on $A$ and $F_\alpha$, and has the right transformation property
under flavor $SU(3)$ transformations. Here we introduce a mass scale
$\Lambda$, which we take $\Lambda=1 \mbox{\rm GeV}$. It is
known\cite{Adkins:1983ya} that the leading order result is given by
\begin{equation}
 \frac{1}{\Lambda}{\cal O}^\alpha_{decay}
  \sim g  {\cal C} D^{(8)}_{\alpha k}(A)p^k,
\end{equation}
where ${\cal C}$ is a dimensionless constant. In the ``traditional''
approach, this is a functional of the profile function $F(r)$, and
obtained by explicitly integrating $J_5^{\alpha k}$ (expressed in terms
of $U_c(\bfx)$, $A$, and $F_\alpha$) over $\bfx$. Note that it has no
dependence on $M_B$ nor on $M_{B'}$.

In the effective theory approach, on the other hand, the coefficient of
the decay operator may be determined by fitting the widths to the
experimental values. We follow this way.

The leading operators are well-known,
\begin{equation}
 {\cal O}^\alpha_{decay}=
  3\left[
  G_0 D^{(8)}_{\alpha k}(A)
  -G_1\sum_{\beta,\gamma\in {\cal J}}
  d_{k \beta \gamma}D^{(8)}_{\alpha\beta}(A)F_\gamma
  -\frac{G_2}{\sqrt{3}}D^{(8)}_{\alpha 8}(A)F_k
 \right]
 p^k,
\end{equation}
where the index $k$ runs $1,2,3$. The constants $G_a\
(a=0,1,2)$ are dimensionless, thanks to the explicit mass scale $\Lambda$.

Once we calculate the amplitude ${\cal M}$, we are ready to obtain the
decay width,
\begin{equation}
 \Gamma_{B\rightarrow B'\phi}=\frac{\left|\bfp\right|^3}{8\pi M_B^2} 
  \frac{1}{\Lambda^2}\left|\overline{\cal M}\right|^2,
\end{equation}
where $\left|\bfp\right|$ stands for the magnitude of the meson
momentum in the rest frame of the initial baryon $B$,
\begin{equation}
 \left|\bfp\right|=\frac{1}{2M_B}
  \sqrt{
  \left[
   (M_B+M_{B'})^2-m^2
  \right]
  \left[
   (M_B-M_{B'})^2-m^2
  \right]
  },
\end{equation}
where $m$ is the mass of the meson.
$\left|\overline{\cal M}\right|^2$ is defined as
\begin{equation}
 \left|\overline{\cal M}\right|^2
  =\frac{1}{\left|\bfp\right|^2}
  \sum_{spin, isospin}\left|{\cal M}\right|^2.
\end{equation}
The symbol $\sum_{spin, isospin}$ denotes the average of the spin and
the isospin for the initial state baryon as well as the sum for the
final state baryon. By extracting the normalization factor,
$4M_BE_{B'}(\left|\bfp\right|)$,
\begin{equation}
 \left|\overline{\cal M}\right|^2=4M_BE_{B'}(\left|\bfp\right|)
  \left|
   \widetilde{\cal M}
   \right|^2,
\end{equation}
we may rewrite it as
\begin{equation}
 \Gamma_{B\rightarrow B'\phi}=\frac{\left|\bfp\right|^3}{2\pi} 
  \frac{E_{B'}(\left|\bfp\right|)}{\Lambda^2 M_B}
  \left|\widetilde{\cal M}\right|^2.
  \label{width}
\end{equation}
This is our formula for the decay width.

The widely used formula,
\begin{equation}
 \Gamma_{B\rightarrow B'\phi}=\frac{\left|\bfp\right|^3}{4\pi M_B M_{B'}}
  \left|\overline{\cal M}\right|^2,
\end{equation}
which corresponds to Eq.~(\ref{width}) seems to be based on the
interaction of Yukawa type
\begin{equation}
 {\cal L}=g_{\phi B'B}
  \overline{\psi}_{B'}i\gamma_5\lambda^\alpha \psi_B \phi^\alpha.
  \label{Yukawa}
\end{equation}
It is this coupling constant $g_{\phi B'B}$ that depends on the initial
and final baryons. Actually, Goldberger-Treiman relation relates
$g_{\phi B'B}$ with $g$ in Eq.~(\ref{Acoupling}). From this point of
view the inverse mass factors in the coefficients in
Ref.~\cite{Diakonov:1997mm} may be understood. Most of the authors do
not seem to care about the normalization (relativistic or
nonrelativistic) of the states.

We have a preference to Eq.~(\ref{Acoupling}) over Eq.~(\ref{Yukawa})
because the derivative coupling is a general sequence of the emission or
absorption of a Nambu-Goldstone boson at low-energies. On the other
hand, the universality (i.e., independence of the initial and final
baryons) of the coupling $g$ is generally less transparent.  It is
however very naturally understood in the Skyrme model, in which, as we
see above, the axial-vector current comes from a single expression for
all the baryons.

It is interesting to note that, although the reasoning seems very
different from ours, the decay width formula with the ratio of baryon
masses in Ref.~\cite{Diakonov:1997mm},
\begin{equation}
 \Gamma=\frac{\left|\bfp\right|^3}{2\pi\left(M_B+M_{B'}\right)^2}
  \frac{M_{B'}}{M_B}\left|\overline{\cal M}\right|^2
\end{equation}
looks similar to ours (\ref{width}), if the factor $(M_B+M_{B'})$ is
identified with our common mass scale $\Lambda$.

\subsection{The best fit values of the couplings and the predictions}
\label{Sec:decay-num}

In this subsection, we calculate the important factor
$\left|\widetilde{\cal M}\right|^2$, and then the decay widths. The
calculation goes as follows.  What we need to calculate is the matrix
element,
\begin{equation}
 \left\langle \Psi_{B'}
  \left|{\cal O}^\alpha_{decay}\right|
  \Psi_{B}\right\rangle
 =\int dA \Psi_{B'}^*(A){\cal O}^\alpha_{decay}\Psi_{B}(A).
\end{equation}
The baryon wave function $\Psi_B(A)$ is a linear combination of the
states in various representations,
\begin{equation}
 \Psi_B(A)=\sum_{i}c^B_i \Psi^{{\cal R}_i}_{FS}(A),
\end{equation}
where $\Psi^{{\cal R}_i}_{FS}(A)$ is the eigenstate of $H_0$ with $F=
(Y,I_3, I)$ and $S=(Y_R=+1, J_3, J)$ in representation ${\cal R}_i$ and
the coefficients $c^B_i$ is those we obtained in the previous
section. So we first calculate $ \left\langle \Psi^{{\cal R}_j}_{F'S'}
\left|{\cal O}^\alpha_{decay}\right| \Psi^{{\cal R}_i}_{FS}
\right\rangle$.  Furthermore, the spin and flavor structure is
completely determined by the Clebsch-Gordan (CG) coefficients. For
example, consider the matrix elements of the $G_0$ decay operator,
\begin{eqnarray}
 &&\int dA \left(\Psi^{{\cal R}_j}_{F'S'}(A)\right)^*
 D_{\alpha k}^{(8)}(A)p^k
 \Psi^{{\cal R}_i}_{FS}(A) \nonumber \\
 &=&\sqrt{\frac{\mbox{\rm dim}{\cal R}_i}{\mbox{\rm dim}{\cal R}_j}}
 P^*(S')P(S)\sum_r 
 \left(
  {\scriptsize
  \begin{array}{ccc}
   \bm{8}& {\cal R}_i& {\cal R}_j\\
   \alpha & F & F'
  \end{array}
  }
 \right)_r
 \left(
  {\scriptsize
  \begin{array}{ccc}
   \bm{8}& {\cal R}_i& {\cal R}_j\\
   k & \tilde{S} & \tilde{S}'
  \end{array}
  }
 \right)_r^* p^k.
\end{eqnarray}
Because
\begin{eqnarray}
 \left(
  {\scriptsize
  \begin{array}{ccc}
   \bm{8}& {\cal R}_i& {\cal R}_j\\
   k & \tilde{S} & \tilde{S}'
  \end{array}
  }
 \right)_r p^k
 &=&\bigg[
   \frac{p^1-ip^2}{\sqrt{2}}\langle J',-J'_3|1,1;J,-J_3\rangle
   \nonumber \\
 &&{}-\frac{p^1+ip^2}{\sqrt{2}}\langle J',-J'_3|1,-1;J,-J_3\rangle
   \nonumber \\
 &&{}-p^3\langle J',-J'_3|1,0;J,-J_3\rangle
  \bigg]
  \times 
  \left(
   {\scriptsize
   \begin{array}{cc}
    \bm{8}& {\cal R}_i\\
    1\ 0 & J\ 1
   \end{array}
   }
  \right|
  \left.
  {\scriptsize
  \begin{array}{c}
   {\cal R}_j\\
   J'\ 1
  \end{array}
  }
  \right)_{r},
\end{eqnarray}
where $\langle J,J_3|j,j_3;j',j'_3\rangle$ is the usual $SU(2)$ CG
coefficient and the last factor is the $SU(3)$ isoscalar factor, the
spin factor $\left[\cdots\right]$ can be extracted. In this way, the
matrix element may be written as
\begin{equation}
 \int dA \left(\Psi^{{\cal R}_j}_{F'S'}(A)\right)^*
 {\cal O}^\alpha_{decay}\Psi^{{\cal R}_i}_{FS}(A)
 =\left( \mbox{\rm isospin CG} \right)\times 
 \left[ \mbox{\rm spin CG with $p^k$}\right]\times 
 \widetilde{\cal M}^{{\cal R}_j{\cal R}_i}_{\alpha,(F',F)},
\end{equation}
where $\widetilde{\cal M}^{{\cal R}_j{\cal R}_i}_{\alpha,(F',F)}$
contains all the other factors, such as the phases, the isoscalar
factors, and so on.  The first two factors are irrelevant when we
calculate the average and sum of the spins and isospins and just give
$\frac{1}{3}\left|\bfp\right|^2$. Only $\widetilde{\cal M}^{{\cal
R}_j{\cal R}_i}_{\alpha,(F',F)}$ matters. Actually, after squaring the
amplitude, averaging the spin and the isospin of the initial baryon, and
summing over the spin and isospin of the final baryon, we have
$\left|\widetilde{\cal M}\right|^2$ as
\begin{equation}
 \left|\widetilde{\cal M}\right|^2
  =\frac{1}{3}
  \left|
   \sum_{i,j} (c^{B'}_j)^* 
   \widetilde{\cal M}^{{\cal R}_j{\cal R}_i}_{\alpha,(F',F)} c^{B}_i
  \right|^2.
\end{equation}

In the following, we give the factors $\widetilde{\cal M}^{{\cal
R}_j{\cal R}_i}_{\alpha,(F',F)}$ for various decays in matrix
forms. Before presenting the factors, we introduce the following
combinations of couplings
\begin{eqnarray}
\begin{array}{lll}
G_{8,10} =  G_0+\frac{1}{2}G_{1} ,&
G_{8,27} =  G_0-\frac{1}{2}G_{1} ,&
G_{\overline{10},27} = G_0+G_1. \\
G_{\overline{10},\overline{35}} = G_0-G_1 ,&
G_{27,10}  = G_0-2 G_1 ,&
G_{27,27}^{(1)}  =G_1,\\
G_{27,27}^{(2)} = G_0-\frac{1}{2}G_{1},& 
G_{27,35}  = G_0+G_1 ,&
G_{27,\overline{35}} = G_0-2 G_1, \\
G_{27,64} = G_0-\frac{1}{2}G_1 ,&
G_{\overline{35},10}  = G_0-G_1 ,& 
G_{\overline{35},27}  = G_0-2 G_1, \\
G_{\overline{35},\overline{35}}^{(1)} = G_0+11 G_1,& 
G_{\overline{35},\overline{35}}^{(2)} = G_0+\frac{4 }{5}G_{1},& 
G_{\overline{35},64}  = G_0+\frac{3 }{2}G_1,  \\
G_{64,27} = G_0-\frac{1}{2}G_1,&
G_{64,35} = G_0-\frac{5 }{2}G_1 ,&
G_{64,\overline{35}} = G_0+\frac{3}{2}G_{1} ,\\
G_{64,64}^{(1)}   =G_1,&
G_{64,64}^{(2)}  = G_0-\frac{1}{2}G_{1}, &
G_{64,81}  =G_0+\frac{3}{2}G_{1},
\end{array}
\end{eqnarray}

\begin{eqnarray}
\begin{array}{lll}
F_{10,8}  = G_0+\frac{1}{2}G_1 ,& 
F_{10,27} = G_0-2 G_1  ,& 
F_{27,8}  = G_0-\frac{1}{2}G_1  , \\
F_{27,\overline{10}} = G_0+G_1,&  
F_{27,27}^{(1)}  = G_1,& 
F_{27,27}^{(2)}  = G_0-\frac{1}{2}G_1 , \\
F_{27,\overline{35}} = G_0-2 G_1 ,& 
F_{27,64} = G_0-\frac{1}{2}G_1,& 
F_{35,27} = G_0+G_1 , \\
F_{35,64} = G_0-\frac{5}{2}G_1 ,&  
F_{\overline{35},\overline{10}}  =G_0-G_1 ,& 
F_{\overline{35},27} = G_0-2 G_1 ,\\
F_{\overline{35},\overline{35}}^{(1)}  = G_0+11 G_1 ,&  
F_{\overline{35},\overline{35}}^{(2)}  =G_0+\frac{4}{5}G_1,&  
F_{\overline{35},64} = G_0+\frac{3}{2}G_1, \\
F_{\overline{35},\overline{81}} = G_0-G_1   ,& 
F_{64,27}  = G_0-\frac{G_1}{2}   ,& 
F_{64,\overline{35}} = G_0+\frac{3}{2}G_1  , \\
F_{64,64}^{(1)}  = G_1 ,& 
F_{64,64}^{(2)} = G_0-\frac{1}{2}G_1 ,& 
F_{64,\overline{81}} = G_0-\frac{5}{2}G_1 ,\\   
F_{81,64}  = G_0+\frac{3}{2}G_1  ,
\end{array} 
\end{eqnarray}

\begin{eqnarray}
\begin{array}{lll}
H_{8,8}^{(1)} = G_0+\frac{1}{2}G_1-\frac{1}{6}G_2,&
H_{8,8}^{(2)} =  G_0 + \frac{1}{2}G_1+ \frac{1}{2}G_2,&
H_{8,\overline{10}}= G_0-G_1-\frac{1}{2}G_2,\\
H_{8,27} =  G_0-2 G_1+\frac{3}{2} G_2, &
H_{\overline{10},8,}  =  G_0-G_1-\frac{1}{2}G_2,&
H_{\overline{10},\overline{10}}  = G_0 - \frac{5}{2}G_1+\frac{1}{2}G_2, \\
H_{\overline{10},27} = G_0+\frac{11}{14}G_1 \frac{3}{14}G_2,&
H_{\overline{10},\overline{35}} = G_0 + \frac{1}{2}G_1 -\frac{3}{2} G_2 ,&
H_{27,8} =  G_0-2 G_1+\frac{3}{2}G_2 ,\\
H_{27,\overline{10}} =  G_0+\frac{11}{14}G_1+\frac{3}{14}G_2 ,&
H_{27,27}^{(1)}  = G_0+\frac{13}{6}G_1+\frac{1}{2}G_2 ,&
H_{27,27}^{(2)}  =  G_0+\frac{29}{38}G_1-\frac{13}{38} G_2  ,\\
H_{27,\overline{35}}   = G_0 - \frac{17}{10}G_1 -\frac{3}{10}G_2 ,&
H_{27,64}  = G_0-2 G_1+\frac{3 }{2}G_2 ,&
H_{\overline{35},\overline{10}} =G_0+ \frac{1}{2}G_1 - \frac{3}{2}G_2,\\
H_{\overline{35},27} = G_0-\frac{17}{10}G_1 -\frac{3}{10}G_2,&  
H_{\overline{35},\overline{35}}^{(1)} = G_0 -\frac{11}{2}G_1+\frac{1}{2}G_2,&
H_{\overline{35},\overline{35}}^{(2)}  = G_0-\frac{19 }{22}G_1+\frac{1}{2}G_2,\\
H_{\overline{35},64}  = G_0+\frac{4}{3}G_1 +\frac{1}{6}G_2,& 
H_{\overline{35},\overline{81}} = G_0 +\frac{1}{2}G_1 -\frac{3}{2}G_2,&
H_{64,27} = G_0-2 G_1+\frac{3 }{2}G_2,  \\
H_{64,\overline{35}} =  G_0+\frac{4}{3}G_1 +\frac{1}{6}G_2,& 
H_{64,64}^{(1)}  = G_0 + \frac{9}{2}G_1 +\frac{1}{2}G_2,& 
H_{64,64}^{(2)} = G_0+\frac{19}{22}G_1 -\frac{9}{22}G_2, \\
H_{64,\overline{81}} = G_0-\frac{16 }{7}G_1-\frac{3 }{14}G_2 ,
\end{array}
\end{eqnarray}
which are useful in representing $\widetilde{\cal M}^{{\cal R}_j{\cal
R}_i}_{\alpha,(F',F)}$. Our naming conventions: $G$ for the decay of the
(mainly) decuplet to the (mainly) octet, $F$ for the decay of the
(mainly) anti-decuplet to the (mainly) decuplet, and $H$ for the decay
of the (mainly) anti-decuplet to the (mainly) octet. The subscript
implies the components. The superscript $(1)$ or $(2)$ distinguishes the
outer degeneracy. For example, $G_{8,10}$ stands for the coupling which
appears in the matrix elements between $\bm{10}$ and $\bm{8}$, and is
needed for the calculation of the decay of a (mainly) decuplet baryon to
a (mainly) octet.

The first order results have been given in
Refs.~\cite{Praszalowicz:2004dn, Ellis:2004uz}. The second order
results, which overlap with some of ours have been given in
Ref.~\cite{Lee:2004in}. Our results are very extended and
lengthy. But they are actually very important part of the present paper,
and useful for $\chi$QSM calculations too.

For the decays of (mainly) decuplet baryons, we have

\begin{eqnarray}
 && \Delta\rightarrow {\rm N}+\pi:\nonumber \\
 &&
\left(
\begin{array}{cccccc}
\frac{3}{\sqrt{5}} G_{8,10}& 
 \frac{\sqrt{2}}{\sqrt{3}} G_{8,27} &0&0&0&0\\
0&\frac{\sqrt{5}}{\sqrt{6}} G_{\overline{10},27}&
 0&\frac{3}{\sqrt{14}}G_{\overline{10},\overline{35}}&0&0 \\
\frac{1}{\sqrt{30}}G_{27,10} &\frac{3}{7} G_{27,27}^{(2)}
&\frac{5}{2} \frac{\sqrt{5}}{\sqrt{21}} G_{27,35}&
\frac{\sqrt{5}}{2\sqrt{21}} G_{27,\overline{35}}
&\frac{25 }{28 {\sqrt{3}}}G_{27,64}&0 \\
0&\frac{1}{2 {\sqrt{105}}}G_{\overline{35},27}&0&
\frac{1}{68}G_{\overline{35},\overline{35}}^{(1)}
+\frac{25}{34} G_{\overline{35},\overline{35}}^{(2)}&
\frac{3\sqrt{5}}{4\sqrt{7}} G_{\overline{35},64}&0\\
0&\frac{1}{56 {\sqrt{3}}}G_{64,27}&\frac{3\sqrt{5}}{8\sqrt{7}} G_{64,35}
&\frac{3 }{8 {\sqrt{35}}}G_{64,\overline{35}}
&\frac{4}{7} G_{64,64}^{(2)}&\frac{9}{8} G_{64,81} \\
\end{array}
\right) \nonumber \\
\end{eqnarray}


\begin{eqnarray}
&&\Sigma^{*}\rightarrow \Lambda+\pi:\nonumber \\
 &&
\left(
\begin{array}{cccccc} 
\frac{3}{{\sqrt{10}}}G_{8,10} & 
 \frac{\sqrt{2}}{\sqrt{5}} G_{8,27}&0&0&0&0 \\
\frac{\sqrt{2}}{3\sqrt{5}} G_{27,10}&\frac{6\sqrt{2}}{7\sqrt{5}} G_{27,27}^{(2)}
&\frac{5\sqrt{2}}{3\sqrt{7}}  G_{27,35}&
\frac{\sqrt{10}}{3\sqrt{7}} G_{27,\overline{35}}&
\frac{25}{42 {\sqrt{2}}} G_{27,64}&0 \\
0&\frac{1}{56} G_{64,27}&
 \frac{3\sqrt{5}}{8\sqrt{7}} G_{64,35}&
 \frac{3 }{8 {\sqrt{7}}}G_{64,\overline{35}}
&\frac{2\sqrt{5}}{7} G_{64,64}^{(2)}&\frac{3\sqrt{5}}{8} G_{64,81} \\
\end{array}
\right) 
\end{eqnarray}


\begin{eqnarray}
&&\Sigma^{*}\rightarrow \Sigma+\pi:\nonumber \\
&&
\left(
\begin{array}{cccccc} 
 \frac{\sqrt{3}}{\sqrt{5}} G_{8,10}&0&0&0&0&0 \\
 0&\frac{\sqrt{3}}{2}G_{\overline{10},27}&0&
  \frac{\sqrt{3}}{2\sqrt{7}} G_{\overline{10},\overline{35}}&0&0 \\
 \frac{\sqrt{3}}{2\sqrt{5}} G_{27,10}&-\frac{\sqrt{15}}{4} G_{27,27}^{(1)}
  &\frac{5\sqrt{3}}{4\sqrt{7}} G_{27,35}&
  -\frac{\sqrt{15}}{4\sqrt{7}}G_{27,\overline{35}}&0&0 \\
 0&\frac{\sqrt{3}}{4\sqrt{7}}G_{\overline{35},27}&0
  &-\frac{1}{8\sqrt{3}}G_{\overline{35},\overline{35}}^{(1)}
  \!+\!\frac{5}{4\sqrt{3}} G_{\overline{35},\overline{35}}^{(2)}
  &\frac{\sqrt{15}}{2\sqrt{7}} G_{\overline{35},64}&0 \\
 0&0&\frac{\sqrt{15}}{2\sqrt{14}} G_{64,35}&
  -\frac{\sqrt{3}}{2\sqrt{14}}G_{64,\overline{35}}
  &-\frac{\sqrt{6}}{\sqrt{5}} G_{64,64}^{(1)}&
  \frac{3\sqrt{3}}{2\sqrt{10}} G_{64,81} 
\end{array}
\right) \nonumber \\
\end{eqnarray} 


\begin{eqnarray}
&&\Xi^{*}\rightarrow \Xi +\pi: \nonumber \\
 &&\left(
\begin{array}{ccccc}
 \frac{3 }{{\sqrt{10}}}G_{8,10}&\frac{1}{{\sqrt{15}}}G_{8,27}&0&0&0 \\
 \frac{7}{4 {\sqrt{15}}} G_{27,10}
  &\frac{57}{28\sqrt{10}}G_{27,27}^{(2)} 
  \!-\!\frac{15}{8\sqrt{10}} G_{27,27}^{(1)}&
  \frac{25}{4 {\sqrt{42}}} G_{27,35} & 
  \frac{5\sqrt{5}}{28\sqrt{3}}  G_{27,64}&0 \\
 0&\frac{\sqrt{5}}{28\sqrt{6}}G_{64,27}&\frac{9}{4 {\sqrt{14}}} G_{64,35}
  &\frac{-3}{2 {\sqrt{5}}} G_{64,64}^{(1)} 
  \!+\!\frac{11}{7 {\sqrt{5}}}G_{64,64}^{(2)}
  &\frac{7\sqrt{3}}{4\sqrt{10}}G_{64,81} \\
\end{array}
   \right) \nonumber \\
\end{eqnarray} 


With the mixing coefficients, one can easily calculate $ \left|
\widetilde{\cal M} \right|^2$. For example, for the decay
$\Delta\rightarrow {\rm N}+\pi$, it is given to first order by
\begin{equation}
 \left|\widetilde{\cal M}\right|^2=\frac{1}{3}
  \left|
   \left(
    \begin{matrix}
     c^{\rm N}_8 & c^{\rm N}_{\overline{10}} & c^{\rm N}_{27_d}
    \end{matrix}
       \right)
   \left( 
    \begin{matrix}
     \frac{3}{\sqrt{5}} G_{8,10}& \frac{\sqrt{2}}{\sqrt{3}} G_{8,27} &0 \\
     0&\frac{\sqrt{5}}{\sqrt{6}} G_{\overline{10},27}& 0 \\
     \frac{1}{\sqrt{30}}G_{27,10} &\frac{3}{7} G_{27,27}^{(2)}
     &\frac{5}{2} \frac{\sqrt{5}}{\sqrt{21}} G_{27,35} \\
    \end{matrix} 
   \right)
  \left(
   \begin{array}{c}
    c^{\Delta}_{10}\\
    c^{\Delta}_{27_q} \\
    c^{\Delta}_{35}
   \end{array}
  \right) 
  \right|^2,
\end{equation}
with $c^{\rm N}_j$ and $c^{\Delta}_i$ being given in Tables~\ref{8mix}
and \ref{10mix}. Note that the size of the matrix depends on the
quantum numbers $(I,Y)$ of the initial and final baryons, and
corresponds to our mixing coefficients given in
Tables~\ref{8mix},\ref{10mix}, and\ref{b10mix}.

The factors for the (mainly) anti-decuplet baryons are obtained
similarly. First, for $\Theta^{+}$,


\begin{eqnarray}
&&\Theta^{+} \rightarrow {\rm N}+{\rm K} :\nonumber \\
 &&\left(
\begin{array}{ccc}
 -\frac{3}{{\sqrt{5}}} H_{8,\overline{10}}&0&0 \\
 -\frac{3}{4} H_{\overline{10},\overline{10}}&
  -\frac{3 }{4 {\sqrt{7}}}H_{\overline{10},\overline{35}}&0 \\
 \frac{7\sqrt{3}}{4\sqrt{10}} H_{27,\overline{10}}&
  -\frac{5\sqrt{15}}{4\sqrt{14}} H_{27,\overline{35}}&0 \\
 \frac{3}{4 {\sqrt{14}}} H_{\overline{35},\overline{10}}&
  -\frac{10}{17{\sqrt{2}}}H_{\overline{35},\overline{35}}^{(1)}
  -\frac{11}{68\sqrt{2}}H_{\overline{35},\overline{35}}^{(2)}
&-\frac{\sqrt{2}}{\sqrt{21}} H_{\overline{35},\overline{81}} \\
 0&\frac{9}{{\sqrt{70}}} H_{64,\overline{35}}&
  -\frac{7}{{\sqrt{30}}}  H_{64,\overline{81}}\\
\end{array}
\right) 
\end{eqnarray} 

There are several interesting decay channels for the (mainly)
anti-decuplet excited nucleon ${\rm N}^{\prime}$, for the (mainly)
anti-decuplet $\Sigma^{\prime}$,
and for the (mainly) anti-decuplet $\phi$ decays,
\begin{turnpage}
 \begin{table}[h]
\begin{eqnarray}
 &&\mbox{N}^{\prime}\rightarrow \mbox{N}+\pi : \nonumber \\
 &&\left(
\begin{array}{cccccc}
 -\frac{27}{20}H_{8,8}^{(1)}-\frac{3}{4}H_{8,8}^{(2)}&
  -\frac{3}{2 {\sqrt{5}}}H_{8,\overline{10}}
  &-\frac{1}{5 {\sqrt{6}}}H_{8,27}&0&0&0 \\
 -\frac{3}{2 {\sqrt{5}}} H_{\overline{10},8}
 &-\frac{3}{8} H_{\overline{10},\overline{10}}&
 -\frac{49}{8 {\sqrt{30}}} H_{\overline{10},27}
 &-\frac{3 }{8 {\sqrt{14}}}H_{\overline{10},\overline{35}}&0&0 \\
 -\frac{1}{5 {\sqrt{6}}}H_{27,8}&-\frac{49}{8 {\sqrt{30}}} H_{27,\overline{10}}
 &-\frac{1083}{1120}H_{27,27}^{(2)}-\frac{9}{32} H_{27,27}^{(1)}
 &-\frac{25\sqrt{5}}{16\sqrt{21}} H_{27,\overline{35}}&
 -\frac{5}{28 {\sqrt{3}}} H_{27,64}&0 \\
 0&-\frac{3}{8 {\sqrt{14}}} H_{\overline{35},\overline{10}}
 &-\frac{25\sqrt{5}}{16\sqrt{21}} H_{\overline{35},27}
 &-\frac{2}{17}H_{\overline{35},\overline{35}}^{(1)}
 -\frac{121}{272} H_{\overline{35},\overline{35}}^{(2)}
 &-\frac{27}{4 {\sqrt{35}}} H_{\overline{35},64}
 &-\frac{2}{3 {\sqrt{35}}}  H_{\overline{35},\overline{81}}\\
 0&0&-\frac{5}{28 {\sqrt{3}}} H_{64,27}
 &-\frac{27}{4 {\sqrt{35}}} H_{64,\overline{35}}
 &-\frac{3}{20}H_{64,64}^{(1)}-\frac{121}{140} H_{64,64}^{(2)}&
 -\frac{49}{60} H_{64,\overline{81}}  
\end{array}
\right) 
\end{eqnarray} 
%
\begin{eqnarray}
 &&{\rm N}^{\prime}\rightarrow {\rm N}+\eta : \nonumber \\
 &&\left(
 \begin{array}{cccccc}
  -\frac{9}{20} H_{8,8}^{(1)}+\frac{3}{4} H_{8,8}^{(2)}&
   -\frac{3}{2 {\sqrt{5}}} H_{8,\overline{10}}
   &\frac{\sqrt{3}}{5\sqrt{2}} H_{8,27}&0&0&0 \\
  -\frac{3}{2 {\sqrt{5}}}H_{\overline{10},8}&
  \frac{3}{8} H_{\overline{10},\overline{10}}
  &\frac{7\sqrt{3}}{8\sqrt{10}}H_{\overline{10},27}&
  -\frac{9}{8 {\sqrt{14}}} H_{\overline{10},\overline{35}}&0&0 \\
  \frac{\sqrt{3}}{5\sqrt{2}} H_{27,8}&
  \frac{7\sqrt{3}}{8\sqrt{10}} H_{27,\overline{10}}&
  -\frac{741}{1120} H_{27,27}^{(2)}+\frac{9}{32} H_{27,27}^{(1)}&
  -\frac{5\sqrt{15}}{16\sqrt{7}}  H_{27,\overline{35}}
  &\frac{5\sqrt{3}}{28} H_{27,64}&0 \\
  0&-\frac{9}{8 {\sqrt{14}}} H_{\overline{35},\overline{10}}
   &-\frac{5\sqrt{15}}{16\sqrt{7}} H_{\overline{35},27}
   &\frac{2}{17} H_{\overline{35},\overline{35}}^{(1)}
   +\frac{121}{272} H_{\overline{35},\overline{35}}^{(2)}
   &\frac{9}{4 {\sqrt{35}}} H_{\overline{35},64}
   &-\frac{2}{{\sqrt{35}}} H_{\overline{35},\overline{81}} \\
  0&0&\frac{5\sqrt{3}}{28} H_{64,27}&
   \frac{9}{4 {\sqrt{35}}} H_{64,\overline{35}}
   &\frac{3}{20} H_{64,64}^{(1)}-\frac{99}{140} H_{64,64}^{(2)}&
   -\frac{7}{20} H_{64,\overline{81}}         
 \end{array}
\right) 
\end{eqnarray} 
%
\begin{eqnarray}
 &&{\rm N}^{\prime}\rightarrow \Delta+\pi : \nonumber \\
 &&\left(
\begin{array}{cccccc}
  \frac{6}{{\sqrt{5}}} F_{10,8}&0&
   \frac{\sqrt{2}}{\sqrt{15}}F_{10,27}&0&0&0 \\
 \frac{2\sqrt{2}}{\sqrt{3}} F_{27,8}&
 \frac{\sqrt{10}}{\sqrt{3}} F_{27,\overline{10}}
 &\frac{6}{7} F_{27,27}^{(2)}&
 \frac{1}{\sqrt{105}}F_{27,\overline{35}}&\frac{1}{28
 {\sqrt{3}}}F_{27,64}&0 \\
 0&0& \frac{5\sqrt{5}}{\sqrt{21}} F_{35,27}&0&
 \frac{3\sqrt{5}}{4\sqrt{7}} F_{35,64}&0 \\
 0&\frac{3\sqrt{2}}{\sqrt{7}} F_{\overline{35},\overline{10}}&
 \frac{\sqrt{5}}{\sqrt{21}}F_{\overline{35},27}
 &\frac{1}{34} F_{\overline{35},\overline{35}}^{(1)}
 +\frac{25}{17} F_{\overline{35},\overline{35}}^{(2)}
 &\frac{3}{4 {\sqrt{35}}} F_{\overline{35},64}&
 \frac{1}{6{\sqrt{35}}}F_{\overline{35},\overline{81}} \\
 0&0&\frac{25}{14 {\sqrt{3}}} F_{64,27}&
 \frac{3\sqrt{5}}{2\sqrt{7}}F_{64,\overline{35}}
 &\frac{8}{7} F_{64,64}^{(2)}&\frac{1}{6} F_{64,\overline{81}} \\
 0&0&0&0&\frac{9}{4} F_{81,64}&0 
\end{array}
\right)
\end{eqnarray} 

 \end{table}
\end{turnpage}

\begin{turnpage}
\begin{table}[h]
\begin{eqnarray}
&&\mbox{N}^{\prime}\rightarrow \Lambda+{\rm K} :\nonumber \\
 &&\left(
\begin{array}{cccccc}
 -\frac{9}{20} H_{8,8}^{(1)}-\frac{3}{4} H_{8,8}^{(2)}&
  \frac{3}{2 {\sqrt{5}}} H_{8,\overline{10}}
  &\frac{\sqrt{3}}{5\sqrt{2}}H_{8,27}&0&0&0 \\
 -\frac{1}{10} H_{27,8}&-\frac{7 }{4 {\sqrt{5}}}H_{27,\overline{10}}
 &-\frac{57\sqrt{3}}{280\sqrt{2}} H_{27,27}^{(2)}
 -\frac{3\sqrt{3}}{8\sqrt{2}}H_{27,27}^{(1)}
 &\frac{5\sqrt{5}}{4\sqrt{14}}H_{27,\overline{35}}&
 \frac{5}{14 {\sqrt{2}}} H_{27,64}&0 \\
 0&0&-\frac{\sqrt{15}}{28} H_{64,27}&
 -\frac{9}{4 {\sqrt{7}}} H_{64,\overline{35}}
 &-\frac{3}{4\sqrt{5}} H_{64,64}^{(1)} -\frac{11}{28\sqrt{5}} H_{64,64}^{(2)}
&\frac{7}{4 {\sqrt{5}}} H_{64,\overline{81}} 
\end{array}
\right)
\end{eqnarray} 
%
\begin{eqnarray}
&&\mbox{N}^{\prime}\rightarrow \Sigma +{\rm K} :\nonumber \\
 &&\left(
\begin{array}{cccccc} 
 \frac{27}{20} H_{8,8}^{(1)}-\frac{3}{4}H_{8,8}^{(2)}&
  -\frac{3}{2 {\sqrt{5}}} H_{8,\overline{10}}&
  \frac{1}{5 {\sqrt{6}}}{H_{8,27}}&0&0&0 \\
 \frac{3}{2 {\sqrt{5}}} H_{\overline{10},8}
 &-\frac{3}{4} H_{\overline{10},\overline{10}}&
 \frac{7}{4 {\sqrt{30}}} H_{\overline{10},27}
 &-\frac{3}{4 {\sqrt{14}}} H_{\overline{10},\overline{35}}&0&0 \\
 -\frac{1}{5} H_{27,8}&\frac{7}{4 {\sqrt{5}}} H_{27,\overline{10}}
 &\frac{171\sqrt{3}}{280\sqrt{2}} H_{27,27}^{(2)} 
 -\frac{3\sqrt{3}}{8\sqrt{2}} H_{27,27}^{(1)}
 &-\frac{5\sqrt{5}}{4\sqrt{14}} H_{27,\overline{35}}&
 \frac{5}{28 {\sqrt{2}}} H_{27,64}&0 \\
 0&\frac{3}{4 {\sqrt{7}}} H_{\overline{35},\overline{10}} &
 \frac{5\sqrt{5}}{4\sqrt{42}} H_{\overline{35},27}
 &-\frac{10}{17\sqrt{2}}H_{\overline{35},\overline{35}}^{(1)}
 -\frac{11}{68\sqrt{2}}H_{\overline{35},\overline{35}}^{(2)}
 &\frac{9}{4 \sqrt{70}} H_{\overline{35},64}&
 -\frac{\sqrt{2}}{\sqrt{35}} H_{\overline{35},\overline{81}} \\
 0&0&-\frac{5\sqrt{5}}{28\sqrt{3}}  H_{64,27}&
 \frac{9}{4 {\sqrt{7}}} H_{64,\overline{35}}
 &-\frac{3}{4\sqrt{5}} H_{64,64}^{(1)}+\frac{33}{28\sqrt{5}} H_{64,64}^{(2)}
 &-\frac{7}{4 {\sqrt{5}}} H_{64,\overline{81}}  
\end{array}
\right) \nonumber \\
\end{eqnarray} 
%
\begin{eqnarray}
 &&\Sigma^{\prime}\rightarrow \mbox{N}+\mbox{K} :\nonumber \\
 &&\left(
\begin{array}{cccccc}
 -\frac{9\sqrt{3}}{10\sqrt{2}}  H_{8,8}^{(1)}
  +\frac{\sqrt{3}}{2\sqrt{2}} H_{8,8}^{(2)}
  &-\frac{\sqrt{3}}{\sqrt{10}} H_{8,\overline{10}}&
  \frac{\sqrt{2}}{5\sqrt{3}} H_{8,27}&0&0&0 \\
 \frac{\sqrt{3}}{\sqrt{10}} H_{\overline{10},8}&
 \frac{\sqrt{3}}{2\sqrt{2}} H_{\overline{10},\overline{10}}
 &-\frac{7}{2 {\sqrt{30}}} H_{\overline{10},27}&
 -\frac{\sqrt{3}}{2\sqrt{14}} H_{\overline{10},\overline{35}}&0&0 \\
 -\frac{1}{15} H_{27,8}&-\frac{7}{12 {\sqrt{5}}} H_{27,\overline{10}}
 &-\frac{171}{280} H_{27,27}^{(2)}
 +\frac{3}{8} H_{27,27}^{(1)}&
 -\frac{5\sqrt{5}}{12\sqrt{7}}H_{27,\overline{35}}
&\frac{5\sqrt{5}}{42\sqrt{2}} H_{27,64}&0 \\
 0&\frac{\sqrt{3}}{4\sqrt{7}}  H_{\overline{35},\overline{10}}&
 \frac{5\sqrt{5}}{4\sqrt{21}}H_{\overline{35},27}
 &\frac{10}{17\sqrt{3}}H_{\overline{35},\overline{35}}^{(1)}
 +\frac{11}{68\sqrt{3}} H_{\overline{35},\overline{35}}^{(2)}
 &-\frac{3\sqrt{3}}{2\sqrt{14}} H_{\overline{35},64}&
 -\frac{2}{3{\sqrt{7}}} H_{\overline{35},\overline{81}} \\
 0&0&-\frac{5}{28 \sqrt{3}} H_{64,27} &
 -\frac{3\sqrt{3}}{4\sqrt{35}}  H_{64,\overline{35}}
 &\frac{\sqrt{3}}{2\sqrt{10}}H_{64,64}^{(1)}
 -\frac{11\sqrt{3}}{14\sqrt{10}} H_{64,64}^{(2)}
 &-\frac{7}{12 \sqrt{5}} H_{64,\overline{81}}
\end{array}
\right) \nonumber \\
\end{eqnarray} 
\end{table}
\end{turnpage}
\begin{turnpage}
\begin{table}[h]
\begin{eqnarray}
 &&\Sigma^{\prime} \rightarrow \Sigma +\pi :\nonumber \\
 &&\left(
\begin{array}{cccccc}
 -\frac{\sqrt{3}}{\sqrt{2}} H_{8,8}^{(2)}&
  -\frac{\sqrt{3}}{\sqrt{10}}H_{8,\overline{10}}&0&0&0&0 \\
 -\frac{\sqrt{3}}{\sqrt{10}} H_{\overline{10},8}&
 -\frac{\sqrt{3}}{2\sqrt{2}} H_{\overline{10},\overline{10}}
 &-\frac{7\sqrt{3}}{4\sqrt{10}}H_{\overline{10},27}&
 -\frac{\sqrt{3}}{4\sqrt{14}}H_{\overline{10},\overline{35}}&0&0 \\
 0&-\frac{7\sqrt{3}}{4\sqrt{10}} H_{27,\overline{10}}&
 -\frac{3\sqrt{3}}{8\sqrt{2}} H_{27,27}^{(1)}
 &-\frac{5\sqrt{15}}{8\sqrt{14}}  H_{27,\overline{35}}&0&0 \\
 0&-\frac{\sqrt{3}}{4\sqrt{14}} H_{\overline{35},\overline{10}}&
 -\frac{5\sqrt{15}}{8\sqrt{14}} H_{\overline{35},27}&
 -\frac{1}{2 {\sqrt{6}}} H_{\overline{35},\overline{35}}^{(1)}
 -\frac{11}{8\sqrt{6}} H_{\overline{35},\overline{35}}^{(2)}
 &-\frac{3\sqrt{3}}{2\sqrt{7}} H_{\overline{35},64}&
 -\frac{1}{2 {\sqrt{14}}}H_{\overline{35},\overline{81}} \\
 0&0&0&-\frac{3\sqrt{3}}{2\sqrt{7}} H_{64,\overline{35}}&
 -\frac{\sqrt{3}}{5\sqrt{2}} H_{64,64}^{(1)}&
-\frac{7}{10} H_{64,\overline{81}} 
\end{array}
\right)
\end{eqnarray} 
%
\begin{eqnarray}
 &&\Sigma^* \rightarrow \Sigma +\eta :\nonumber \\
 &&\left(
\begin{array}{cccccc}
\frac{9}{10} H_{8,8}^{(1)}&-\frac{3}{2 {\sqrt{5}}}H_{8,\overline{10}}&
 \frac{1}{5} H_{8,27}&0&0&0 \\
 -\frac{3}{2 {\sqrt{5}}} H_{\overline{10},8}&0&
 \frac{7}{4 {\sqrt{5}}}H_{\overline{10},27}
 &-\frac{3}{4 {\sqrt{7}}}H_{\overline{10},\overline{35}}&0&0 \\
 \frac{1}{5} H_{27,8}&\frac{7}{4 {\sqrt{5}}}H_{27,\overline{10}}&
 -\frac{57}{280} H_{27,27}^{(2)}
 &-\frac{5\sqrt{5}}{8\sqrt{7}} H_{27,\overline{35}}&
 \frac{5\sqrt{5}}{28\sqrt{2}} H_{27,64}&0 \\
 0&-\frac{3}{4 {\sqrt{7}}}H_{\overline{35},\overline{10}}&
 -\frac{5\sqrt{5}}{8\sqrt{7}} H_{\overline{35},27}
 &-\frac{1}{34}H_{\overline{35},\overline{35}}^{(1)}
 +\frac{55}{136} H_{\overline{35},\overline{35}}^{(2)}
 &\frac{9}{4 {\sqrt{14}}}H_{\overline{35},64}&
 -\frac{\sqrt{3}}{2\sqrt{7}} H_{\overline{35},\overline{81}} \\
 0&0&\frac{5\sqrt{5}}{28\sqrt{2}} H_{64,27}&
 \frac{9}{4 {\sqrt{14}}}H_{64,\overline{35}}
 &-\frac{33}{70} H_{64,64}^{(2)}&
 -\frac{7\sqrt{3}}{20\sqrt{2}} H_{64,\overline{81}}
\end{array}
\right) 
\end{eqnarray} 
%
\begin{eqnarray}
 &&\Sigma^{\prime} \rightarrow \Lambda +\pi :\nonumber \\
&&\left(
\begin{array}{cccccc}
\frac{9}{10} H_{8,8}^{(1)}&
 \frac{3}{2 {\sqrt{5}}}H_{8,\overline{10}}&\frac{1}{5} H_{8,27}&0&0&0 \\
 \frac{1}{30} H_{27,8}&\frac{7}{6 {\sqrt{5}}} H_{27,\overline{10}}
 &\frac{57}{70} H_{27,27}^{(2)}&
 \frac{5\sqrt{5}}{6\sqrt{7}} H_{27,\overline{35}}&
 \frac{5\sqrt{5}}{42\sqrt{2}}H_{27,64}&0 \\
 0&0&\frac{\sqrt{5}}{28\sqrt{2}}  H_{64,27}&
 \frac{9}{4 {\sqrt{14}}} H_{64,\overline{35}}
 &\frac{11}{14} H_{64,64}^{(2)}&\frac{7}{4 {\sqrt{6}}} H_{64,\overline{81}}
\end{array}
\right)
\end{eqnarray} 
\end{table}
\end{turnpage}
\begin{turnpage}
\begin{table}[h]
\begin{eqnarray}
&&\Sigma^{\prime} \rightarrow \Xi + {\rm K} :\nonumber \\
 &&\left(
\begin{array}{cccccc}
 -\frac{9\sqrt{3}}{10\sqrt{2}} H_{8,8}^{(1)}
  -\frac{\sqrt{3}}{2\sqrt{2}} H_{8,8}^{(2)}&
  {\sqrt{\frac{3}{10}}} H_{8,\overline{10}}&
  \frac{\sqrt{2}}{5\sqrt{3}} H_{8,27}&0&0&0 \\
 -\frac{1}{15} H_{27,8}&-\frac{7}{3 {\sqrt{5}}}H_{27,\overline{10}}&
 -\frac{171}{280} H_{27,27}^{(2)}
 -\frac{3}{8}H_{27,27}^{(1)}&
 \frac{5\sqrt{5}}{6\sqrt{7}} H_{27,\overline{35}}
 &\frac{5\sqrt{5}}{42\sqrt{2}} H_{27,64}&0 \\
 0&0&-\frac{5}{28 {\sqrt{3}}} H_{64,27}&
 -\frac{3\sqrt{15}}{4\sqrt{7}}H_{64,\overline{35}}
 &-\frac{\sqrt{3}}{2\sqrt{10}}H_{64,64}^{(1)}
 -\frac{11\sqrt{3}}{14\sqrt{10}} H_{64,64}^{(2)}
 &\frac{7}{4 {\sqrt{5}}}H_{64,\overline{81}}
\end{array}
\right)
\end{eqnarray} 
%
\begin{eqnarray}
&&\Sigma^{\prime} \rightarrow \Sigma^{*} + \pi : \nonumber \\
 &&\left(
\begin{array}[c]{cccccc}
 -\frac{\sqrt{6}}{\sqrt{5}} F_{10,8}&0&
  -\frac{\sqrt{3}}{\sqrt{10}} F_{10,27}&0&0&0 \\
 0&-\frac{\sqrt{3}}{\sqrt{2}} F_{27,\overline{10}}&
 \frac{\sqrt{15}}{2\sqrt{2}} F_{27,27}^{(1)}
 &-\frac{\sqrt{3}}{2\sqrt{14}} F_{27,\overline{35}}&0&0 \\
 0&0&-\frac{5\sqrt{3}}{2\sqrt{14}} F_{35,27}&0&
 -\frac{\sqrt{15}}{2\sqrt{7}}F_{35,64}&0 \\
 0&-\frac{\sqrt{3}}{\sqrt{14}} F_{\overline{35},\overline{10}}&
 \frac{\sqrt{15}}{2\sqrt{14}}F_{\overline{35},27}
 &\frac{1}{4\sqrt{6}}F_{\overline{35},\overline{35}}^{(1)}
 -\frac{5}{2{\sqrt{6}}}F_{\overline{35},\overline{35}}^{(2)}
 &\frac{\sqrt{3}}{2\sqrt{7}} F_{\overline{35},64}&
 -\frac{1}{4 {\sqrt{14}}}F_{\overline{35},\overline{81}} \\
 0&0&0&-\frac{\sqrt{15}}{\sqrt{14}} F_{64,\overline{35}}& 
 \frac{2\sqrt{3}}{\sqrt{5}}F_{64,64}^{(1)}
 &-\frac{1}{{\sqrt{10}}}F_{64,\overline{81}} \\
 0&0&0&0&-\frac{3\sqrt{3}}{2\sqrt{5}} F_{81,64}&0 
\end{array}
\right)
\end{eqnarray} 
%
\begin{eqnarray}
 &&\Sigma^{\prime} \rightarrow \Delta + \mbox{K} :\nonumber \\
 &&\left(
\begin{array}{cccccc}
 \frac{ 2 \sqrt{6}}{\sqrt{5}}
  F_{10,8}&0&\frac{1}{{\sqrt{30}}}F_{10,27}&0&0&0 \\
 -\frac{4}{3} F_{27,8}&\frac{2\sqrt{5}}{3} F_{27,\overline{10}}
 &\frac{9}{14} F_{27,27}^{(2)}+\frac{5}{4} F_{27,27}^{(1)}
 &\frac{1}{3 {\sqrt{35}}}F_{27,\overline{35}}&
 \frac{\sqrt{5}}{42\sqrt{2}}F_{27,64}&0 \\
 0&0&\frac{5\sqrt{5}}{2\sqrt{21}} F_{35,27}&0&
 \frac{\sqrt{3}}{2\sqrt{14}} F_{35,64}&0 \\
 0&-\frac{2\sqrt{3}}{\sqrt{7}} F_{\overline{35},\overline{10}}&
 -\frac{\sqrt{5}}{\sqrt{21}} F_{\overline{35},27}
 &\frac{13}{68\sqrt{3}}F_{\overline{35},\overline{35}}^{(1)}
 +\frac{35}{17\sqrt{3}}F_{\overline{35},\overline{35}}^{(2)}
 &\frac{\sqrt{3}}{2\sqrt{14}} F_{\overline{35},64}&
 \frac{1}{12\sqrt{7}}F_{\overline{35},\overline{81}} \\
 0&0&-\frac{25}{14\sqrt{3}} F_{64,27}&
 \frac{\sqrt{15}}{2\sqrt{7}} F_{64,\overline{35}}
 &\frac{2\sqrt{6}}{\sqrt{5}}F_{64,64}^{(1)}
 +\frac{4\sqrt{6}}{7\sqrt{5}} F_{64,64}^{(2)}&
 \frac{1}{6 {\sqrt{5}}}F_{64,\overline{81}} \\
 0&0&0&0&\frac{3\sqrt{3}}{2\sqrt{10}}F_{81,64}&0 
\end{array} 
\right)
\end{eqnarray} 
\end{table}
\end{turnpage}
\begin{turnpage}
\begin{table}[h]
\begin{eqnarray}
&&\phi \rightarrow \Sigma + {\rm K} :\nonumber \\
 &&\left(
\begin{array}{ccccc}
 -\frac{3}{{\sqrt{10}}} H_{8,\overline{10}}&
  \frac{1}{{\sqrt{15}}}H_{8,27}&0&0&0 \\
 \frac{3}{4 {\sqrt{2}}} H_{\overline{10},\overline{10}}&
 -\frac{7}{8 {\sqrt{3}}} H_{\overline{10},27}
 &-\frac{3\sqrt{5}}{8\sqrt{7}} H_{\overline{10},\overline{35}}&0&0 \\
 -\frac{7}{8 {\sqrt{10}}} H_{27,\overline{10}}
 &-\frac{171\sqrt{3}}{224\sqrt{5}} H_{27,27}^{(2)}
 +\frac{15\sqrt{3}}{32\sqrt{5}} H_{27,27}^{(1)}
 &-\frac{25}{16 {\sqrt{7}}} H_{27,\overline{35}}&
 \frac{5\sqrt{5}}{28\sqrt{2}} H_{27,64}&0 \\
 \frac{3}{8 {\sqrt{14}}} H_{\overline{35},\overline{10}}&
 \frac{25}{16 {\sqrt{21}}} H_{\overline{35},27}
 &\frac{5\sqrt{5}}{34} H_{\overline{35},\overline{35}}^{(1)}
 +\frac{11\sqrt{5}}{272} H_{\overline{35},\overline{35}}^{(2)}
 &-\frac{9}{4 {\sqrt{14}}} H_{\overline{35},64}&
 -\frac{\sqrt{3}}{2\sqrt{7}} H_{\overline{35},\overline{81}} \\
 0&-\frac{5}{28 {\sqrt{6}}} H_{64,27}&
 -\frac{9}{4 {\sqrt{70}}} H_{64,\overline{35}}
 &\frac{3}{10} H_{64,64}^{(1)}-\frac{33}{70} H_{64,64}^{(2)}&
 -\frac{7\sqrt{3}}{20\sqrt{2}} H_{64,\overline{81}}
\end{array}
\right)
\end{eqnarray} 
%
\begin{eqnarray}
 &&\phi \rightarrow \Xi + \pi : \nonumber \\
 &&\left(
\begin{array}{ccccc}
 \frac{3}{\sqrt{10}} H_{8,\overline{10}}&\frac{1}{{\sqrt{15}}}H_{8,27}&0&0&0 \\
 \frac{7 }{4 {\sqrt{15}}}H_{27,\overline{10}} &
 \frac{57}{28 {\sqrt{10}}}H_{27,27}^{(2)}
 &\frac{25}{4 {\sqrt{42}}} H_{27,\overline{35}}
 &\frac{5\sqrt{5}}{28\sqrt{3}}  H_{27,64}&0 \\
 0&\frac{\sqrt{5}}{28\sqrt{6}} H_{64,27}&
 \frac{9 }{4 {\sqrt{14}}}H_{64,\overline{35}}
 &\frac{11}{7 {\sqrt{5}}} H_{64,64}^{(2)}
 &\frac{7\sqrt{3}}{4\sqrt{10}} H_{64,\overline{81}} 
\end{array}
\right) 
\end{eqnarray} 
%
\begin{eqnarray}
&& \phi \rightarrow \Xi^{*} + \pi : \nonumber \\
 &&\left(
\begin{array}{ccccc}
 0&\frac{1}{{\sqrt{3}}}F_{10,27}  &0&0&0 \\
 \frac{1}{\sqrt{3}}F_{27,\overline{10}}
 &-\frac{3\sqrt{2}}{7} F_{27,27}^{(2)}&
 \frac{\sqrt{5}}{\sqrt{42}} F_{27,\overline{35}}&
 -\frac{5}{28 {\sqrt{3}}} F_{27,64}&0 \\
 0&\frac{\sqrt{5}}{\sqrt{42}} F_{35,27}&0&\frac{3}{4}
 {\sqrt{\frac{5}{7}}} F_{35,64}&0 \\
 0&-\frac{5}{28 {\sqrt{3}}} F_{64,27}
 &\frac{3\sqrt{5}}{4\sqrt{7}} F_{64,\overline{35}}&
 -\frac{4\sqrt{2}}{7}  F_{64,64}^{(2)}
 &\frac{\sqrt{3}}{4} F_{64,\overline{81}} \\
 0&0&0&\frac{\sqrt{3}}{4} F_{81,64}&0 \\
\end{array}
\right)
\end{eqnarray} 

\end{table}
\end{turnpage}

Let us assemble all of these ingredients. In perturbation theory, it is
important to keep the order of the expansion. In order to obtain the
widths to second order in $\delta m$, we  need to
drop the higher order terms consistently.

Because the mixings among the representations are large, however, we
often get \textit{negative} decay widths if we drop higher order terms
in the squares of the amplitudes. Of course such results are
unacceptable. Considering that the perturbative contributions are large,
we also show the results in which the squares are \textit{not}
expanded. (In Appendix~\ref{Sec:DPP}, we show that negative decay widths
can appear even the mixings are smaller.)

We perform the second order calculation as well as the first order one.
In the following, the case $(a)$ is the first order result with the
square of the amplitude being expanded, the case $(a')$ with it being
not expanded.  The cases $(b)$ and $(b')$ are corresponding second order
results.  The second order results are drastically different from the
first order ones. But, before discussing the results, let us explain the
procedure.

First of all, we best fit the couplings, $G_0$ and $G_1$, to the
experimental decay widths of the (mainly) decuplet baryons. The
procedure is similar to that in the previous section. We minimize
\begin{equation}
 \chi^2=\sum_i\frac{\left(\Gamma_i-\Gamma_i^{exp}\right)^2}{\sigma_i^2},
\end{equation}
where $\Gamma_i$ stands for the calculated value of channel $i$,
$\Gamma_i^{exp}$ is its experimental value, and $\sigma_i$ represents
experimental uncertainty. The values we use are given in
Table~\ref{decayex}. We give the best fit set of couplings $G_0$ and
$G_1$ in Table~\ref{Ga}. We then determine $G_2$ by using the $F/D$
ratio,
\begin{equation}
 F/D=\frac{5}{9}\frac{H_{88}^{(2)}}{H_{88}^{(1)}}.
\end{equation}
The experimental value of $F/D$ is $0.56\pm 0.02$, though many authors
prefer to use $F/D=0.59$. The values of $G_2$ for various cases are also
given in Table~\ref{Ga}.

\begin{table}[h]
 \caption{\label{decayex}Experimental values of the widths of the
 (mainly) decuplet baryons and their uncertainty. } 
 \begin{ruledtabular}
  \begin{tabular}{c|cccc}
   (MeV)&$\Delta \rightarrow {\rm N}+\pi$ & 
   $\Sigma^*\rightarrow \Lambda+\pi$ & 
   $\Sigma^*\rightarrow \Sigma +\pi$ & 
   $\Xi^*\rightarrow \Xi+\pi$ \\ \hline
   $\Gamma^{exp}$ & 120 & 32.6 & 4.45 & 9.50 \\
   $\sigma$       & 5   & 0.74 & 0.74 & 1.8
 \end{tabular}
 \end{ruledtabular}
\end{table}

\begin{table}[h]
 \caption{\label{Ga}Coefficients of the decay operators. $G_0$ and $G_1$
 are obtained by fitting the calculated decay widths of the (mainly)
 decuplet baryons to the experimental values, while $G_2$ is fixed by
 the $F/D$ ratio. The results $(a)$ and $(a')$ are those in the first
 order with the squares of the amplitudes being expanded, and not
 expanded, respectively. The results $(b)$ and $(b')$ are those in the
 second order. The $\chi^2$ values for these cases are also given.}
 \begin{ruledtabular}
  \begin{tabular}{c|cccc}
   &$(a)$& $(a')$& $(b)$& $(b')$ \\ \hline
   $G_0$    & 4.74  & 5.29  & 5.64  & 5.77  \\
   $G_1$    & 13.5  & 8.81  & 15.4  & 9.93  \\
   $\chi^2$ & 38.8  & 4.73  & 13.1  & 6.43  \\ \hline
   $G_2$    & 0.07  & 0.08  & 0.08  & 0.09
 \end{tabular}
 \end{ruledtabular}
\end{table}

The best fit values of the (mainly) decuplet decay widths are summarized
in Table~\ref{decay10}, where we also give the phase space factor $K$,
\begin{equation}
 \Gamma=K\left|\widetilde{\cal M}\right|^2,\qquad
  K\equiv \frac{\left|\bfp\right|^3}{2\pi \Lambda^2}\frac{E_{B'}}{M_B}.
  \label{kin}
\end{equation}
We use the experimental values for the baryon and the meson masses, if
they are known, in all decay width calculations. For (mainly)
anti-decuplet ${\rm N}'$ and $\Sigma'$, we use our predicted values
(\ref{massprediction}).

\begin{table}[h]
 \caption{\label{decay10}Decay widths for the (mainly) decuplet
 baryons. Calculations are done with the coupling constants given in
 Table~\ref{Ga}. The kinematical factor $K$ is explained in
 Eq.~(\ref{kin}). }
\begin{ruledtabular}
 \begin{tabular}{c||ccccc}
  (MeV)& $K$ & 
  $\Gamma_{(a)}$ & 
  $\Gamma_{(a')}$ &
  $\Gamma_{(b)}$ &
  $\Gamma_{(b')}$ 
  \\ \hline
  $\Delta\rightarrow {\rm N}\pi$    & 1.47 & 92.1 & 114  & 110  & 112 \\
  $\Sigma^*\rightarrow \Lambda \pi$ & 1.18 & 33.9 & 32.8 & 32.8 & 32.9 \\
  $\Sigma^*\rightarrow \Sigma \pi$  & 0.26 & 5.08 & 5.50 & 5.73 & 5.31 \\
  $\Xi^*\rightarrow \Xi \pi$        & 0.49 & 13.0 & 11.4 & 13.9 & 12.2
 \end{tabular}
\end{ruledtabular}
\end{table}

Once we determine the best fit values of $G_a$, the decay widths of
(mainly) anti-decuplet baryons can be calculated. The results are given
in Table~\ref{decay10bar}.

\begin{table}[h]
 \caption{\label{decay10bar}Predictions for the decay widths for the 
 (mainly) anti-decuplet  baryons.}
\begin{ruledtabular}
 \begin{tabular}{c||ccccc}
  (MeV)& $K$ & 
  $\Gamma_{(a)}$ & 
  $\Gamma_{(a')}$ &
  $\Gamma_{(b)}$ &
  $\Gamma_{(b')}$ 
  \\ \hline
  $\Theta^+\rightarrow {\rm N} {\rm K}$ & 1.91 & 217  & 74.7 & 378  & 147 \\
  ${\rm N}'\rightarrow {\rm N}\pi$      & 20.9 & 870  & 246  & 1417 & 106 \\
  ${\rm N}'\rightarrow {\rm N}\eta$     & 6.71 &-43.1 & 0.07 & -84.7& 0.61\\
  ${\rm N}'\rightarrow \Delta \pi$      & 7.36 & 0    & 72.6 & 144  & 188 \\
  ${\rm N}'\rightarrow \Lambda {\rm K}$ & 2.04 & 25.1 & 8.37 & 58.5 & 34.6\\
  ${\rm N}'\rightarrow \Sigma {\rm K}$  & 0.22 & 7.04 & 2.50 & 13.6 & 5.10\\
  $\Sigma'\rightarrow {\rm N} {\rm K}$  & 15.1 & -1747& 22.1 & -164 & 85.7\\
  $\Sigma'\rightarrow \Sigma \pi$       & 14.6 & 443  & 86.6 & 757  & 50.7\\
  $\Sigma'\rightarrow \Sigma \eta$      & 1.62 &-4.11 & 0.61 & -13.4& 0.13\\
  $\Sigma'\rightarrow \Lambda \pi$      & 18.7 & 476  & 114  & 721  & 194 \\
  $\Sigma'\rightarrow \Xi {\rm K}$      & 0.03 & 0.10 & 0.04 & 0.11 & 0.04\\
  $\Sigma'\rightarrow \Sigma^*\pi$      & 5.94 & 0    & 6.71 & 2.25 & 81.1\\
  $\Sigma'\rightarrow \Delta {\rm K}$   & 1.93 & 0    & 41.2 & 25.2 & 116 \\
  $\phi\rightarrow \Sigma {\rm K}$      & 5.12 & -137 & 14.9 & -181 & 44.5\\
  $\phi\rightarrow \Xi \pi$             & 10.8 & 416  & 62.9 & 294  & 58.6\\
  $\phi\rightarrow \Xi^* \pi$           & 2.68 & 0    & 8.63 & 47.2 & 12.5
 \end{tabular}
\end{ruledtabular}
\end{table}

By looking at the Table~\ref{decay10bar}, one can easily see that the
widths change rather randomly by going to the second order from the
first order. It gets wider in a channel, while narrower in another. The
behavior also depends on whether we expand the amplitude or not.

We have also done the calculations with the calculated baryon masses
even for $\mbox{N}$, $\Delta$, etc., and have found that the results
change drastically.

Our result for the width of $\Theta^+$ is of order 100 MeV, which
clearly contradicts with reported experimental results. 

Still, one may get some insights from our calculations. First of all, we
see the smaller the coupling $G_1$ is, the narrower the widths
are. Second, our results are almost insensitive to the value of
$G_2$. We did the similar calculations with $F/D=0.59$, but the results
are only slightly changed and qualitatively the same despite the fact
that the value of $G_2$ changes considerably. 

\section{Summary and Discussions}
\label{Sec:Sum}

In this paper, we reconsider the Skyrme model from an effective theory
point of view. In this approach, the Skyrme model parameters, which
appear in the collective coordinate quantized Hamiltonian, are
determined by fitting the calculated baryon masses to the experimental
values. Once the Skyrme model parameters have been fixed, various
physical quantities can be calculated. In particular, we make a
prediction for the masses of ${\rm N}^{\prime}$ and $\Sigma^{\prime}$.

The idea behind this approach is that the $\chi$PT provides a framework
that represents QCD at low energies in which the soliton picture of
baryons, which is a consequence of the large-$N_c$ limit, emerges. We
start with the action up to ${\cal O}(p^4)$ and keep only the terms
which are of leading order in $N_c$. After quantization, we calculate
everything as a systematic expansion in powers of $\delta
m$. Note however that we keep in mind that there are
infinitely many terms which contribute to the Skyrme model
parameters. Thus the number of parameters appears to increase from the
starting action to the Hamiltonian. From the effective theory point of
view, it is not the number of
independent parameters of the ``model'' but the symmetry and the power
counting that matter.  The
``derivation'' given in Appendix~\ref{Sec:Trad} is, however, convenient to
generate relevant operators which respect them.

The basic idea that the higher order contributions improve the Skyrme
model picture seem to be justified in Appendix~\ref{Sec:Trad} by
comparing the conventional Skyrme model with the one with $L_5$ and
$L_8$ terms in the ``traditional'' approach.

We have performed the complete second-order calculations
for the masses, and determined the Skyrme model parameters. We find that,
although the octet behaves good, the decuplet and the anti-decuplet
have large mixings and the perturbative treatment may be questioned.

We also re-examine the decay calculation in the Skyrme models, by
deriving the formula for the decay width from the derivative coupling
interaction to Nambu-Goldstone bosons. A careful derivation reveals how
the decay widths depend on the initial and final baryons. 

We calculate the widths of several interesting decays by using the
formula. In particular, our calculation predicts a wide
decay width for $\Theta^{+}$, in contradiction to the
experiments. If $\Theta^{+}$ has really a very narrow width, as
reported, our theory fails to reproduce it. A possible explanation of
this failure is that our perturbative treatment is poor for the (mainly)
decuplet and the (mainly) anti-decuplet states. Because the decay
parameters are determined by using the (mainly) decuplets, this could
influence very much. Another explanation comes from the very subtle
nature of the decay width calculations. The results heavily on the
kinematics, i.e., the masses of the baryons and the factors in the
formula. A few percent change of the mass can often cause a hundred
percent (even more) change of the decay width. The theoretical ambiguity
is extremely large. The results in Appendix~\ref{Sec:DPP} seem to
support this explanation.

Is our fitting procedure appropriate? We vary all of the parameters as
free parameters and treat then equally. But there must be a natural
hierarchy in them: leading order parameters must be fitted to the bulk
structure, and subleading parameters should account for fine
structures. We tried to find such a systematic procedure, but so far,
the presented method is the most satisfactory.

What should we do to improve the results?  It is difficult
to go to the next order in perturbation theory, because in the next
order we need to include more operators, thus more parameters to be
fitted. A diagonalization, rather than perturbative expansion, may be an
option, though it somehow goes beyond the controlled effective theory
framework.

Does $\Theta^{+}$ really exist? Is it narrow? Why so? Is the narrowness
a general feature of the Skyrme model? We do not have definite answers
yet. But our results suggest that if it really exists and is
really narrow, it seems very peculiar even from the Skyrme model point
of view. As shown in Appendix~\ref{Sec:DPP}, it is not just because of
the difference of the symmetry breaking interactions.

Prasza\l owicz\cite{Praszalowicz:2003tc} discussed how $\Theta^{+}$
becomes narrow in the large-$N_c$ limit, and showed that the narrowness
comes from the interplay between the cancellation in
$G_{\overline{10}}\equiv H_{8,\overline{10}}$ (which becomes exact in
the nonrelativistic limit) and the phase space volume dependence. The
important factor of his argument is of course the cancellation in
$G_{\overline{10}}$, but it comes from the $\chi$QSM calculations. In
our effective theory treatment, on the other hand, the couplings $G_a$
are parameters to be fitted. The only possible way to understand such
``cancellation'' in the effective theory context seems symmetry. We do
not know if it exists, nor what it is.

\appendix

\section{Mathematical Tools}
\label{Sec:Math}

In this section, we summarize some basic mathematical formulae for the
calculations of the matrix elements. 

\subsection{Properties of the wave function and basic formulae for
  matrix elements}

Let us first introduce the notations. A baryon wave function $\Psi$ has
the flavor index $F=(Y,I_3,I)$ and the ``spin'' index $S=(Y_R,J_3,J)$.
The eigenstate wave function (\ref{eigenstate}) of $H_0$ may be written
as
\begin{equation}
 \Psi^{\cal R}(g)_{FS}=\sqrt{\mbox{\rm dim}{\cal R}}P(S) 
  \left(D^{\cal R}_{F,\tilde{S}}(g)\right)^*, \qquad 
  g\in SU(3),
\end{equation}
where $ P(S)=(-1)^{J_3-Y_R/2} $, $\tilde{S}=(Y_R,-J_3,J)$, and $D^{\cal
R}(g)$ is the representation matrix of $g$ for representation ${\cal
R}$. Note that physical states must satisfy the constraint
(\ref{WZWconstraint}), $Y_R=1$, but this is irrelevant to most of the
results in this section. The wave function is normalized in the sense,
\begin{equation}
 \left\langle\Psi^{\cal R}_{FS}\big|\Psi^{{\cal R}'}_{F'S'}\right\rangle
  =\delta_{{\cal R}{\cal R}'}\delta_{FF'}\delta_{SS'}.
\end{equation}

Flavor transformation $f\in SU(3)$ acts {\em from the left}, $g\mapsto
f g$. The corresponding unitary operator $U_{flavor}(f)$ acts as
\begin{equation}
 U_{flavor}(f)\Psi^{\cal R}(g)_{FS}=\Psi^{\cal R}(f^{-1}g)_{FS}
  =\Psi^{\cal R}(g)_{F'S}D^{\cal R}_{F'F}(f).
\end{equation}
Here and hereafter, the summation over repeated indices is
understood. On the other hand, ``spin'' transformation\footnote{Only the
$SU(2)$ subgroup corresponds to the usual spatial rotation.} $s \in
SU(3)$ acts {\em from the right}, $g\mapsto gs^{-1}$. The
corresponding unitary operator $U_{spin}(s)$ acts as
\begin{eqnarray}
 U_{spin}(s)\Psi^{\cal R}(g)_{FS}&=&\Psi^{\cal R}(gs)_{FS}
  =\sqrt{\mbox{\rm dim}{\cal R}}P(S) 
  \left(D^{\cal R}_{F\tilde{S}'}(g)\right)^*
  \left(D^{\cal R}_{\tilde{S}'\tilde{S}}(s)\right)^* \nonumber \\
 &=&\sqrt{\mbox{\rm dim}{\cal R}}P(S') 
  \left(D^{\cal R}_{F\tilde{S}'}(g)\right)^*
  D^{\overline{\cal R}}_{{S'}^*{S}^*}(s),
\end{eqnarray}
where $\overline{\cal R}$ stands for the conjugate representation to
${\cal R}$ and ${S}^*$ stands for $(-Y_R, J_3,J)$. We have used the
phase convention of Ref.~\cite{deSwart:1963gc}. When $s$ is restricted
to the ``upper-left'' $SU(2)$ subgroup, it reduces to the usual spin
transformation law.

The infinitesimal transformation (Lie derivative) of the ``spin''
transformation defines the operator $F_\alpha$ introduced in
Eq.~(\ref{F}),
\begin{equation}
 F_\alpha \Psi^{\cal R}(g)_{FS}=
  \sqrt{\mbox{\rm dim}{\cal R}}P(S)
  \left(D^{\cal R}_{F\tilde{S}'}(g)\right)^*
  \left(-T_\alpha^{\cal R}\right)_{\tilde{S},\tilde{S}'},
\end{equation}
where $T_\alpha^{\cal R}$ is the $SU(3)$ generator in the representation
${\cal R}$. In particular, because
\begin{equation}
 \left(-T^{\cal R}_8\right)_{\tilde{S},\tilde{S}'}
  =-\frac{\sqrt{3}}{2}Y_R \delta_{Y_RY_R'}\delta_{JJ'}\delta_{J_3J_3'},
\end{equation}
we have
\begin{equation}
 F_8 \Psi^{\cal R}(g)_{FS}=-\frac{\sqrt{3}}{2}Y_R \Psi^{\cal R}(g)_{FS}.
\end{equation}


The basic calculational tool for various matrix elements is the
orthogonality of irreducible representations,
\begin{equation}
 \int dg \left(D^{\cal R}_{ij}(g)\right)^*D^{{\cal R}'}_{kl}(g)=
  \frac{1}{\mbox{\rm dim}{\cal R}}\delta_{\cal R,R'}\delta_{ik}\delta_{jl},
\end{equation}
where $dg$ is a normalized Haar measure. For a compact group such as
$SU(3)$ it is left- and right-invariant.

Another important tool is the $SU(3)$ Clebsch-Gordan (CG)
coefficients\cite{deSwart:1963gc,Williams,KaedingWilliams}. It enables
us to calculate the following integral,
\begin{equation}
 \int dg \left(D^{\cal R}_{ij}(g)\right)^*
  D^{{\cal R}_1}_{i_1,j_1}(g)D^{{\cal R}_2}_{i_2,j_2}(g)
  =\frac{1}{\mbox{\rm dim}{\cal R}}\sum_{r=1}^m
  \left(
   {\scriptsize
   \begin{array}{ccc}
    {\cal R}_1 & {\cal R}_2 & {\cal R} \\
    i_1 & i_2 & i
   \end{array}
   }
  \right)^*_r
  \left(
   {\scriptsize
   \begin{array}{ccc}
    {\cal R}_1 & {\cal R}_2 & {\cal R} \\
   j_1 & j_2 & j
   \end{array}
   }
  \right)_r,
\end{equation}
where the subscript $r$ counts the multiplicity $m$ of the
representation ${\cal R}$ in the direct product representation ${\cal
R}_1\otimes {\cal R}_2$, or, equivalently, the multiplicity of
$\overline{\cal R}_1$ in $\overline{\cal R}\otimes
{\cal R}_2$. Note that in the following we do not always
work with the ``physical basis'' which diagonalizes the
(right-)hypercharge and (iso-)spin, the CG coefficients are not
necessarily real.

All of the operators whose matrix elements we need to evaluate involve
the octet (adjoint) representation. Thus our first formula is
\begin{eqnarray}
 &&\left\langle \Psi^{{\cal R}}_{FS}
    \left|D^{(8)}_{\alpha\beta}\right|
  \Psi^{{\cal R}'}_{F'S'}\right\rangle
  =\int dg \left(\Psi^{\cal R}_{FS}(g)\right)^*D^{(8)}_{\alpha\beta}(g)
  \Psi^{{\cal R}'}_{F'S'}(g) \nonumber \\
 &=&\sqrt{\mbox{\rm dim}{\cal R}}\sqrt{\mbox{\rm dim}{\cal R}'}
  P^*(S)P(S')
  \left(
   \int dg \left(D^{\cal R}_{F\tilde{S}}(g)\right)^*
   D^{(8)}_{\alpha\beta}(g)
   D^{{\cal R}'}_{F'\tilde{S}'}(g)
  \right)^* \nonumber \\
 &=&\sqrt{\frac{{\mbox{\rm dim}}{{\cal R}'}}{{\mbox{\rm dim}}{\cal R}}}
  P^*(S)P(S')\sum_{r=1}^m
  \left(
   {\scriptsize
   \begin{array}{ccc}
    \bm{8} & {\cal R}' & {\cal R} \\
    \alpha & F' & F
   \end{array}
   }
  \right)_r
  \left(
   {\scriptsize
   \begin{array}{ccc}
    \bm{8} & {\cal R}' & {\cal R} \\
    \beta & \tilde{S}' & \tilde{S}
   \end{array}
   }
  \right)^*_r.
\end{eqnarray}
A decay operator contains (at most) an $F_\delta$ with $D^{(8)}$, thus
our second formula is
\begin{eqnarray}
 &&\left\langle \Psi^{{\cal R}}_{FS}
  \left|
   D^{(8)}_{\alpha\beta}F_\delta
  \right|
  \Psi^{{\cal R}'}_{F'S'}\right\rangle
  =\int dg \left(\Psi^{\cal R}_{FS}(g)\right)^*
  D^{(8)}_{\alpha\beta}(g)
  \left(F_\delta\Psi^{{\cal R}'}_{F'S'}(g)\right) \nonumber \\
 &=&\sqrt{\mbox{\rm dim}{\cal R}}\sqrt{\mbox{\rm dim}{\cal R}'}
  P^*(S)P(S')
  \left(
   \int dg \left(D^{\cal R}_{F\tilde{S}}(g)\right)^*
   D^{(8)}_{\alpha\beta}(g)
   D^{{\cal R}'}_{F'\tilde{S}''}(g) 
  \right)^*
  \left(-T_\delta^{{\cal R}'}\right)_{\tilde{S}',\tilde{S}''}
  \nonumber \\
 &=&-\sqrt{\frac{{\mbox{\rm dim}}{{\cal R}'}}{{\mbox{\rm dim}}{\cal R}}}
  P^*(S)P(S')\sum_{r=1}^m
  \left(
   {\scriptsize
   \begin{array}{ccc}
    \bm{8} & {\cal R}' & {\cal R} \\
    \alpha & F' & F
   \end{array}
   }
  \right)_r
  \left(
   {\scriptsize
   \begin{array}{ccc}
    \bm{8} & {\cal R}' & {\cal R} \\
    \beta & \tilde{S}'' & \tilde{S}
   \end{array}
   }
  \right)^*_r 
  \left(T_\delta^{{\cal R}'}\right)_{\tilde{S}',\tilde{S}''}.
\end{eqnarray}

Most of the symmetry breaking operators in $H_1$ contains two
$F_\alpha$'s with $D^{(8)}$. The third formula is useful in evaluating
the matrix elements of them,
\begin{eqnarray}
 &&\left\langle \Psi^{{\cal R}}_{FS}
  \left|
  F_\delta D^{(8)}_{\alpha\beta}F_\eta
  \right|
  \Psi^{{\cal R}'}_{F'S'}\right\rangle
  =\int dg \left(F_\delta\Psi^{\cal R}_{FS}(g)\right)^*
  D^{(8)}_{\alpha\beta}(g)
  \left(F_\eta\Psi^{{\cal R}'}_{F'S'}(g)\right) \nonumber \\
 &=&\sqrt{\mbox{\rm dim}{\cal R}}\sqrt{\mbox{\rm dim}{\cal R}'}
  P^*(S)P(S') \nonumber \\
 &&\quad \times
  \left(-T_\delta^{{\cal R}}\right)_{\tilde{S}_1,\tilde{S}}
  \left(
  \int dg 
  \left(D^{\cal R}_{F\tilde{S}_1}(g)\right)^*
  D^{(8)}_{\alpha\beta}(g)
  D^{{\cal R}'}_{F'\tilde{S}_1'}(g) 
  \right)^*
  \left(-T_\eta^{{\cal R}'}\right)_{\tilde{S}',\tilde{S}_1'}\nonumber \\
 &=&\sqrt{\frac{{\mbox{\rm dim}}{{\cal R}'}}{{\mbox{\rm dim}}{\cal R}}}
  P^*(S)P(S')\sum_{r=1}^m
  \left(
   {\scriptsize
   \begin{array}{ccc}
    \bm{8} & {\cal R}' & {\cal R} \\
    \alpha & F' & F
   \end{array}
   }
  \right)_r
  \left(T_\delta^{{\cal R}}\right)_{\tilde{S}_1,\tilde{S}}
  \left(
   {\scriptsize
   \begin{array}{ccc}
    \bm{8} & {\cal R}' & {\cal R} \\
    \beta & \tilde{S}_1' & \tilde{S}_1
   \end{array}
   }
  \right)^*_r
  \left(T_\eta^{{\cal R}'}\right)_{\tilde{S}',\tilde{S}_1'}.
\end{eqnarray}
The sum of the last three factor may be rewritten as
\begin{equation}
 \left(T_\delta^{{\cal R}}\right)_{{S}_1,\tilde{S}}
  \left(
   {\scriptsize
   \begin{array}{ccc}
    \bm{8} & {\cal R}' & {\cal R} \\
    \beta & {S}_1' & {S}_1
   \end{array}
   }
  \right)^*_r
  \left(T_\eta^{{\cal R}'}\right)_{\tilde{S}',{S}_1'}.
\end{equation}

\subsection{Simplification in the diagonal case}

Certain simplification occurs in the case ${\cal R}={\cal R}'$. In this
case the CG coefficients may be considered as  $\mbox{\rm dim}{\cal
R}\times \mbox{\rm dim}{\cal R}$ matrices,
\begin{equation}
 \left( M^{({\cal R},r)}_{\alpha}\right)_{ij}\equiv
  \left(
   {\scriptsize
   \begin{array}{ccc}
    \bm{8} & {\cal R} & {\cal R} \\
    \alpha & j & i
   \end{array}
   }
  \right)_r.
\end{equation}
In the following, we derive several useful formulae by giving the
explicit forms of the matrices $M^{({\cal R},r)}_{\alpha}$. The
analysis may be easily generalized to more general compact groups (at
least to $SU(n)$).

First of all, we will show that $M^{({\cal R},r)}_{\alpha}$
satisfies the commutation relation,
\begin{equation}
 \left[T^{\cal R}_\alpha, M^{({\cal R},r)}_{\beta}\right]
 =i f_{\alpha\beta\gamma} M^{({\cal R},r)}_{\gamma}.
 \label{adjointaction}
\end{equation}
Consider the integral
\begin{equation}
 \int dg 
  \left(D^{\cal R}_{ij}(g)\right)^*
  D^{(8)}_{\alpha\beta}(g)
  D^{{\cal R}}_{kl}(g)
  =\frac{1}{\mbox{\rm dim}{\cal R}}\sum_{r=1}^m
  \left(
   {\scriptsize
   \begin{array}{ccc}
    \bm{8} & {\cal R} & {\cal R} \\
    \alpha & k & i
   \end{array}
   }
  \right)^*_r
  \left(
   {\scriptsize
   \begin{array}{ccc}
    \bm{8} & {\cal R} & {\cal R} \\
    \beta & l & j
   \end{array}
   }
  \right)_r.
  \label{R8R}
\end{equation}
Let us make a change of integration variable $g\rightarrow gh$. Since
the measure is right invariant, we have
\begin{equation}
\int dg 
  \left(D^{\cal R}_{ij}(gh)\right)^*
  D^{(8)}_{\alpha\beta}(gh)
  D^{{\cal R}}_{kl}(gh)
  =\int dg 
  \left(D^{\cal R}_{ij}(g)\right)^*
  D^{(8)}_{\alpha\beta}(g)
  D^{{\cal R}}_{kl}(g).
\end{equation}
For an infinitesimal $h$, it leads to
\begin{equation}
 -\left(T^{\cal R}_\eta\right)_{jj'}
  \left( M^{({\cal R},r)}_{\beta}\right)_{j'l}
  +\left(T^{(8)}_\eta\right)_{\beta'\beta}
  \left( M^{({\cal R},r)}_{\beta'}\right)_{jl}
  +\left( M^{({\cal R},r)}_{\beta}\right)_{jl'}
  \left(T^{\cal R}_\eta\right)_{l'l}
  =0,
\end{equation}
where the independence of the CG coefficients has been used. From this,
the commutation relation (\ref{adjointaction}) directly follows.

Next, we show that such a matrix that satisfies
Eq.~(\ref{adjointaction}) is a linear combination of $T^{\cal R}_\alpha$
and $D^{\cal R}_\alpha\equiv d_{\alpha\beta\gamma}T^{\cal
R}_{\beta}T^{\cal R}_{\gamma}$. (Do not confuse $D^{\cal R}_\alpha$ with
the representation matrix $D^{\cal R}_{ij}(g)$.) If the Dynkin index of
${\cal R}$ is $(0,q)$ or $(p,0)$, they are not independent: $D^{\cal
R}_\alpha$ is proportional to $T^{\cal R}_\alpha$. The proof goes as
follows. It is easy to show that $T^{\cal R}_\alpha$ and $D^{\cal
R}_\alpha$ satisfies Eq.~(\ref{adjointaction}). The point is that the
multiplicity $m$ of $\bm{8}$ in ${\cal R}\otimes\overline{\cal R}$ is at
most 2 as we shortly show, so that $T^{\cal R}_\alpha$ and $D^{\cal
R}_\alpha$ span the complete set.

\begin{figure}
 \includegraphics[width=0.8\linewidth, clip]{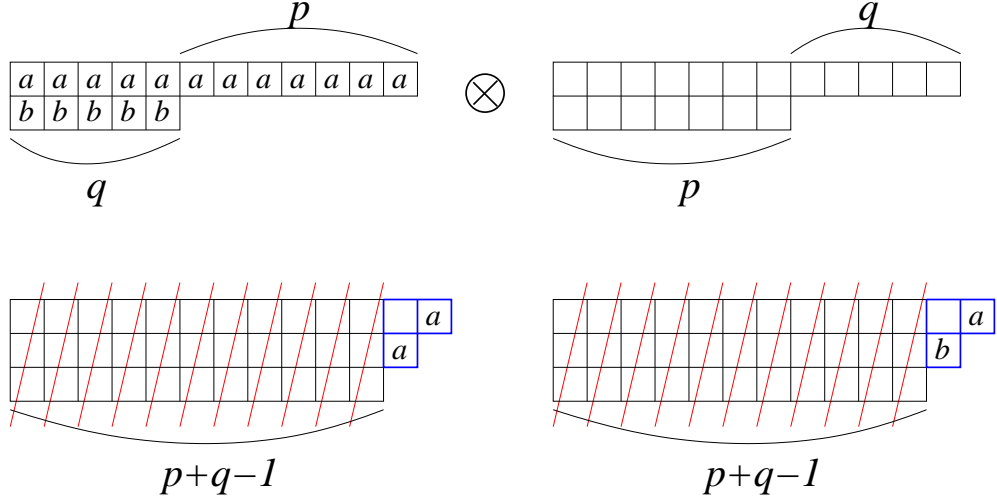}%
 \caption{\label{YT} The Young tableaux method of forming the adjoint
 (octet) representation in the direct product ${\cal R}\otimes
 \overline{\cal R}$. After making $p+q-1$ singlet combinations, there
 are two ways of forming an adjoint representation. If $q=0$, the way
 shown on the right is impossible.}
\end{figure}

Let us consider the CG decomposition of ${\cal R}\otimes\overline{\cal
R}$ using Young tableaux (Littlewood's method). (See, for example,
Ref.~\cite{Cheng:1985bj}.)  Suppose the Dynkin index of the
representation ${\cal R}$ is $(p,q)$, and, without loss of generality,
$p\ge q$. There are $(p+2q)+(q+2p)=3(p+q)$ boxes in the product ${\cal
R}\otimes \overline{\cal R}$. As is shown in Fig.~\ref{YT}, there are at
most two ways to form an adjoint representation.
This explicitly shows that the multiplicity $m$ of $\bm{8}$ in ${\cal
R}\otimes\overline{\cal R}$ is at most 2.

Similarly, we can show that for $SU(n)$ there are at most $n-1$ ways of
forming an adjoint representation from the direct product ${\cal
R}\otimes\overline{\cal R}$. This number is just the rank of the group,
$\mbox{\rm rank}(G)$, and it is also the number of invariant
tensors. The matrix $M^{({\cal R},r)}_{\alpha}$ may be written as a
linear combination of $\mbox{\rm rank}(G)$ quantities,
\begin{equation}
 T^{({\cal R},s)}_\alpha\equiv g_{\alpha\alpha_1\alpha_2\cdots\alpha_s}
  T^{\cal R}_{\alpha_1}T^{\cal R}_{\alpha_2}\cdots T^{\cal R}_{\alpha_s},
  \quad (s=1,\cdots, \mbox{\rm rank}(G))
\end{equation}
where $g_{\alpha\alpha_1\alpha_2\cdots\alpha_s}$ is a real symmetric invariant
tensor of $SU(n)$, and therefore $T^{({\cal R},s)}_\alpha$ is
Hermitian. For $SU(3)$, there are two symmtric invariant tensors,
$\delta_{\alpha\beta}$ and $d_{\alpha\beta\gamma}$.  Independence of
$T^{({\cal R},s)}_\alpha$ may be examined by defining a $\mbox{\rm
rank}(G)\times \mbox{\rm rank}(G)$ matrix $C^{st}$,
\begin{equation}
 C^{st}id_{\cal R}=T^{({\cal R},s)}_\alpha T^{({\cal R},t)}_\alpha.
\end{equation}
Note that the right hand side commutes with $T^{\cal R}_\beta$ for all
$\beta$, so that it is proportional to $id_{\cal R}$ by Schur's lemma.
In general, $m=\mbox{\rm rank}(C) \le \mbox{\rm rank}(G)$. We arrange
that the first $m$ $T^{({\cal R},s)}_\alpha$'s are independent.  


We may now write the matrix $M^{({\cal R}, r)}_{\alpha}$ as
a linear combination of independent $T^{({\cal R},s)}_\alpha$'s,
\begin{equation}
 M^{({\cal R},r)}_{\alpha}=
  \sum_{s=1}^{m}V_{rs}
  T^{({\cal R},s)}_\alpha .
  \label{M-T}
\end{equation}
Note that $V_{rs}$ is a regular $m\times m$ matrix, but it is {\em not}
orthogonal. 

The normalization of $M^{({\cal R}, r)}_{\alpha}$ is determined by the
orthogonality condition of the CG coefficients,
\begin{equation}
 \left(
  \left(M^{({\cal R},r)}_{\alpha}\right)^\dagger
     M^{({\cal R},s)}_{\alpha}
    \right)_{ij}
  =\delta_{ij}\delta_{rs}.
\end{equation}
By substituting (\ref{M-T}) into the above expression, we get
\begin{equation}
 \left(VCV^\dagger\right)_{rs}=\delta_{rs},
\end{equation}
thus,
\begin{equation}
 \left(V^\dagger V\right)_{rs}=\left(C^{-1}\right)_{rs}.
  \label{VdaggerV}
\end{equation}

We can write down (\ref{R8R}) in terms of $T^{({\cal R},r)}_\alpha$,
\begin{eqnarray}
 \int dg 
  \left(D^{\cal R}_{ij}(g)\right)^*
  D^{(8)}_{\alpha\beta}(g)
  D^{{\cal R}}_{kl}(g)
  &=&\frac{1}{\mbox{\rm dim}{\cal R}}\sum_{r=1}^m
  \big(M^{({\cal R},r)}_\alpha\big)_{ik}^*
  \big(M^{({\cal R},r)}_\beta\big)_{jl} \nonumber \\
 &=&\frac{1}{\mbox{\rm dim}{\cal R}}\sum_{s,t=1}^m
  \left(V^\dagger V\right)_{st} 
  \big(T^{({\cal R},s)}_\alpha\big)_{ki} 
  \big(T^{({\cal R},t)}_\beta\big)_{jl}.
\end{eqnarray}
By substituting (\ref{VdaggerV}) into this, we finally get the
expression,
\begin{equation}
 \int dg 
  \left(D^{\cal R}_{ij}(g)\right)^*
  D^{(8)}_{\alpha\beta}(g)
  D^{{\cal R}}_{kl}(g)
  =\frac{1}{\mbox{\rm dim}{\cal R}}\sum_{s,t=1}^m
  \left(C^{-1}\right)_{st}
  \big(T^{({\cal R},s)}_\alpha\big)_{ki} 
  \big(T^{({\cal R},t)}_\beta\big)_{jl}.
  \label{int-diagonal}
\end{equation}
For $SU(n)$, $D^{(8)}(g)$ should be replaced by $D^{Ad}(g)$.

\subsection{Formulae for the matrix elements diagonal in representation}
\label{repdiagonal}

Let us now calculate the matrix elements of various operators, which are
diagonal in representation, by using the formulae derived in the
previous subsection. 

For $SU(3)$, there are two symmetric invariant tensors
$\delta_{\alpha\beta}$ and $d_{\alpha\beta\gamma}$, we have
\begin{equation}
 T^{({\cal R},1)}_\alpha\equiv T^{\cal R}_\alpha,
  \quad T^{({\cal R},2)}_\alpha \equiv d_{\alpha\beta\gamma} 
  T^{\cal R}_\beta T^{\cal R}_\gamma
  = D^{\cal R}_\alpha.
\end{equation}
The matrix $C^{rs}$ may be written as
\begin{equation}
 C=\left(
    \begin{array}{cc}
     C_2 & C_3\\
     C_3 & D_2
    \end{array}
   \right),
\end{equation}
where $C_2$ and $C_3$ are quadratic and cubic
Casimirs\cite{Barut:1986dd} respectively,
\begin{eqnarray}
 C_2&=&\frac{1}{3}\left(p^2+q^2+pq+3(p+q)\right), \\
 C_3&=&\frac{1}{18}(p-q)\left(2p^2+2q^2+5pq+9(p+q+1)\right),
\end{eqnarray}
for representation ${\cal R}=(p,q)$, while $D_2\equiv
\sum_{\alpha}D^{\cal R}_\alpha D^{\cal R}_\alpha$ may be written as
\begin{equation}
 D_2=\left(\frac{1}{3}C_2+\frac{1}{4}\right)C_2.
\end{equation}
By looking at the determinant of $C$,
\begin{equation}
 \det C =\frac{1}{12}pq(p+2)(q+2)(p^2+q^2+2pq+4(p+q)+3),
\end{equation}
we see that $T^{\cal R}_\alpha$ and $D^{\cal R}_\alpha$ are not
independent for $p=0$ or $q=0$. Actually,
\begin{equation}
 D^{\cal R}_\alpha=\frac{C_3}{C_2}T^{\cal R}_\alpha.
\end{equation}
In this case, the integral (\ref{int-diagonal}) gets simpler,
\begin{equation}
 \int dg 
  \left(D^{\cal R}_{ij}(g)\right)^*
  D^{(8)}_{\alpha\beta}(g)
  D^{{\cal R}}_{kl}(g)
  =\frac{1}{\mbox{\rm dim}{\cal R}}\frac{1}{C_2({\cal R})}
  \big(T^{\cal R}_\alpha\big)_{ki} 
  \big(T^{\cal R}_\beta\big)_{jl}.
\end{equation}

When $\det C\ne 0$, $T^{\cal R}_\alpha$ and $D^{\cal R}_\alpha$ are
independent and the matrix $C$ has the inverse,
\begin{equation}
 C^{-1}=\frac{1}{C_2^2(3+4C_2)-12C_3^2}
  \left(
   \begin{array}{cc}
    C_2(3+4C_2)& -12C_3\\
    -12C_3 & 12 C_2
   \end{array}
  \right).
\end{equation}

\subsubsection{Hamiltonian operators}

Here we explicitly use $Y_R=1$ and present the results for the various
flavor $SU(3)$ operators appeared in the Hamiltonian. They are expressed
in terms of flavor $T^{\cal R}_8$ and $D^{\cal R}_8$, i.e., in the
Gell-Mann-Okubo\cite{Okubo:1961jc} form.

Let us first show explicitly the simplest case.
\begin{eqnarray}
 \left\langle \Psi^{{\cal R}}_{F_2S_2}
  \left|D^{(8)}_{\alpha\beta}\right|
  \Psi^{\cal R}_{F_1S_1}\right\rangle
 &=& \mbox{\rm dim}{\cal R} P^*(S_2)P(S_1)
  \left(
   \int dg  \left(D^{\cal R}_{F_2\tilde{S}_2}(g)\right)^*
   D^{(8)}_{\alpha\beta}(g)
   D^{\cal R}_{F_1\tilde{S}_1}(g)
  \right)^* \nonumber \\
 &=&P^*(S_2)P(S_1) \sum_{s,t=1}^2
  \left(C^{-1}\right)_{st}
  \big(T^{{\cal R},s}_\alpha\big)_{F_2F_1} 
  \big(T^{{\cal R},t}_\beta\big)_{\tilde{S}_1\tilde{S}_2}.
  \label{Dab}
\end{eqnarray}
Since
\begin{eqnarray}
 \left(T^{\cal R}_8\right)_{\tilde{S}_1\tilde{S}_2}
  &=&\frac{\sqrt{3}}{2}\delta_{\tilde{S}_1\tilde{S}_2}, \label{T8}\\
 \left(D^{\cal R}_8\right)_{\tilde{S}_1\tilde{S}_2}
  &=&\frac{\sqrt{3}}{2}
  \left(\bfJ^2-\frac{1}{3}C_2({\cal R})-\frac{1}{4}\right)
  \delta_{\tilde{S}_1\tilde{S}_2},\label{D8}
\end{eqnarray}
we have
\begin{eqnarray}
 \left\langle \Psi^{{\cal R}}_{F_2S_2}
  \left|D^{(8)}_{88}\right|
  \Psi^{\cal R}_{F_1S_1}\right\rangle
 &=&\frac{\sqrt{3}}{2}
 \bigg\{
 \big(T_8\big)_{F_2F_1} 
 \left(
  C^{-1}_{11}+C^{-1}_{12}
  \left(
   \bfJ^2-\frac{1}{3}C_2-\frac{1}{4}
  \right)
 \right) \nonumber \\
 &&{} +\big(D_8\big)_{F_2F_1} 
  \left(
  C^{-1}_{21}+C^{-1}_{22}
  \left(
   \bfJ^2-\frac{1}{3}C_2-\frac{1}{4}
  \right)
 \right)
 \bigg\}\delta_{\tilde{S}_1\tilde{S}_2},
\end{eqnarray}
where $\bfJ^2$ stands for the spin $SU(2)$ quadratic Casimir. We dropped
all the ${\cal R}$ dependence on the right hand side for notational
simplicity.

Other matrix elements may be calculated in a similar way. The results
are summarized as follows,
\begin{eqnarray}
 \left\langle \Psi^{{\cal R}}_{F_2S_2}
  \left|{\cal O}_i\right|
  \Psi^{\cal R}_{F_1S_1}\right\rangle
 &=& P^*(S_2)P(S_1)
 \bigg\{
 \big(T_8\big)_{F_2F_1} 
 \left(
  C^{-1}_{11}O^{(1)}_i+C^{-1}_{12}O^{(2)}_i
 \right)_{\tilde{S}_1\tilde{S}_2} \nonumber \\
 &&{} +\big(D_8\big)_{F_2F_1} 
  \left(
  C^{-1}_{21}O^{(1)}_i+C^{-1}_{22}O^{(2)}_i
 \right)_{\tilde{S}_1\tilde{S}_2}
 \bigg\},\label{opdiag}
\end{eqnarray}
where
\begin{eqnarray}
 O^{(1)}_x&=&\frac{\sqrt{3}}{2}\bfJ^2, \\
 O^{(2)}_x&=&\frac{\sqrt{3}}{6}\left(3\bfJ^2-C_2-\frac{3}{4}\right)\bfJ^2, \\
 O^{(1)}_y&=&\frac{3}{8}+\half C_2 
  +\frac{2}{3}C_3 -\frac{3}{2}\bfJ^2, \\
 O^{(2)}_y&=&-\frac{3}{16}-\frac{1}{4}C_2-\frac{1}{3}C_3+\frac{1}{2}\bfJ^2
  +C_2\bfJ^2 - \left(\bfJ^2\right)^2, \\
 O^{(1)}_z&=&\frac{\sqrt{3}}{2}\left(C_2-\bfJ^2-\frac{9}{4}\right), \\
 O^{(2)}_z&=&\frac{\sqrt{3}}{6}\left(3\bfJ^2-C_2-\frac{3}{4}\right)
  \left(C_2-\bfJ^2-\frac{9}{4}\right), \\
 O^{(1)}_w&=&-\frac{3}{8}+\frac{1}{3}C_3 -\frac{1}{2}\bfJ^2, \\
 O^{(2)}_w&=&-\frac{3}{32}+\frac{1}{8}C_2+\frac{1}{3}\left(C_2\right)^2
  +\frac{1}{3}C_3-\frac{3}{4}\bfJ^2-\frac{5}{6}C_2\bfJ^2
  +\half \left(\bfJ^2\right)^2.
\end{eqnarray}
The usefulness of these formulae rests on the fact that they are easily
calculated for {\em arbitrary} representation ${\cal R}$.

\subsubsection{Decay operators}

The matrix elements of the $G_0$ operator may be calculated by using
Eq.~(\ref{Dab}),
\begin{eqnarray}
 \left\langle \Psi^{{\cal R}}_{F_2S_2}
  \left|D^{(8)}_{\alpha i}\right|
  \Psi^{\cal R}_{F_1S_1}\right\rangle
 &=&P^*(S_2)P(S_1) \sum_{s,t=1}^2
  \left(C^{-1}\right)_{st}
  \big(T^{{\cal R},s}_\alpha\big)_{F_2F_1} 
  \big(T^{{\cal R},t}_i\big)_{\tilde{S}_1\tilde{S}_2} \nonumber \\
 &=&P^*(S_2)P(S_1)
  \bigg\{
 \big(T_\alpha\big)_{F_2F_1} 
  \left(
  C^{-1}_{11}\left(T_i\right)_{\tilde{S}_1\tilde{S}_2}
  +C^{-1}_{12}\left(D_i\right)_{\tilde{S}_1\tilde{S}_2}
 \right) \nonumber \\
 &&\qquad{} +\big(D_\alpha\big)_{F_2F_1} 
  \left(
  C^{-1}_{21}\left(T_i\right)_{\tilde{S}_1\tilde{S}_2}
  +C^{-1}_{22}\left(D_i\right)_{\tilde{S}_1\tilde{S}_2}
 \right)
 \bigg\}.
\end{eqnarray}
Since $T_i$ is the usual spin $SU(2)$ generator, the matrix elements
between the states with different spins vanish and only the $D_i$ terms
contribute. When the both states have the same spin, from the
transformation property, $D_i$ is proportional to $T_i$. By using
\begin{equation}
 \sum_{i=1}^3 T_iD_i= \frac{1}{3}C_3+\frac{1}{2\sqrt{3}}C_2T_8
  +\frac{1}{2\sqrt{3}}\bfJ^2 T_8-\frac{1}{6\sqrt{3}}(T_8)^3
  +\frac{1}{2\sqrt{3}}T_8,
\end{equation}
we have for the same spin states (with $Y_R=1$),
\begin{equation}
 \left(D_i\right)_{\tilde{S}_1\tilde{S}_2}
  =A  \left(T_i\right)_{\tilde{S}_1\tilde{S}_2},
\end{equation}
with
\begin{equation}
 A=\frac{1}{\bfJ^2}
  \left(
   \frac{1}{3}C_3+\frac{1}{4}C_2+\frac{1}{4}\bfJ^2+\frac{3}{16}
  \right).
\end{equation}

The matrix elements, diagonal in representation, of the decay
operators which are linear in $F_\alpha$ may be calculated in a similar
way. For the $G_2$ operator, we have
\begin{eqnarray}
 \left\langle \Psi^{{\cal R}}_{F_2S_2}
  \left|D^{(8)}_{\alpha8}  F_i \right|
  \Psi^{\cal R}_{F_1S_1}\right\rangle
 &=&P^*(S_2)P(S_1) \sum_{s,t=1}^2
  \left(C^{-1}\right)_{st}
  \big(T^{{\cal R},s}_\alpha\big)_{F_2F_1} 
  \big(T^{{\cal R},t}_8\big)_{\tilde{S}'_1\tilde{S}_2}
  \left(-T^{\cal R}_i\right)_{\tilde{S}_1\tilde{S}'_1}.\nonumber \\
\end{eqnarray}
By using Eqs.~(\ref{T8}) and (\ref{D8}), this can be rewrite as
\begin{eqnarray}
 &&\left\langle \Psi^{{\cal R}}_{F_2S_2}
  \left|D^{(8)}_{\alpha8}  F_i \right|
  \Psi^{\cal R}_{F_1S_1}\right\rangle
 =-\frac{\sqrt{3}}{2}P^*(S_2)P(S_1) 
 \bigg\{
 \big(T_\alpha\big)_{F_2F_1} 
  \left(
  C^{-1}_{11}+C^{-1}_{12}
  \left(
   \bfJ^2-\frac{1}{3}C_2-\frac{1}{4}
  \right)
 \right) \nonumber \\
 &&\qquad{} +\big(D_\alpha\big)_{F_2F_1} 
  \left(
  C^{-1}_{21}+C^{-1}_{22}
  \left(
   \bfJ^2-\frac{1}{3}C_2-\frac{1}{4}
  \right)
 \right)
 \bigg\}
 \left(T_i\right)_{\tilde{S}_1\tilde{S}_2}.
\end{eqnarray}
Similarly, we can calculate the matrix elements for the $G_1$ operator,
\begin{eqnarray}
 &&\left\langle \Psi^{{\cal R}}_{F_2S_2}
  \left|
   \sum_{\beta,\gamma \in{\cal J}}
   d_{i\beta\gamma}D^{(8)}_{\alpha\beta}  F_\gamma 
  \right|
  \Psi^{\cal R}_{F_1S_1}\right\rangle \nonumber \\
  &=&-P^*(S_2)P(S_1) \sum_{s,t=1}^2
  \left(C^{-1}\right)_{st}
  \big(T^{{\cal R},s}_\alpha\big)_{F_2F_1} 
  \sum_{\beta,\gamma \in{\cal J}}d_{i\beta\gamma}
  \big(T^{\cal R}_\gamma T^{{\cal R},t}_\beta\big)_{\tilde{S}_1\tilde{S}_2}.
\end{eqnarray}
Note that
\begin{eqnarray}
 \sum_{\beta,\gamma\in{\cal J}}d_{i\beta\gamma}T_\beta T_\gamma&=&
  D_i-\frac{2}{\sqrt{3}}T_iT_8, \\
 \sum_{\beta,\gamma\in{\cal J}}d_{i\beta\gamma}T_\beta D_\gamma&=&
  \left(\frac{1}{3}C_2+\frac{1}{4}\right)T_i
  -\frac{1}{\sqrt{3}}\left(D_iT_8+D_8 T_i\right),
\end{eqnarray}
we have
\begin{eqnarray}
 &&\left\langle \Psi^{{\cal R}}_{F_2S_2}
  \left|
   \sum_{\beta,\gamma \in{\cal J}}
   d_{i\beta\gamma}D^{(8)}_{\alpha\beta}  F_\gamma 
  \right|
  \Psi^{\cal R}_{F_1S_1}\right\rangle \nonumber \\
  &=&-P^*(S_2)P(S_1)
   \bigg\{
   \big(T_\alpha\big)_{F_2F_1} 
   \left(
    C^{-1}_{11}(D_i-T_i)
    +C^{-1}_{12}
    \left(
     \left(\frac{1}{2}C_2-\half \bfJ^2+\frac{3}{8}\right)T_i-\half D_i
    \right)
   \right)_{\tilde{S}_1\tilde{S}_2}\nonumber \\
 &&{} +\big(D_\alpha\big)_{F_2F_1} 
  \left(
   C^{-1}_{21}(D_i-T_i)
   +C^{-1}_{22}
   \left(
    \left(\frac{1}{2}C_2-\half \bfJ^2+\frac{3}{8}\right)T_i-\half D_i
   \right)
  \right)_{\tilde{S}_1\tilde{S}_2}
  \bigg\}. 
\end{eqnarray}

When $\det C=0$, the corresponding formulae become much simpler.

\section{Tables of various matrix elements}
\label{Sec:Tables}

Those matrix elements of the symmetry breaking operators which are
needed to calculated the baryon masses are given in
Sec.~\ref{symbreak}. 
In this Appendix, we summarize other matrix elements which are necessary
to calculate the mixings to second order.

These matrix elements are calculated with the method explained in the
previous section.

To second order, the (mainly) octet states can mix with
$\bm{\overline{10}}$, $\bm{27}_d$, $\bm{\overline{35}}_d$, and
$\bm{64}_d$. The (mainly) anti-decuplet states can mix with
$\bm{\overline{81}}_d$ in addition to them. We therefore need 
$\left\langle \bm{27}_d\left|{\cal O}_i\right|\bm{27}_d\right\rangle$,
$ \left\langle \bm{\overline{35}}_d\left|
{\cal O}_i\right|\bm{\overline{35}}_d\right\rangle$,
$ \left\langle\bm{\overline{35}}_d\left|{\cal
 O}_i\right| \bm{27}_d\right\rangle$,
$ \left\langle\bm{64}_d\left|{\cal
 O}_i\right| \bm{27}_d\right\rangle$,
$ \left\langle\bm{64}_d\left|{\cal
 O}_i\right| \bm{\overline{35}}_d\right\rangle$, and
$ \left\langle\bm{\overline{81}}_d
 \left|{\cal O}_i\right| \bm{\overline{35}}_d\right\rangle$.
\begin{table}[h!]
 \caption{\label{27d27d}$ \left\langle \bm{27}_d\left|{\cal
 O}_i\right|\bm{27}_d\right\rangle$} 
 \begin{ruledtabular}
  \begin{tabular}{c|ccccc}
   $(I,Y)$ & $\gamma$ & $x$ & $y$ & $z$ & $w$ 
   \\ \hline
   $(\half,+1)$ & $\frac{137}{560}$ & $\frac{3}{4}\frac{137}{560}$ & 
   $\frac{39\sqrt{3}}{1120}$ & $\frac{137}{112}$ &
   $-\frac{2803}{1120\sqrt{3}}$ \\
   $(1,0)$ & $\frac{13}{280}$ & $\frac{3}{4}\frac{13}{280}$ & 
   $-\frac{29\sqrt{3}}{560}$ & $\frac{13}{56}$ &
   $-\frac{407}{560\sqrt{3}}$ \\
   $(\half, -1)$ & $\frac{2}{35}$ & $\frac{3}{4}\frac{2}{35}$ & 
   $-\frac{13\sqrt{3}}{35}$ & $\frac{2}{7}$ & 
   $-\frac{311}{140\sqrt{3}}$ \\
   $(0,0)$ & $\frac{13}{70}$ & $\frac{3}{4}\frac{13}{70}$ & 
   $-\frac{29\sqrt{3}}{140}$ & $\frac{13}{14}$ &
   $-\frac{407}{140\sqrt{3}}$ \\
   $(\frac{3}{2}, -1)$ & $-\frac{17}{112}$ & $-\frac{3}{4}\frac{17}{112}$ & 
   $-\frac{31\sqrt{3}}{224}$ & $-\frac{85}{112}$ &
   $\frac{235}{224\sqrt{3}}$ \\
  \end{tabular}
 \end{ruledtabular}
\end{table}

\begin{table}[h!]
 \caption{\label{35bd35bd}$ \left\langle \bm{\overline{35}}_d\left|{\cal
 O}_i\right|\bm{\overline{35}}_d\right\rangle$} 
 \begin{ruledtabular}
  \begin{tabular}{c|ccccc}
   $(I,Y)$ & $\gamma$ & $x$ & $y$ & $z$ & $w$
   \\ \hline
   $(0, +2)$ & $\frac{1}{4}$ & $\frac{3}{4}\frac{1}{4}$ &
   $-\frac{5\sqrt{3}}{8}$ & $\frac{9}{4}$ & $-\frac{13\sqrt{3}}{8}$ \\
   $(\half, +1)$ & $\frac{3}{16}$ & $\frac{3}{4}\frac{3}{16}$ & 
   $-\frac{11\sqrt{3}}{32}$ & $\frac{27}{16}$ & $-\frac{43\sqrt{3}}{32}$\\
   $(1, 0)$ & $\frac{1}{8}$ & $\frac{3}{4}\frac{1}{8}$ & 
   $-\frac{\sqrt{3}}{16}$ & $\frac{9}{8}$ & $-\frac{17\sqrt{3}}{16}$ \\
   $(\frac{3}{2}, -1)$ & $\frac{1}{16}$ & $\frac{3}{4}\frac{1}{16}$ &
   $\frac{7\sqrt{3}}{32}$ & $\frac{9}{16}$ & $-\frac{25\sqrt{3}}{32}$ \\
  \end{tabular}
 \end{ruledtabular}
\end{table}


\begin{table}[h!]
 \caption{\label{35bd27d}$ \left\langle\bm{\overline{35}}_d\left|{\cal
 O}_i\right| \bm{27}_d\right\rangle$}
 \begin{ruledtabular}
  \begin{tabular}{c|ccccc}
   $(I,Y)$ & $\gamma$ & $x$ & $y$ & $z$ & $w$ \\ \hline
   $(\half, +1)$ &
   $\frac{\sqrt{15}}{16\sqrt{7}}$ & 
   $\frac{3}{4} \frac{\sqrt{15}}{16\sqrt{7}}$ & 
   $\frac{17\sqrt{5}}{32\sqrt{7}}$ & 
   $\frac{\sqrt{105}}{16}$ &
   $-\frac{11\sqrt{5}}{32\sqrt{7}}$ \\
   $(1,0)$ &
   $\frac{\sqrt{5}}{8\sqrt{7}}$ & 
   $\frac{3}{4} \frac{\sqrt{5}}{8\sqrt{7}}$ &
   $\frac{17\sqrt{5}}{16\sqrt{21}}$ & 
   $\frac{\sqrt{35}}{8}$ &
   $-\frac{11\sqrt{5}}{16\sqrt{21}}$ \\
   $(\frac{3}{2},-1)$ &
   $\frac{\sqrt{15}}{16\sqrt{7}}$ & 
   $\frac{3}{4} \frac{\sqrt{15}}{16\sqrt{7}}$ &
   $\frac{17\sqrt{5}}{32\sqrt{7}}$ & 
   $\frac{\sqrt{105}}{16}$ &
   $-\frac{11\sqrt{5}}{32\sqrt{7}}$ \\
  \end{tabular}
 \end{ruledtabular}
\end{table}

\begin{table}[h!]
 \caption{\label{64d27d}$ \left\langle\bm{64}_d\left|{\cal
 O}_i\right| \bm{27}_d\right\rangle$}
 \begin{ruledtabular}
  \begin{tabular}{c|ccccc}
   $(I,Y)$ & $\gamma$ & $x$ & $y$ & $z$ & $w$ \\ \hline
   $(\half, +1)$ &
   $\frac{5\sqrt{3}}{28}$ & 
   $\frac{3}{4}\frac{5\sqrt{3}}{28}$ & 
   $-\frac{5}{14}$ & 
   $\frac{85\sqrt{3}}{56}$ &
   $\frac{5}{56}$ \\
   $(1,0)$ &
   $\frac{5\sqrt{5}}{28\sqrt{2}}$ & 
   $\frac{3}{4} \frac{5\sqrt{5}}{28\sqrt{2}}$ & 
   $-\frac{5\sqrt{5}}{14\sqrt{6}}$ & 
   $\frac{85\sqrt{5}}{56\sqrt{2}}$ &
   $\frac{5\sqrt{5}}{56\sqrt{6}}$ \\
   $(\frac{3}{2},-1)$ &
   $\frac{5\sqrt{3}}{28\sqrt{2}}$ & 
   $\frac{3}{4} \frac{5\sqrt{3}}{28\sqrt{2}}$ & 
   $-\frac{5}{14\sqrt{2}}$ & 
   $\frac{85\sqrt{3}}{56\sqrt{2}}$ &
   $\frac{5}{56\sqrt{2}}$ \\
   $(\frac{1}{2},-1)$ &
   $\frac{5\sqrt{3}}{28}$ & 
   $\frac{3}{4} \frac{5\sqrt{3}}{28}$ & 
   $-\frac{5}{14}$ & 
   $\frac{85\sqrt{3}}{56}$ &
   $\frac{5}{56}$ \\
   $(0,0)$ &
   $\frac{3\sqrt{5}}{14\sqrt{2}}$ & 
   $\frac{3}{4} \frac{3\sqrt{5}}{14\sqrt{2}}$ & 
   $-\frac{\sqrt{15}}{7\sqrt{2}}$ & 
   $\frac{51\sqrt{5}}{28\sqrt{2}}$ &
   $\frac{\sqrt{15}}{28\sqrt{2}}$ \\
  \end{tabular}
 \end{ruledtabular}
\end{table}

\begin{table}[h!]
 \caption{\label{64d35bd}$ \left\langle\bm{64}_d\left|{\cal
 O}_i\right| \bm{\overline{35}}_d\right\rangle$}
 \begin{ruledtabular}
  \begin{tabular}{c|ccccc}
   $(I,Y)$ & $\gamma$ & $x$ & $y$ & $z$ & $w$ \\ \hline
   $(\half, +1)$ &
   $\frac{1}{4\sqrt{35}}$ & 
   $\frac{3}{4} \frac{1}{4\sqrt{35}}$ & 
   $\sqrt{\frac{3}{35}}$ & 
   $\frac{3\sqrt{7}}{8\sqrt{5}}$ &
   $-\frac{13\sqrt{3}}{8\sqrt{35}}$ \\
   $(1,0)$ &
   $\frac{1}{4\sqrt{14}}$ & 
   $\frac{3}{4} \frac{1}{4\sqrt{14}}$ & 
   $\sqrt{\frac{3}{14}}$ &
   $\frac{3\sqrt{7}}{8\sqrt{2}}$ &
   $-\frac{13\sqrt{3}}{8\sqrt{14}}$ \\
   $(\frac{3}{2},-1)$ &
   $\frac{3}{4\sqrt{70}}$ & 
   $\frac{3}{4} \frac{3}{4\sqrt{70}}$ & 
   $\frac{3\sqrt{3}}{\sqrt{70}}$ & 
   $\frac{9\sqrt{7}}{8\sqrt{10}}$ &
   $-\frac{39\sqrt{3}}{8\sqrt{70}}$ \\
  \end{tabular}
 \end{ruledtabular}
\end{table}

\begin{table}[h!]
 \caption{\label{81bd35bd}
 $ \left\langle\bm{\overline{81}}_d
 \left|{\cal O}_i\right| \bm{\overline{35}}_d\right\rangle$}
 \begin{ruledtabular}
  \begin{tabular}{c|ccccc}
   $(I,Y)$ & $\gamma$ & $x$ & $y$ & $z$ & $w$ \\ \hline
   $(0,+2)$ &
   $\frac{\sqrt{3}}{2\sqrt{7}}$ &
   $\frac{3}{4}\frac{\sqrt{3}}{2\sqrt{7}}$ &
   $-\frac{1}{4\sqrt{7}}$ &
   $\frac{13\sqrt{3}}{2\sqrt{7}}$ &
   $\frac{1}{4\sqrt{7}}$ \\
   $(\half, +1)$ &
   $\frac{2}{\sqrt{35}}$ & 
   $\frac{3}{4} \frac{2}{\sqrt{35}}$ & 
   $-\frac{1}{\sqrt{105}}$ & 
   $\frac{26}{\sqrt{35}}$ &
   $\frac{1}{\sqrt{105}}$ \\
   $(1,0)$ &
   $\frac{\sqrt{3}}{2\sqrt{7}}$ & 
   $\frac{3}{4} \frac{\sqrt{3}}{2\sqrt{7}}$ & 
   $-\frac{1}{4\sqrt{7}}$ &
   $\frac{13\sqrt{3}}{2\sqrt{7}}$ &
   $\frac{1}{4\sqrt{7}}$ \\
   $(\frac{3}{2},-1)$ &
   $\sqrt{\frac{3}{35}}$ & 
   $\frac{3}{4}\sqrt{\frac{3}{35}}$ & 
   $-\frac{1}{2\sqrt{35}}$ & 
   $\frac{13\sqrt{3}}{\sqrt{35}}$ &
   $\frac{1}{2\sqrt{35}}$ \\
  \end{tabular}
 \end{ruledtabular}
\end{table}

The (mainly) decuplet states can mix with $\bm{27}_q$, $\bm{35}$,
$\bm{\overline{35}}_q$, $\bm{64}_q$, and $\bm{81}$. Thus we need
$ \left\langle \bm{27}_q\left|{\cal O}_i
 \right|\bm{27}_q\right\rangle$,
$ \left\langle \bm{35}\left|{\cal O}_i
 \right|\bm{35}\right\rangle$,
$ \left\langle \bm{35}\left|{\cal
 O}_i\right|\bm{27}_q\right\rangle$,
$ \left\langle
 \bm{\overline{35}}_q \left|{\cal O}_i\right|\bm{27}_q\right\rangle$,
$ \left\langle \bm{64}_q\left|{\cal
 O}_i\right|\bm{27}_q\right\rangle$,
$ \left\langle \bm{64}_q\left|{\cal
 O}_i\right|\bm{35}\right\rangle$, and
$ \left\langle \bm{81}\left|{\cal
 O}_i\right|\bm{35}\right\rangle$.

\begin{table}[h!]
 \caption{\label{27q27q}$ \left\langle \bm{27}_q\left|{\cal O}_i
 \right|\bm{27}_q\right\rangle$}
 \begin{ruledtabular}
  \begin{tabular}{c|ccccc}
   $(I,Y)$ & $\gamma$ & $x$ & $y$ & $z$ & $w$ 
   \\ \hline
   $(\frac{3}{2}, +1)$ & 
   $\frac{13}{112}$ & $\frac{15}{4}\frac{13}{112}$ & 
   $\frac{45\sqrt{3}}{224}$ & $\frac{13}{56}$ & $-\frac{11}{32\sqrt{3}}$
   \\
   $(1,0)$ & $-\frac{1}{56}$ & $-\frac{15}{4}\frac{1}{56}$ & 
   $-\frac{25\sqrt{3}}{112}$ & $-\frac{1}{28}$ & $-\frac{1}{16\sqrt{3}}$ 
   \\
   $(\half, -1)$ & $-\frac{17}{112}$ & $-\frac{15}{4}\frac{17}{112}$ & 
   $-\frac{145\sqrt{3}}{224}$ & $-\frac{17}{56}$ &
   $\frac{7}{32\sqrt{3}}$ 
   \\
  \end{tabular}
 \end{ruledtabular}
\end{table}

\begin{table}[h!]
 \caption{\label{3535}$ \left\langle \bm{35}\left|{\cal O}_i
 \right|\bm{35}\right\rangle$}
 \begin{ruledtabular}
  \begin{tabular}{c|ccccc}
   $(I,Y)$ & $\gamma$ & $x$ & $y$ & $z$ & $w$ 
   \\ \hline
   $(\frac{3}{2}, +1)$ & 
   $\frac{3}{16}$ & $\frac{15}{4}\frac{3}{16}$ & 
   $\frac{11\sqrt{3}}{32}$ & $\frac{9}{8}$ & $-\frac{17\sqrt{3}}{32}$
   \\
   $(1,0)$ & 
   $\frac{1}{8}$ & $\frac{15}{4}\frac{1}{8}$ & 
   $-\frac{3\sqrt{3}}{16}$ & $\frac{3}{4}$ & $-\frac{11\sqrt{3}}{16}$ 
   \\
   $(\half, -1)$ & 
   $\frac{1}{16}$ & $\frac{15}{4}\frac{1}{16}$ & 
   $-\frac{23\sqrt{3}}{32}$ & 
   $\frac{3}{8}$ &
   $-\frac{27\sqrt{3}}{32}$  \\
   $(0,-2)$ & 
   $0$ & $0$ & 
   $-\frac{5\sqrt{3}}{4}$ & 
   $0$ & $-\sqrt{3}$ \\
  \end{tabular}
 \end{ruledtabular}
\end{table}


\begin{table}[h!]
 \caption{\label{3527q}$ \left\langle \bm{35}\left|{\cal
 O}_i\right|\bm{27}_q\right\rangle$}
 \begin{ruledtabular}
  \begin{tabular}{c|ccccc}
   $(I,Y)$ & $\gamma$ & $x$ & $y$ & $z$ & $w$ \\ \hline
   $(\frac{3}{2}, +1)$ &
   $\frac{\sqrt{15}}{16\sqrt{7}}$ & 
   $\frac{15}{4}\frac{\sqrt{15}}{16\sqrt{7}}$ & 
   $-\frac{\sqrt{5}}{32\sqrt{7}}$ & 
   $\frac{\sqrt{15}}{4\sqrt{7}}$ &
   $-\frac{17\sqrt{5}}{32\sqrt{7}}$ \\
   $(1,0)$ &
   $\frac{\sqrt{5}}{8\sqrt{7}}$ & 
   $\frac{15}{4} \frac{\sqrt{5}}{8\sqrt{7}}$ & 
   $-\frac{\sqrt{5}}{16\sqrt{21}}$ & 
   $\frac{\sqrt{5}}{2\sqrt{7}}$ &
   $-\frac{17\sqrt{5}}{16\sqrt{21}}$ \\
   $(\half,-1)$ &
   $\frac{\sqrt{15}}{16\sqrt{7}}$ & 
   $\frac{15}{4} \frac{\sqrt{15}}{16\sqrt{7}}$ & 
   $-\frac{\sqrt{5}}{32\sqrt{7}}$ & 
   $\frac{\sqrt{15}}{4\sqrt{7}}$ &
   $-\frac{17\sqrt{5}}{32\sqrt{7}}$ \\
  \end{tabular}
 \end{ruledtabular}
\end{table}

\begin{table}[h!]
 \caption{\label{35bq27q} $ \left\langle
 \bm{\overline{35}}_q \left|{\cal O}_i\right|\bm{27}_q\right\rangle$}
 \begin{ruledtabular}
  \begin{tabular}{c|ccccc}
   $(I,Y)$ & $\gamma$ & $x$ & $y$ & $z$ & $w$ \\ \hline
   $(\frac{3}{2}, +1)$ &
   $\sqrt{\frac{3}{35}}$ & 
   $\frac{15}{4}\sqrt{\frac{3}{35}}$ & 
   $\frac{\sqrt{5}}{2\sqrt{7}}$ & 
   $\frac{4\sqrt{3}}{\sqrt{35}}$ &
   $-\frac{\sqrt{5}}{2\sqrt{7}}$ \\
   $(1,0)$ &
   $\frac{1}{2\sqrt{7}}$ & 
   $\frac{15}{4} \frac{1}{2\sqrt{7}}$ &
   $\frac{5}{4\sqrt{21}}$ & 
   $\frac{2}{\sqrt{7}}$ &
   $-\frac{5}{4\sqrt{21}}$ \\
  \end{tabular}
 \end{ruledtabular}
\end{table}

\begin{table}[h!]
 \caption{\label{64q27q} $ \left\langle \bm{64}_q\left|{\cal
 O}_i\right|\bm{27}_q\right\rangle$}
 \begin{ruledtabular}
  \begin{tabular}{c|ccccc}
   $(I,Y)$ & $\gamma$ & $x$ & $y$ & $z$ & $w$ \\ \hline
   $(\frac{3}{2}, +1)$ &
   $\frac{5\sqrt{3}}{56}$ & 
   $\frac{15}{4} \frac{5\sqrt{3}}{56}$ &
   $-\frac{5}{7}$ & 
   $\frac{55\sqrt{3}}{112}$ &
   $\frac{5}{16}$ \\
   $(1,0)$ &
   $\frac{5\sqrt{5}}{56}$ & 
   $\frac{15}{4} \frac{5\sqrt{5}}{56}$ & 
   $-\frac{5\sqrt{5}}{7\sqrt{3}}$ & 
   $\frac{55\sqrt{5}}{112}$ &
   $\frac{5\sqrt{5}}{16\sqrt{3}}$ \\
   $(\half,-1)$ &
   $\frac{5\sqrt{3}}{28\sqrt{2}}$ & 
   $\frac{15}{4} \frac{5\sqrt{3}}{28\sqrt{2}}$ & 
   $-\frac{5\sqrt{2}}{\sqrt{7}}$ & 
   $\frac{55\sqrt{3}}{56\sqrt{2}}$ &
   $\frac{5}{8\sqrt{2}}$ \\
   $(0,0)$ &
   $\frac{3\sqrt{5}}{28}$ & 
   $\frac{15}{4} \frac{3\sqrt{5}}{28}$ & 
   $-\frac{2\sqrt{15}}{7}$ & 
   $\frac{33\sqrt{5}}{56}$ &
   $\frac{\sqrt{15}}{8}$ 
  \end{tabular}
 \end{ruledtabular}
\end{table}

\begin{table}[h!]
 \caption{\label{64q35}$ \left\langle \bm{64}_q\left|{\cal
 O}_i\right|\bm{35}\right\rangle$}
 \begin{ruledtabular}
  \begin{tabular}{c|ccccc}
   $(I,Y)$ & $\gamma$ & $x$ & $y$ & $z$ & $w$ \\ \hline
   $(\frac{3}{2}, +1)$ &
   $\frac{9}{8\sqrt{35}}$ & 
   $\frac{15}{4} \frac{9}{8\sqrt{35}}$ & 
   $\frac{3\sqrt{15}}{4\sqrt{7}}$ & 
   $\frac{27\sqrt{5}}{16\sqrt{7}}$ &
   $-\frac{51\sqrt{3}}{16\sqrt{35}}$ \\
   $(1,0)$ &
   $\frac{3}{8\sqrt{7}}$ & 
   $\frac{15}{4} \frac{3}{8\sqrt{7}}$ & 
   $\frac{5\sqrt{3}}{4\sqrt{7}}$ & 
   $\frac{45}{16\sqrt{7}}$ &
   $-\frac{17\sqrt{3}}{16\sqrt{7}}$ \\
   $(\half,-1)$ &
   $\frac{3}{4\sqrt{70}}$ & 
   $\frac{15}{4} \frac{3}{4\sqrt{70}}$ & 
   $\frac{\sqrt{15}}{2\sqrt{14}}$ & 
   $\frac{9\sqrt{5}}{8\sqrt{14}}$ &
   $-\frac{17\sqrt{3}}{8\sqrt{70}}$ \\
  \end{tabular}
 \end{ruledtabular}
\end{table}

\begin{table}[h!]
 \caption{\label{8135}$ \left\langle \bm{81}\left|{\cal
 O}_i\right|\bm{35}\right\rangle$}
 \begin{ruledtabular}
  \begin{tabular}{c|ccccc}
   $(I,Y)$ & $\gamma$ & $x$ & $y$ & $z$ & $w$ \\ \hline
   $(\frac{3}{2}, +1)$ &
   $\frac{3}{2\sqrt{35}}$ & 
   $\frac{15}{4} \frac{3}{2\sqrt{35}}$ & 
   $-\frac{3\sqrt{15}}{4\sqrt{7}}$ & 
   $\frac{3\sqrt{5}}{\sqrt{7}}$ &
   $\frac{3\sqrt{3}}{4\sqrt{35}}$ \\
   $(1,0)$ &
   $\frac{3}{4\sqrt{7}}$ & 
   $\frac{15}{4} \frac{3}{4\sqrt{7}}$ & 
   $-\frac{15\sqrt{3}}{8\sqrt{7}}$ & 
   $\frac{15}{2\sqrt{7}}$ &
   $\frac{3\sqrt{3}}{8\sqrt{7}}$ \\
   $(\half,-1)$ &
   $\sqrt{\frac{3}{35}}$ & 
   $\frac{15}{4} \sqrt{\frac{3}{35}}$ &
   $-\frac{3\sqrt{5}}{2\sqrt{7}}$ & 
   $\frac{2\sqrt{15}}{\sqrt{7}}$ &
   $\frac{3}{2\sqrt{35}}$ \\
   $(0,-2)$ &
   $\frac{3}{4\sqrt{7}}$ & 
   $\frac{15}{4} \frac{3}{4\sqrt{7}}$ & 
   $-\frac{15\sqrt{3}}{8\sqrt{7}}$ & 
   $\frac{15}{2\sqrt{7}}$ &
   $\frac{3\sqrt{3}}{8\sqrt{7}}$ \\
  \end{tabular}
 \end{ruledtabular}
\end{table}

Because these matrix elements listed above only contribute to the
second order calculations, those of ${\cal O}_{v_1}$ and ${\cal
O}_{v_2}$ are not necessary, because these operators themselves are of
second order.  There are, however, extra matrix elements that we need to
calculate; 
$ \left\langle \bm{\overline{35}}_d \left|{\cal O}_v
 \right|\bm{8} \right\rangle$,
$ \left\langle \bm{64}_d \left|{\cal O}_v
 \right|\bm{8} \right\rangle$,
$ \left\langle \bm{64}_d \left|{\cal O}_v
 \right|\bm{\overline{10}} \right\rangle$,
$ \left\langle \bm{\overline{81}}_d \left|{\cal O}_v
 \right|\bm{\overline{10}} \right\rangle$,
$ \left\langle \bm{\overline{35}}_q \left|{\cal O}_v
 \right|\bm{10} \right\rangle$,
$ \left\langle \bm{64}_q \left|{\cal O}_v
 \right|\bm{10} \right\rangle$, and
$ \left\langle \bm{81} \left|{\cal O}_v
 \right|\bm{10} \right\rangle$.
\begin{table}[h!]
 \caption{\label{v8}$ \left\langle \bm{\overline{35}}_d \left|{\cal O}_v
 \right|\bm{8} \right\rangle$ \& 
 $ \left\langle \bm{64}_d \left|{\cal O}_v
 \right|\bm{8} \right\rangle$}
 \begin{ruledtabular}
  \begin{tabular}{c|cccc}
   & \multicolumn{2}{c}{$\left\langle \bm{\overline{35}}_d \left|{\cal O}_v
   \right|\bm{8} \right\rangle$} 
   & \multicolumn{2}{c}{$ \left\langle \bm{64}_d \left|{\cal O}_v
   \right|\bm{8} \right\rangle$} \\ \hline
   (I,Y) & $v_1$ & $v_2$ & $v_1$ & $v_2$\\ \hline

   $(\frac{1}{2},1)$ & $\frac{3}{4\sqrt{70}}$ & $\frac{1}{4\sqrt{70}}$ 
   & $\frac{3}{28\sqrt{2}}$ & $\frac{1}{28\sqrt{2}}$\\

   $(1,0)$ & $\frac{1}{2\sqrt{35}}$ & $\frac{1}{6\sqrt{35}}$ 
   & $\frac{\sqrt{5}}{28\sqrt{2}}$ & $\frac{\sqrt{5}}{84\sqrt{2}}$\\

   $(\frac{1}{2},-1)$ & 0 & 0
   & $\frac{3}{28\sqrt{2}}$ & $\frac{1}{28\sqrt{2}}$\\

   $(0,0)$ & 0 & 0 
   & $\frac{9}{28\sqrt{10}}$ & $\frac{3}{28\sqrt{10}}$
 \end{tabular}
 \end{ruledtabular}
\end{table}

\begin{table}[h!]
 \caption{\label{v10b}$ \left\langle \bm{64}_d \left|{\cal O}_v
 \right|\bm{\overline{10}} \right\rangle$ \&
 $ \left\langle \bm{\overline{81}}_d \left|{\cal O}_v
 \right|\bm{\overline{10}} \right\rangle$}
 \begin{ruledtabular}
  \begin{tabular}{c|cccc}
   & \multicolumn{2}{c}{$ \left\langle \bm{64}_d \left|{\cal O}_v
   \right|\bm{\overline{10}} \right\rangle$}
   & \multicolumn{2}{c}{$ \left\langle \bm{\overline{81}}_d \left|{\cal O}_v
   \right|\bm{\overline{10}} \right\rangle$}
   \\ \hline
   (I,Y) & $v_1$ & $v_2$ & $v_1$ & $v_2$\\ \hline

   $(0,2)$ & 0 & 0  
   & $\frac{3\sqrt{3}}{56}$ & $\frac{\sqrt{3}}{56}$\\

   $(\frac{1}{2},1)$ & $\frac{3}{28\sqrt{10}}$ & $\frac{1}{28\sqrt{10}}$ 
   & $\frac{9}{28\sqrt{10}}$ & $\frac{3}{28\sqrt{10}}$\\

   $(1,0)$ & $\frac{1}{14\sqrt{2}}$ & $\frac{1}{42\sqrt{2}}$ 
   & $\frac{3\sqrt{3}}{56}$ & $\frac{\sqrt{3}}{56}$\\

   $(\frac{3}{2},-1)$ & $\frac{3}{56}$ & $\frac{1}{56}$
   & $\frac{3\sqrt{3}}{56\sqrt{2}}$ & $\frac{\sqrt{3}}{56\sqrt{2}}$
 \end{tabular}
 \end{ruledtabular}
\end{table}

\begin{table}[h!]
 \caption{\label{v10}$ \left\langle \bm{\overline{35}}_q \left|{\cal O}_v
 \right|\bm{10} \right\rangle$, 
 $ \left\langle \bm{64}_q \left|{\cal O}_v
 \right|\bm{10} \right\rangle$ \&
 $ \left\langle \bm{81} \left|{\cal O}_v
 \right|\bm{10} \right\rangle$ }
 \begin{ruledtabular}
  \begin{tabular}{c|cccccc}
   & \multicolumn{2}{c}{$ \left\langle \bm{\overline{35}}_q \left|{\cal O}_v
   \right|\bm{10} \right\rangle$}
   & \multicolumn{2}{c}{$ \left\langle \bm{64}_q \left|{\cal O}_v
   \right|\bm{10} \right\rangle$}
   & \multicolumn{2}{c}{$ \left\langle \bm{81} \left|{\cal O}_v
   \right|\bm{10} \right\rangle$}
   \\ \hline
   (I,Y) & $v_1$ & $v_2$ & $v_1$ & $v_2$ & $v_1$ & $v_2$\\ \hline

   $(\frac{3}{2},1)$ & $\frac{\sqrt{5}}{6\sqrt{2}}$ 
   & $\frac{\sqrt{5}}{18\sqrt{2}}$  
   & $\frac{3\sqrt{5}}{56\sqrt{2}}$ 
   & $\frac{\sqrt{5}}{56\sqrt{2}}$
   & $\frac{3\sqrt{5}}{112\sqrt{2}}$ 
   & $\frac{\sqrt{5}}{112\sqrt{2}}$ \\

   $(1,0)$ &  $\frac{\sqrt{5}}{6\sqrt{2}}$ 
   & $\frac{\sqrt{5}}{18\sqrt{2}}$  
   & $\frac{\sqrt{5}}{28}$ & $\frac{\sqrt{5}}{84}$
   & $\frac{3\sqrt{5}}{112}$ 
   & $\frac{\sqrt{5}}{112}$\\

   $(\frac{1}{2},-1)$ & 0 & 0
   & $\frac{3}{56}$ & $\frac{1}{56}$
   & $\frac{3\sqrt{3}}{56\sqrt{2}}$ 
   & $\frac{\sqrt{3}}{56\sqrt{2}}$\\

   $(0,-2)$ & 0 & 0 & 0 & 0
   & $\frac{3\sqrt{5}}{112}$ 
   & $\frac{\sqrt{5}}{112}$
 \end{tabular}
 \end{ruledtabular}
\end{table}

\section{``Traditional'' Approach to the Skyrme Model}
\label{Sec:Trad}

In this section, we consider the ``traditional'' approach to the action
(\ref{action}). Namely, we calculate the profile function $F(r)$ of the
soliton, then all of the Skyrme model parameters are determined by the
$\chi$PT parameters and the integrations involving the profile function.

It has been known that, in the conventional Skyrme model where ${\cal
O}_\gamma$ is the only symmetry breaking interaction, the physical
values of the $\chi$PT parameters do not reproduce the baryon mass
spectrum. In particular, the best fit value of the pion decay constant
$F_\pi$ is typically less than one third of the physical value (e.g.,
$46$ MeV\cite{Praszalowicz:1985bt} for the physical value $184.8$ MeV),
while the Kaon mass (if it is treated as a parameter) becomes quite
large (around $800$ MeV).

From the effective theory point of view, this discrepancy can be
understood easily. There are an infinite number of operators in the
$\chi$PT action and they contribute to the Skyrme model parameters. The
conventional Skyrme model ignores all of such contributions from the
higher order terms. Furthermore, we do not know those coupling constants
at all. In this paper, we therefore give up to ``calculate''
the Skyrme model parameters from the $\chi$PT parameters, and fit them
directly to the experimental values. 

An important question is whether the best fit values of the $\chi$PT
parameters ``improve'' as we systematically take the higher order
contributions into account. In this Appendix, we address this
question. Starting with the action (\ref{action}), we calculate the
profile function and the Skyrme model parameters. By fitting them to the
experimental values of the baryon masses, we obtain the best fit values
of the $\chi$PT parameters.

\subsection{Profile function $F(r)$}

First of all, it is useful to define the subtracted action $S_{sub}[U]$
as
\begin{equation}
 S_{sub}[U] \equiv S[U]
  -  \frac{F_0^2 B_0}{8}\int d^4x
  \Tr\left({\cal M}^\dagger +{\cal M}\right) 
  -L_8B_0^2\int d^4x
  \Tr\left({\cal M}^\dagger {\cal M}^\dagger 
      +{\cal M}{\cal M}\right),
\end{equation}
so that $S_{sub}[U=1]=0$. 

The profile function $F(r)$ minimizes $M[F]$ defined by
\begin{equation}
 S_{sub}[U_c]=-M[F]\int dt,
\end{equation}
subject to the boundary conditions,
\begin{equation}
 F(0)=\pi,\quad F(\infty)=0.
\end{equation}
The minimum is called $M_{cl}$,
\begin{equation}
 M_{cl}= \min_{F(r)} M[F].
\end{equation}
The profile function satisfies the Euler-Lagrange equation,
\begin{equation}
 \frac{d}{d{r}}\pd{M}{(dF/dr)}-\pd{M}{F}=0.
\end{equation}
By substituting the solution $F(r)$, we obtain $M_{cl}$,
\begin{eqnarray}
 M_{cl}&=&\frac{4\pi F_0}{e}\int_0^{\infty}\tilde{r}^2d\tilde{r}
  \bigg[
  \frac{1}{8}
  \left\{
   \left(F'\right)^2+\frac{2\sin^2F}{\tilde{r}^2}
  \right\}+\frac{\sin^2F}{2\tilde{r}^2}
  \left\{
   2\left(F'\right)^2+\frac{\sin^2F}{\tilde{r}^2}
  \right\} \nonumber \\
 &&{}+\frac{mB_0}{2F_0^2e^2}\left(1-\cos F\right)
  +\frac{4mB_0L_5}{F_0^2}\cos F
  \left\{
    \left(F'\right)^2+\frac{2\sin^2F}{\tilde{r}^2}
  \right\} \nonumber \\
 &&{}+\frac{8m^2B_0^2L_8}{F_0^4e^2}\sin^2F
  \bigg],
\end{eqnarray}
where we have introduced $\tilde{r}=F_0er$ and $F'=dF/d\tilde{r}$. Note
that there are higher order contributions and they affect the behavior
of the profile function, and hence the values of the other parameters.
The existence of the higher order interactions affects the values
of the parameters in two ways; through the behavior of the profile
function and the new terms from the higher order interactions.

For a large value of $r$, $F(r)$ and its derivative behave like $F\sim
0$, $F'\sim 0$, and thus $\sin F\sim F$, $\cos F\sim 1$, we have
\begin{eqnarray}
 \pd{M}{F'}&\sim&\frac{4\pi F_0}{e}
  \left(
   \frac{1}{4}+\frac{8mB_0L_5}{F_0^2}
  \right)\tilde{r}^2  F', \\
 \pd{M}{F}&\sim&\frac{4\pi F_0}{e}
  \left(
   \frac{mB_0}{2F_0^2e^2}+\frac{16m^2B_0^2L_8}{F_0^4e^2}
  \right) \tilde{r}^2 F.
\end{eqnarray}
By solving the Euler-Lagrange equation, the asymptotic behavior,
\begin{equation}
  F(\tilde{r})\sim \frac{e^{-\mu \tilde{r}}}{\tilde{r}},
\end{equation}
is obtained, where $\mu$ is given by
\begin{eqnarray}
 \mu&=&\frac{\sqrt{2mB_0}}{F_0e}
  \sqrt{
  \frac{1+\frac{32mB_0L8}{F_0^2}}{1+\frac{32mB_0L_5}{F_0^2}}
  } \nonumber \\
 &\approx&
  \frac{\sqrt{2mB_0}}{F_0e}\left(1+\frac{16mB_0}{F_0^2}(L_8-L_5)\right).
\end{eqnarray}
Note that $\mu$ is close to the pion mass but not exactly the same.

\subsection{Inertia tensor $I_{\alpha\beta}(A)$}

After substituting $U=AU_cA^\dagger$ into $S_{sub}[U]$, the inertia
tensor can be easily read off as the coefficients of the terms quadratic
in the ``angular velocity'' $\omega^\alpha$,
\begin{eqnarray}
 I_{\alpha\beta}(A)&=&-\frac{F_0^2}{16}\int d^3x \half\Tr
  \left[
   \lambda_\alpha U_c^\dagger\lambda_\beta U_c
   +\lambda_\beta U_c^\dagger\lambda_\alpha U_c 
   -\{\lambda_\alpha,\lambda_\beta\}
  \right] \nonumber \\
 &&+{}\frac{1}{128e^2}\int d^3x \Tr
  \left(
   [U_c^\dagger \lambda_\alpha U_c-\lambda_\alpha,U_c^\dagger\der_iU_c]
   [U_c^\dagger \lambda_\beta U_c-\lambda_\beta,U_c^\dagger\der_iU_c]
  \right) \nonumber \\
 &&{}-\frac{2L_5B_0}{3}\frac{2m+m_s}{3}
  \int d^3x \left(1+2\cos F\right) 
  \half\Tr
  \left[
   \lambda_\alpha U_c^\dagger\lambda_\beta U_c
   +\lambda_\beta U_c^\dagger\lambda_\alpha U_c 
   -\{\lambda_\alpha,\lambda_\beta\}
  \right] 
  \nonumber \\
 &&{}+\frac{L_5B_0}{\sqrt{3}}\frac{2m+m_s}{3}
  \int d^3x \left(1-\cos F\right)
  \half\Tr
  \bigg[
  \lambda_8
  \big(
  \lambda_\alpha U_c^\dagger \lambda_\beta U_c
  +\lambda_\beta U_c^\dagger \lambda_\alpha U_c \nonumber \\
 &&{}\qquad\qquad\qquad\qquad\qquad
  +U_c^\dagger \lambda_\alpha U_c \lambda_\beta
  +U_c^\dagger \lambda_\beta U_c \lambda_\alpha  
  -2\{\lambda_\alpha,\lambda_\beta\}
  \big)
  \bigg] \nonumber \\
 &&{}-\frac{L_5B_0}{2}\frac{m_s-m}{\sqrt{3}}\int d^3x
  \half \Tr
  \big[
  (A^\dagger \lambda_8 A) 
  \big(
  \left[\lambda_\alpha,U_c\right]U_c^\dagger\left[\lambda_\beta,U_c\right]
  +\left[\lambda_\beta,U_c\right]U_c^\dagger\left[\lambda_\alpha,U_c\right]
  \nonumber \\
 &&{}\qquad\qquad\qquad\qquad
  +\left[\lambda_\alpha,U_c^\dagger\right]
  U_c\left[\lambda_\beta,U_c^\dagger\right]
  +\left[\lambda_\beta,U_c^\dagger\right]
  U_c\left[\lambda_\alpha,U_c^\dagger\right]
  \big)
  \big] \nonumber  \\
 &=&I_{\alpha\beta}^0+I'_{\alpha\beta}(A).
\end{eqnarray}

The $A$-independent part $I_{\alpha\beta}^0$ may be written as
Eq.~(\ref{Izero}) with
\begin{eqnarray}
 I_1&=&\frac{2\pi}{3e^3F_0}\int_0^\infty \tilde{r}^2 d\tilde{r}\sin^2F
  \left[
   1+4\left\{\left(F'\right)^2+\frac{\sin^2F}{\tilde{r}^2}\right\}
   +\frac{32\overline{m}B_0}{F_0^2}L_5\cos F
  \right], \\
 I_2&=&\frac{\pi}{2e^3F_0}\int_0^\infty \tilde{r}^2 d\tilde{r}
 (1-\cos F) 
 \left[
  1+\left(F'\right)^2+\frac{2\sin^2F}{\tilde{r}^2}
  +\frac{8\overline{m}B_0}{F_0^2}L_5\left(1+\cos F\right)
 \right],
\end{eqnarray}
where 
\begin{equation}
 \overline{m}=\frac{2m+m_s}{3}
\end{equation}
has been introduced.

The $A$-dependent part $I'_{\alpha\beta}(A)$ has the complicated
structure as Eq.~(\ref{Iprime}). The parameters are now expressed as
\begin{eqnarray}
 \overline{x}&=&-\frac{64\pi L_5B_0}{9e^3F_0^3}\delta m
  \int_0^\infty \tilde{r}^2 d\tilde{r}
  \cos F\sin^2F, \\
 \overline{y}&=&-\frac{32\pi L_5B_0}{3\sqrt{3}e^3F_0^3}\delta m
  \int_0^\infty \tilde{r}^2 d\tilde{r} 
  \sin^2F, \\
 \overline{z}&=&\frac{32\pi L_5B_0}{9e^3F_0^3}\delta m
  \int_0^\infty \tilde{r}^2 d\tilde{r} 
(1-\cos F)(2-\cos F), \\
 \overline{w}&=&\frac{16\pi L_5B_0}{3\sqrt{3}e^3F_0^3}\delta m
  \int_0^\infty  \tilde{r}^2 d\tilde{r} 
(1-\cos F)(2-\cos F).
\end{eqnarray}

\subsection{Potential}

The ``angular velocity'' independent part (except for $M_{cl}$) of the
Lagrangian is the potential part Eq.~(\ref{potential}), $-V(A)$. The
parameter $\gamma$ in the first-order term (\ref{pot1}) is given by
\begin{equation}
 \gamma=\gamma_0+\delta_1\gamma+\delta_2\gamma,
\end{equation}
with
\begin{eqnarray}
 \gamma_0&=&\delta m\frac{4\pi B_0}{3e^3F_0}
  \int_0^\infty \tilde{r}^2 d\tilde{r} (1-\cos F), \\
 \delta_1\gamma&=&\delta m\frac{32\pi B_0L_5}{3eF_0}
  \int_0^\infty \tilde{r}^2 d\tilde{r} \cos F 
  \left(
   \left(F'\right)^2+\frac{2\sin^2F}{\tilde{r}^2}
  \right), \\
 \delta_2\gamma&=&\delta m\frac{64\pi \overline{m}B_0^2L_8}{3e^3F_0^3}
  \int_0^\infty \tilde{r}^2 d\tilde{r} \left(1-\cos 2F\right).
\end{eqnarray}
The parameter $v$ in the second-order term (\ref{pot2}) is given by
\begin{eqnarray}
 v&=&\left(\delta m\right)^2\frac{16\pi B_0^2 L_8}{9e^3F_0^3}
  \int_0^\infty \tilde{r}^2 d\tilde{r}
  \left(1-\cos F\right)\left(1-2\cos F\right).
\end{eqnarray}

\subsection{Numerical calculations}

Starting with the $\chi$PT parameters $F_0$, $B_0$, $e$, $L_5$, $L_8$,
and quark masses, $m$ and $m_s$, we can first calculate $F(r)$ and then,
by using it, the Skyrme model parameters, $M_{cl}$, $I_1$, $I_2$,
$\gamma$, $x$, $y$, $z$, $w$, and $v$. Once these parameters are
determined, the baryon masses can be easily calculated in the
second-order perturbation theory. In order to best fit the $\chi$PT
parameters, we need to solve {\em reversely}. The procedure is similar
to that discussed in Sec.~\ref{Sec:Num}, but a bit more complicated.
In order to simplify the calculation, we make the following things; 
\begin{enumerate}
 \item The quark masses are fixed. As a reference, we adopt the
       following values
       \begin{equation}
	m=6\ \mbox{\rm MeV},\quad m_s = 150\ \mbox{\rm MeV}.
       \end{equation}
       Actually, it only fixes the ratio, because the change of the
       magnitude can be absorbed in $B_0$. (Note that ${\cal M}$ appears
       only in the combination $B_0{\cal M}$.) The ratio is better
       determined experimentally and known to be\cite{Eidelman:2004wy,
       Leutwyler:1996qg}
       \begin{equation}
	\frac{m_s}{(m_u+m_d)/2}=\frac{2(m_s/m_d)}{1+(m_u/m_d)}\approx 25.8
       \end{equation}
       which is close to the value $150/6=25$.
 \item The values of $L_5$ and $L_8$ are fixed. When we vary these
       parameters too, we find that numerical calculation becomes very
       unstable. In reality, all of the formulation in this paper
       assumes these parameters to be small. In searching the ``valley''
       numerically, this assumption is often ignored, and we believe that
       it is the reason of the instability. Instead, we fix these
       parameters to be the central values determined
       experimentally\cite{Pich:1995bw}, 
       \begin{equation}
	L_5=1.4\times 10^{-3},\quad L_8=0.9\times 10^{-3}.
       \end{equation}
\end{enumerate}

There is another important point. As Yabu and Ando\cite{Yabu:1987hm}
discussed, there is a kind of ``zero-point energy'' contribution
universal to all of the calculated baryon masses. This contribution may
be calculated as the symmetry breaking effects to the fictitious
(unphysical) singlet baryon mass,
\begin{eqnarray}
 M_{vac}&=&\frac{\gamma}{2}
  \left\langle1\left|\left(1-D_{88}^{(8)}(A)\right)\right|1\right\rangle
  +v
  \left\langle 1
  \left|
   \left(
    1\!-\!\sum_{\alpha\in{\cal I}}\left(D_{8\alpha}^{(8)}(A)\right)^2
    \!\!-\left(D_{88}^{(8)}(A)\right)^2
   \right)
  \right| 1
  \right\rangle \nonumber \\
 &&{}-\frac{1}{M_8-M_1}
  \left|
   \left\langle 8
   \left|
   \left(
    -\frac{\gamma}{2}D_{88}^{(8)}(A)
   \right)
   \right| 1
   \right\rangle
  \right|^2 \nonumber \\
 &=& \frac{\gamma}{2}+\frac{v}{2}-\frac{\gamma^2I_2}{48},
\end{eqnarray}
where we have introduced
\begin{equation}
 M_1=M_{cl}+\frac{3}{8}\left[\frac{1}{I_1}-\frac{2}{I_2}\right].
\end{equation}
We subtract it from all of the calculated masses. 

Note that, in the effective theory approach, this contribution is
renormalized in the parameter $M_{cl}$, and, therefore, does not need to
be considered separately. It has been implicitly taken into account.

Our numerical results are
\begin{equation}
 F_0=82.7\ \mbox{\rm MeV},\ B_0=2697,\ 
  e=4.51,
\end{equation}
which lead to the following values for the physical parameters,
\begin{equation}
 F_\pi= 91.4\ \mbox{\rm MeV},\ M_\pi=185.3\ \mbox{\rm MeV}, \ 
  M_{\rm K}=867.2\ \mbox{\rm MeV}
\end{equation}
and the baryon masses for these values are given in
Table~\ref{traditional}.
\begin{table}[h]
 \caption{\label{traditional}Baryon masses for the best fit values in
 the ``traditional'' approach.  The results in the first low is with the
 higher order contributions, and in the second low without. The
 ``zero-point energy'' contribution has been subtracted.}
 \begin{ruledtabular}
  \begin{tabular}{c||cccccccccc}
   Baryon &$M_{\rm N}$& $M_\Sigma$& $M_\Xi$& $M_\Lambda$& $M_\Delta$& 
   $M_{\Sigma^*}$& $M_{\Xi^*}$& $M_\Omega$&$M_\Theta$& 
   $M_{\phi}$\\
   \hline
   $\mbox{Mass}_{w}$(MeV)
   &915 & 1287 & 1411 & 1116 & 1185 & 1358 & 1518 & 1666 & 1563 & 1965 \\
   $\mbox{Mass}_{w/o}$(MeV)&898 & 1269 & 1405 & 1116 & 1118 & 1321 & 1501 & 1660 
   & 1639 & 1948
  \end{tabular}
 \end{ruledtabular}
\end{table}
These values should be compared with those calculated with
$L_5=L_8=0$, that is, those without the contributions from higher order
terms. 
\begin{equation}
 F_0=F_\pi=58.2\ \mbox{\rm MeV},\ B_0=7825,\ 
  e=4.06,
\end{equation}
which lead to
\begin{equation}
 M_\pi=306.4\ \mbox{\rm MeV}, \   M_{\rm K}=1104.9\ \mbox{\rm MeV}.
\end{equation}
The baryon masses are also given in Table~\ref{traditional}.

It is interesting to note that the values of the physical parameters
shift {\em in the right direction}.  Even though these values are still
far from the experimental values, we think that this is an explicit
demonstration that our basic strategy is right.

\section{Symmetry breaking interactions in the chiral quark-soliton model
 }
\label{Sec:DPP}

In this Appendix, we perform a similar ``best fit'' analysis with the
symmetry breaking terms (\ref{DPP}) which appear in the $\chi$QSM. In
the derivation of these terms\cite{Blotz:1992pw}, they are related to
the $\pi$-N $\sigma$ term, soliton moments of inertia, and so on, but we
ignore this fact and just treat the couplings as free parameters. The
reason is that it makes the comparison with our approach transparent and
reveals how the $\chi$QSM predictions depend on the detailed form of the
parameters. 

\subsection{Best fit to the baryon masses}

To obtain the masses in the second order perturbation theory, we need
the matrix elements of $H_1^{DPP}$. The matrix elements of
$D_{88}^{(8)}$ are given in Sec.~\ref{symbreak} and in
Appendix~\ref{Sec:Tables}, and those for $Y$ are trivial. It is only
${\cal O}_{DPP}\equiv D_{8i}^{(8)}F^i$ of which the matrix elements need
to be calculated. We first present the matrix elements for the spin
$J=\half$ states in Tables~\ref{dpp1} and \ref{dpp2}.  

\begin{table}[h!]
 \caption{\label{dpp1}The matrix elements of ${\cal O}_{DPP}$,
 $\left\langle{\cal R}_i \left|{\cal O}_{DPP}\right|{\cal
 R}_j\right\rangle$, for the spin doublet states, which are abbreviated
 as $({\cal R}_i,{\cal R}_j)$.}
 \begin{ruledtabular}
  \begin{tabular}{c|cccccc}
   $\ (I,Y)\ $ & $(\bm{8},\bm{8})$ & $(\bm{8},\bm{\overline{10}})$ & 
   $(\bm{8},\bm{27}_d)$ & 
   $(\bm{\overline{10}},\bm{\overline{10}})$ &
   $(\bm{\overline{10}},\bm{27}_d)$ &
   $(\bm{\overline{10}},\bm{\overline{35}}_d)$ \\ \hline
   $(0,+2)$ & 
   $0$ & $ 0$ & $ 0 $ &
   $ -\frac{{\sqrt{3}}}{8}$ & $ 0$ & $ \frac{\sqrt{3}}{8\sqrt{7}}$ \\
   $(\half, +1)$ &
   $-\frac{{\sqrt{3}}}{20}$ & $ \frac{\sqrt{3}}{4\sqrt{5}}$ & 
   $ -\frac{1}{10 {\sqrt{2}}}$ &
   $ -\frac{{\sqrt{3}}}{16}$ & $ -\frac{7}{16 {\sqrt{10}}}$ & 
   $ \frac{3 \sqrt{3}}{16\sqrt{14}}$ \\
   $(1,0)$ &
   $-\frac{3 {\sqrt{3}}}{20}$ & $ \frac{\sqrt{3}}{4\sqrt{5}}$ & 
   $ -\frac{1}{10 {\sqrt{3}}}$ &
   $ 0$ & $ -\frac{7}{8 {\sqrt{15}}}$ & $ \frac{\sqrt{3}}{8\sqrt{7}}$ \\
   $(\half, -1)$ &
   $\frac{{\sqrt{3}}}{5}$ & $ 0$ & $ -\frac{1}{10 {\sqrt{2}}} $ &
   $ 0$ & $ 0$ & $ 0$ \\
   $(\frac{3}{2},-1)$ &
   $0$ & $ 0$ & $ 0 $ &
   $ \frac{{\sqrt{3}}}{16}$ & $ -\frac{7}{16 {\sqrt{2}}}$ & 
   $ \frac{\sqrt{15}}{16\sqrt{14}}$ \\
   $(0,0)$ &
   $\frac{3 {\sqrt{3}}}{20}$ & $ 0$ & $ -\frac{{\sqrt{3}}}{20}$ &
   $ 0$ & $ 0$ & $ 0$ \\
  \end{tabular}
 \end{ruledtabular}
\end{table}

\begin{table}[h!]
 \caption{\label{dpp2} The same as in Table~\ref{dpp1}.}
 \begin{ruledtabular}
  \begin{tabular}{c|cccccc}
   $\ (I,Y)\ $ & $(\bm{27}_d,\bm{27}_d)$ & $(\bm{27}_d,\bm{\overline{35}}_d)$ & 
   $(\bm{27}_d,\bm{64}_d)$ & 
   $(\bm{\overline{35}}_d,\bm{\overline{35}}_d)$ &
   $(\bm{\overline{35}}_d,\bm{64}_d)$ &
   $(\bm{\overline{35}}_d,\bm{\overline{81}}_d)$ \\ \hline
   $(0,+2)$ & 
   $0$ & $ 0$ & $ 0 $ &
   $-\frac{{\sqrt{3}}}{8}$ & $ 0$ & $ -\frac{3\sqrt{3}}{8\sqrt{7}} $ \\
   $(\half, +1)$ &
   $\frac{71 {\sqrt{3}}}{1120}$ & $ \frac{5 \sqrt{5}}{32\sqrt{7}}$ & 
   $ -\frac{5}{56}$ &
   $-\frac{3 {\sqrt{3}}}{32}$ & $ -\frac{3\sqrt{3}}{8\sqrt{35}}$ & 
   $ -\frac{3}{2 {\sqrt{35}}} $ \\
   $(1,0)$ &
   $\frac{19 {\sqrt{3}}}{560}$ & $ \frac{5 {\sqrt{5}}}{16\sqrt{21}}$ & 
   $ -\frac{5 \sqrt{5}}{56\sqrt{6}} $ &
   $-\frac{{\sqrt{3}}}{16}$ & $ -\frac{3\sqrt{3}}{8\sqrt{14}}$ & 
   $ -\frac{3\sqrt{3}}{8\sqrt{7}} $ \\
   $(\half, -1)$ &
   $\frac{11 {\sqrt{3}}}{70}$ & $ 0$ & $ -\frac{5}{56} $ &
   $0$ & $ 0$ & $ 0 $ \\
   $(\frac{3}{2},-1)$ &
   $\frac{{\sqrt{3}}}{224}$ & $ \frac{5\sqrt{5}} {32\sqrt{7}}$ & 
   $ -\frac{5}{56 {\sqrt{2}}} $ &
   $-\frac{{\sqrt{3}}}{32}$ & $ -\frac{9\sqrt{3}}{8\sqrt{70}}$ & 
   $ -\frac{3\sqrt{3}}{4\sqrt{35}} $ \\
   $(0,0)$ &
   $\frac{19 {\sqrt{3}}}{140}$ & $ 0$ & 
   $ -\frac{\sqrt{15}}{28\sqrt{2}}$ &
   $0$ & $ 0$ & $ 0 $    
  \end{tabular}
 \end{ruledtabular}
\end{table}

Those for the
spin $J=\frac{3}{2}$ states in Tables~\ref{dpp3} and \ref{dpp4}. 
\begin{table}[h!]
 \caption{\label{dpp3}The matrix elements of ${\cal O}_{DPP}$,
 $\left\langle{\cal R}_i \left|{\cal O}_{DPP}\right|{\cal
 R}_j\right\rangle$, for the spin quartet states, which are abbreviated
 as $({\cal R}_i,{\cal R}_j)$.}
 \begin{ruledtabular}
  \begin{tabular}{c|ccccc}
   $\ (I,Y)\ $ & $(\bm{10},\bm{10})$ & $(\bm{10},\bm{27}_q)$ & 
   $(\bm{10},\bm{35})$ & 
   $(\bm{27}_q,\bm{27}_q)$ & $(\bm{27}_q,\bm{35})$ 
   \\ \hline
   $(\frac{3}{2},+1)$ & 
   $-\frac{5 {\sqrt{3}}}{16}$ & $\frac{5\sqrt{5}}{16\sqrt{2}}$ & 
   $-\frac{5\sqrt{3}}{16\sqrt{14}}$ &
   $-\frac{65 {\sqrt{3}}}{224}$ & $-\frac{{\sqrt{35}}}{32}$ \\
   $(1, 0)$ &
   $0$ & $\frac{5}{8 {\sqrt{3}}}$ & $-\frac{\sqrt{15}}{8\sqrt{7}}$ &
   $\frac{5 {\sqrt{3}}}{112}$ & $-\frac{\sqrt{35}}{16\sqrt{3}}$  \\
   $(\half,-1)$ &
   $\frac{5 {\sqrt{3}}}{16}$ & $\frac{5}{16 \sqrt{2}}$ & 
   $-\frac{3\sqrt{15}}{16\sqrt{14}}$ &
   $\frac{85 {\sqrt{3}}}{224}$ & $-\frac{\sqrt{35}}{32}$\\
   $(0, -2)$ &
   $\frac{5 {\sqrt{3}}}{8}$ & $0$ & $-\frac{\sqrt{15}}{8\sqrt{7}}$ &
   $0$ & $0$ \\
  \end{tabular}
 \end{ruledtabular}
\end{table}

\begin{table}[h!]
 \caption{\label{dpp4}The same as in Table~\ref{dpp3}.}
 \begin{ruledtabular}
  \begin{tabular}{c|ccccc}
   $\ (I,Y)\ $  & $(\bm{27}_q,\bm{\overline{35}}_q)$ & 
   $(\bm{27}_q,\bm{64}_q)$ & $(\bm{35},\bm{35})$ &
   $(\bm{35},\bm{64}_q)$ & 
   $(\bm{35},\bm{81})$ 
   \\ \hline
   $(\frac{3}{2},+1)$ & 
   $\frac{\sqrt{5}}{2\sqrt{7}}$ & $-\frac{5}{112}$ &
   $\frac{{\sqrt{3}}}{32}$ &  $\frac{3\sqrt{21}}{16\sqrt{5}}$ & 
   $-\frac{3\sqrt{3}}{4\sqrt{35}}$   \\
   $(1, 0)$ &
   $\frac{5}{4 {\sqrt{21}}}$  & $-\frac{5 \sqrt{5}}{112\sqrt{3}}$ &
   $\frac{3 {\sqrt{3}}}{16}$ &  $\frac{{\sqrt{21}}}{16}$ & 
   $-\frac{3\sqrt{3}}{8\sqrt{7}}$ 
   \\
   $(\half,-1)$ &
   $0$ & $-\frac{5}{56 {\sqrt{2}}}$ &
   $\frac{11 {\sqrt{3}}}{32}$ &   $\frac{\sqrt{21}}{8\sqrt{10}}$ & 
   $-\frac{3}{2 {\sqrt{35}}}$ 
   \\
   $(0, -2)$ &
   $0$ & $0$ &$\frac{{\sqrt{3}}}{2}$ & 
   $0$ &  $-\frac{3 \sqrt{3}}{8\sqrt{7}}$ 
   \\
  \end{tabular}
 \end{ruledtabular}
\end{table}

Most of
them have never been given in the literature. (Incidentally, the matrix
elements diagonal in representation are given by similar formulae to
(\ref{opdiag}) given in Appendix~\ref{Sec:Math} with
\begin{eqnarray}
 {\cal O}_{DPP}^{(1)}&=&-\bfJ^{2}, \\
 {\cal O}_{DPP}^{(2)}&=&
  -\left( 
    \frac{1}{3}C_{3} + \frac{1}{4}C_{2} + \frac{1}{4}\bfJ^2 + \frac{3}{16}
   \right).
\end{eqnarray}
See Sec.~\ref{repdiagonal} for the notation.)

By using these matrix elements, we can calculate the baryon masses, and
by best fitting the calculated values to the observed ones, we can
determined the parameters $\alpha$, $\beta$, and $\gamma$ as well as
$M_{cl}$, $I_1$, and $I_2$.
The procedure is the same as that employed in Sec~\ref{Sec:Num} so that
we do not explain it again. 
The best fit set of parameters is
\begin{eqnarray}
 M_{cl}&=& 837\ \mbox{MeV},\ I_1^{-1}=163\ \mbox{MeV},\ 
  I_2^{-1}=394\ \mbox{MeV}, \nonumber \\
 \alpha&=&-554 \ \mbox{MeV},\ \beta=-40.9\ \mbox{MeV}, \ 
  \gamma=42.0 \ \mbox{MeV},
  \label{dpp:bestfit}
\end{eqnarray}
which leads to the masses given in Table~\ref{dpp:mass}, with $\chi^2 =
6.4\times 10^1$.

\begin{table}[h]
 \caption{\label{dpp:mass}Baryon masses calculated by using the best fit
 set of parameters (\ref{dpp:bestfit}) with the $\chi$QSM Hamiltonian.}
 \begin{ruledtabular}
  \begin{tabular}{c|cccccccccc}
   (MeV)&$\mbox{\rm N}$& $\Sigma$& $\Xi$& $\Lambda$& $\Delta$& ${\Sigma^*}$&
   ${\Xi^*}$& $\Omega$&$\Theta$& $\phi$\\
   \hline
   $M^{DPP}_i$ & 
   942 & 1206 & 1335 & 1116 & 1227 & 1383 & 1531 & 1672 & 1538 & 1868
  \end{tabular}
 \end{ruledtabular}
\end{table}
 
Considering the number of parameters is small, the fit is very good. It
is also remarkable that the parameters have expected magnitudes. The
best fit values predict the masses of the other members of pentaquarks,
\begin{equation}
M_{{\rm N}^{\prime}} =1668\ {\rm MeV},
 \quad M_{\Sigma^{\prime}}=1777\ {\rm MeV}.
 \label{dpp:massprediction}
\end{equation}
The results are not very different from the ones in (\ref{EKPprediction}).

\subsection{Decays}

Next we turn to the calculation of the decay widths. Again the procedure
is the same as in Sec~\ref{Sec:Widths}. Actually the necessary matrix
elements are the same. The only difference comes from the mixings.
The mixing coefficients which correspond to our results in
Tables~\ref{8mix}, \ref{10mix}, and \ref{b10mix} are given in
Tables~\ref{dpp8mix}, \ref{dpp10mix}, and \ref{dppb10mix}. 

\begin{table}[h!]
\caption{\label{dpp8mix} Mixing coefficients for the (mainly) octet
 states with the $\chi$QSM symmetry breaking terms. They correspond to
 those in Table~\ref{8mix}.}
 \begin{ruledtabular}
\begin{tabular}{c|cccc}
 $\quad{\cal R}_i\quad $ & $\mbox{N}$ & $\Sigma$ & $\Xi$ & $\Lambda$ 
 \\ \hline
 $\bm{8}$ & 
 $1$ $(-0.030)$ & $1$ $(-0.027)$ & $1$ $(-0.010)$ & $1$ $(-0.015)$ \\
 $\bm{\overline{10}}$ & 
 $0.202$ $(-0.023)$ & $0.202$ $(0.030)$ & $0$ $(0)$ & $0$ $(0)$ \\
 $\bm{27}_d$ & 
 $0.140$ $(0.003)$ & $0.114$ $(0.022)$ & $0.140$ $(0.020)$ & 
 $0.171$ $(0.008)$ \\
 $\bm{\overline{35}}_d$ & 
 $0$ $(0.022)$ & $0$ $(0.021)$ & $0$ $(0)$ & $0$ $(0)$ \\
 $\bm{64}_d$ & 
 $0$ $(0.010)$ & $0$ $(0.008)$ & $0$ $(0.010)$ & $0$ $(0.014)$ 
\end{tabular}
 \end{ruledtabular}
\end{table}

\begin{table}[h]
\caption{\label{dpp10mix} Mixing coefficients for the (mainly) decuplet
 states with the $\chi$QSM symmetry breaking terms. They correspond to
 those in Table~\ref{10mix}.}
 \begin{ruledtabular}
\begin{tabular}{c|cccc}
 $\quad{\cal R}_i\quad $ & $\Delta$ & $\Sigma^*$ & $\Xi^*$ & $\Omega$ 
 \\ \hline
 $\bm{10}$ & 
 $1$ $(-0.105)$ & $1$ $(-0.060)$ & $1$ $(-0.026)$ & $1$ $(-0.005)$ \\
 $\bm{27}_q$ & 
 $0.451$ $(0.005)$ & $0.329$ $(0.007)$ & $0.202$ $(0.006)$ & $0$ $(0)$ \\
 $\bm{35}$ & 
 $0.081$ $(0.022)$ & $0.103$ $(0.023)$ & $0.109$ $(0.019)$ & 
 $0.103$ $(0.013)$ \\
 $\bm{\overline{35}}_q$ & 
 $0$ $(0.058)$ & $0$ $(0.027)$ & $0$ $(0)$ & $0$ $(0)$ \\
 $\bm{64}_q$ & 
 $0$ $(0.026)$ & $0$ $(0.025)$ & $0$ $(0.017)$ & $0$ $(0)$ \\
 $\bm{81}$ & 
 $0$ $(0.004)$ & $0$ $(0.006)$ & $0$ $(0.007)$ & $0$ $(0.006)$
\end{tabular}
 \end{ruledtabular}
\end{table}

\begin{table}[h]
\caption{\label{dppb10mix} Mixing coefficients for the (mainly)
 anti-decuplet states with the $\chi$QSM symmetry breaking terms. They
 correspond to those in Table~\ref{b10mix}.}
 \begin{ruledtabular}
\begin{tabular}{c|cccc}
 $\quad{\cal R}_i\quad $ & $\Theta$ & $\mbox{N}'$ & $\Sigma'$ & $\phi$ 
 \\ \hline
 $\bm{8}$ & 
 $0$ $(0)$ & $-0.202$ $(-0.009)$ & $-0.202$ $(-0.049)$ & $0$ $(0)$ \\
 $\bm{\overline{10}}$ & 
 $1$ $(-0.009)$ & $1$ $(-0.036)$ & $1$ $(-0.044)$ & $1$ $(-0.033)$ \\
 $\bm{27}_d$ & 
 $0$ $(0)$ & $0.105$ $(-0.037)$ & $0.171$ $(-0.029)$ & $0.234$ $(0.005)$ \\
 $\bm{\overline{35}}_d$ & 
 $0.131$ $(0)$ & $0.139$ $(0.008)$ & $0.131$ $(0.016)$ 
 & $0.104$ $(0.019)$ \\
 $\bm{64}_d$ & 
 $0$ $(0)$ & $0$ $(0.012)$ & $0$ $(0.018)$ & $0$ $(0.019)$ \\
 $\bm{\overline{81}}_d$ & 
 $0$ $(0.009)$ & $0$ $(0.010)$ & $0$ $(0.009)$ & $0$ $(0.006)$
\end{tabular}
 \end{ruledtabular}
\end{table}

One can
easily see that the mixings are much smaller for the $\chi$QSM breaking
terms. From the goodness of the fit to the masses and the smallness of
the mixings, one may think that the perturbative treatment of the
symmetry breaking terms is justified for this model. 

With these mixing coefficients, the decay widths are readily calculated,
by using our width formula (\ref{width}). First we present the best
fitted values of $G_0$, $G_1$, and $G_2$ in Table~\ref{dppGa}.

\begin{table}[h]
 \caption{\label{dppGa}Coefficients of the decay operators for the
 $\chi$QSM symmetry breaking terms. The notation is the same as in
 Table~\ref{Ga}.}
 \begin{ruledtabular}
  \begin{tabular}{c|cccc}
   &$(a)$& $(a')$& $(b)$& $(b')$ \\ \hline
   $G_0$    & 5.33  & 4.11  & 3.38  & 3.95  \\
   $G_1$    & 9.98  & 13.7  & 21.4  & 17.5  \\
   $\chi^2$ & 43.1  & 14.6  & 23.9  & 36.0  \\ \hline
   $G_2$    & 0.08  & 0.06  & 0.04  & 0.05
 \end{tabular}
 \end{ruledtabular}
\end{table}

These values lead to the decay widths for the (mainly)
decuplet baryons given in Table~\ref{dppdecay10}.

\begin{table}[h]
 \caption{\label{dppdecay10}Decay widths for the (mainly) decuplet
 baryons with the coupling constants in Table~\ref{dppGa}.}
\begin{ruledtabular}
 \begin{tabular}{c||ccccc}
  (MeV)& $K$ & 
  $\Gamma_{(a)}$ & 
  $\Gamma_{(a')}$ &
  $\Gamma_{(b)}$ &
  $\Gamma_{(b')}$ 
  \\ \hline
  $\Delta\rightarrow {\rm N}\pi$    & 1.47 & 90.2 & 106  & 105  & 95.8 \\
  $\Sigma^*\rightarrow \Lambda \pi$ & 1.18 & 34.1 & 33.1 & 32.9 & 33.5 \\
  $\Sigma^*\rightarrow \Sigma \pi$  & 0.26 & 4.69 & 5.85 & 6.63 & 6.19 \\
  $\Xi^*\rightarrow \Xi \pi$        & 0.49 & 12.9 & 12.7 & 13.8 & 13.7
 \end{tabular}
\end{ruledtabular}
\end{table}

Finally, we obtain the various decay widths for the (mainly)
anti-decuplet baryons, which are summarized in
Table~\ref{dppdecay10bar}.

\begin{table}[h]
 \caption{\label{dppdecay10bar}Predictions for the decay widths for the 
 (mainly) anti-decuplet  baryons with the $\chi$QSM symmetry breaking terms.}
\begin{ruledtabular}
 \begin{tabular}{c||ccccc}
  (MeV)& $K$ & 
  $\Gamma_{(a)}$ & 
  $\Gamma_{(a')}$ &
  $\Gamma_{(b)}$ &
  $\Gamma_{(b')}$ 
  \\ \hline
  $\Theta^+\rightarrow {\rm N} {\rm K}$ & 1.91 & 63.0 & 250  & 727  & 429 \\
  ${\rm N}'\rightarrow {\rm N}\pi$      & 18.2 & 210  & 669  & 1793 & 1030\\
  ${\rm N}'\rightarrow {\rm N}\eta$     & 4.82 &-0.30 & 26.2 & 104  & 56.9\\
  ${\rm N}'\rightarrow \Delta \pi$      & 5.69 & 0    & 132  & 242  & 259 \\
  ${\rm N}'\rightarrow \Lambda {\rm K}$ & 0.86 & 2.19 & 14.7 & 48.1 & 26.7\\
  ${\rm N}'\rightarrow \Sigma {\rm K}$  & ---  & ---  & ---  & ---  & ---\\
  $\Sigma'\rightarrow {\rm N} {\rm K}$  & 12.5 &-21.5 & 8.47 & 102  & 36.8\\
  $\Sigma'\rightarrow \Sigma \pi$       & 12.2 & 101  & 346  & 1175 & 676 \\
  $\Sigma'\rightarrow \Sigma \eta$      & 0.50 & 0.40 & 4.63 & 14.1 & 7.79\\
  $\Sigma'\rightarrow \Lambda \pi$      & 16.0 & 93.0 & 328  & 1030 & 619 \\
  $\Sigma'\rightarrow \Xi {\rm K}$      & ---  & ---  & ---  & ---  & ---\\
  $\Sigma'\rightarrow \Sigma^*\pi$      & 4.40 & 0    & 4.01 & 15.0 & 5.37\\
  $\Sigma'\rightarrow \Delta {\rm K}$   & 0.74 & 0    & 13.8 & 20.9 & 17.1\\
  $\phi\rightarrow \Sigma {\rm K}$      & 5.11 &-17.7 & 25.1 & 5.83 & 35.8\\ 
  $\phi\rightarrow \Xi \pi$             & 10.8 & 72.7 & 328  & 1095 & 614 \\ 
  $\phi\rightarrow \Xi^* \pi$           & 2.68 & 0    & 0.91 & 5.37 & 2.32
 \end{tabular}
\end{ruledtabular}
\end{table}

Despite the good perturbative behavior in the baryon mass fitting, the
decay widths vary considerably. In most cases, the second order results
are very different from the first order ones. Even though perturbation
theory seems to work good, some negative values appear when we expand
the amplitudes. In particular, the width of $\Theta^{+}$ is predicted
much larger than the experimental values.

\begin{acknowledgments}
 We would like to thank K.~Inoue for discussions on group theoretical
 problems.  One of the author (K.~H.) would like to thank H.~Yabu for
 the discussions on several aspects of the Skyrme model. He is also
 grateful to T.~Kunihiro for informing him of Ref.~\cite{Kindo:1987dc}.
 This work is partially supported by Grant-in-Aid for Scientific
 Research on Priority Area, Number of Area 763, ``Dynamics of Strings
 and Fields,'' from the Ministry of Education, Culture, Sports, Science
 and Technology, Japan.
\end{acknowledgments}



\end{document}